\documentclass[aps,prc,twocolumn,superscriptaddress]{revtex4}

\usepackage{amsmath,amssymb,graphicx}
\usepackage{url}
\usepackage[colorlinks,urlcolor=blue]{hyperref}
\usepackage{graphicx,color}
\allowdisplaybreaks

\newcommand{\ba}{\begin{eqnarray}}
\newcommand{\ea}{\end{eqnarray}}

\def\lsim{\,\raise0.3ex\hbox{$<$\kern-0.75em\raise-1.1ex\hbox{$\sim$}}\,}
\def\gsim{\,\raise0.3ex\hbox{$>$\kern-0.75em\raise-1.1ex\hbox{$\sim$}}\,}

\begin{document}

\title{Linear Boltzmann transport for jet propagation in the quark-gluon plasma: Inelastic processes and jet modification}


\author{Tan Luo}
\email[]{tan.luo@usc.es}
\affiliation{Key Laboratory of Quark and Lepton Physics (MOE) and Institute of Particle Physics, Central China Normal University, Wuhan 430079, China}
\affiliation{Instituto Galego de F\'isica de Altas Enerx\'ias IGFAE, Universidade de Santiago de Compostela, E-15782 Galicia, Spain}

\author{Yayun He}
\email[]{yyhe@m.scnu.edu.cn}
\affiliation{Key Laboratory of Quark and Lepton Physics (MOE) and Institute of Particle Physics, Central China Normal University, Wuhan 430079, China}
\affiliation{Guangdong Provincial Key Laboratory of Nuclear Science, Institute of Quantum Matter, South China Normal University, Guangzhou 510006, China}
\affiliation{Guangdong-Hong Kong Joint Laboratory of Quantum Matter, Southern Nuclear Science Computing Center, South China Normal University, Guangzhou 510006, China}

\author{Shanshan Cao}
\email[]{shanshan.cao@sdu.edu.cn}
\affiliation{Institute of Frontier and Interdisciplinary Science, Shandong University, Qingdao, Shandong 266237, China}

\author{Xin-Nian Wang}
\email[]{xnwang@lbl.gov}
\thanks{present address is $^6$}
\affiliation{Key Laboratory of Quark and Lepton Physics (MOE) and Institute of Particle Physics, Central China Normal University, Wuhan 430079, China}
\affiliation{Nuclear Science Division Mailstop 70R0319,  Lawrence Berkeley National Laboratory, Berkeley, CA 94740}

\begin{abstract}
A Linear Boltzmann Transport (LBT) Monte Carlo model has been developed to describe jet propagation and interaction with the quark-gluon plasma (QGP) in relativistic heavy-ion collisions. A complete set of elastic scattering processes and medium-induced gluon emissions based on the higher-twist formalism are incorporated for both jet shower and medium recoil partons. It has been employed to describe experimental data on large transverse momentum hadron and jet spectra, correlation and jet substructures in high-energy heavy-ion collisions. We document in detail the structure of the model and validation of the Monte Carlo implementations of the physics processes in LBT, in particular, the inelastic process of medium-induced gluon radiation. We carry out a comprehensive examination of the jet-medium interaction as implemented in LBT through energy loss and momentum broadening of a single hard parton, the energy and transverse momentum transfer from leading partons to medium-induced gluons and jet-induced medium excitation, and medium modification of reconstructed jets in a static and uniform medium. With realistic and event-by-event hydrodynamic medium in heavy-ion collisions, we compute and compare to experimental data on the jet cone-size dependence of  the single inclusive jet suppression at both Relativistic Heavy-Ion Collider (RHIC) and the Large Hadron Collider (LHC), the dijet asymmetry at LHC and $\gamma$-jet correlation at RHIC.  Effects of medium-induced gluon emissions and jet-induced medium excitation on jet observables are systematically examined. Rescatterings of the radiated gluons and recoil partons with the QGP are found essential to account for the enhancement of soft particle yield toward the edge of the jet cone. 
\end{abstract}


\maketitle

\section{Introduction}

Exploring properties of the color deconfined state of nuclear matter, known as the quark-gluon plasma (QGP), is the central goal of relativistic heavy-ion collision experiments at the Relativistic Heavy-Ion Collider (RHIC) and the CERN Large Hadron Collider (LHC)~\cite{Shuryak:2014zxa,STAR:2005gfr,PHENIX:2004vcz,Muller:2012zq}. Among various probes of the nuclear matter, jet quenching reveals the fine structure of the QGP at  short distance scales, since the momentum transfer between energetic jet partons and the medium can be much larger than the thermal scale given by the local temperature~\cite{Akiba:2015jwa,Mehtar-Tani:2011hma,Armesto:2011ir,Casalderrey-Solana:2011ule,Kumar:2019uvu}. Many observables have been proposed to study how jets are modified  by the QGP as they traverse the hot and dense medium,  from the nuclear modification factor of high transverse momentum ($p_\mathrm{T}$) hadrons and fully reconstructed jets, their anisotropic flow coefficients, to the inner structures of jets and jet-related correlations~\cite{Connors:2017ptx,Wang:1992qdg,Majumder:2010qh,Qin:2015srf,Blaizot:2015lma,Cao:2020wlm,Cao:2022odi}. Over the past three decades, studies on jets have transitioned from understanding their interaction dynamics with the hot subatomic matter to utilizing them to extract transport properties of the QGP~\cite{JET:2013cls,JETSCAPE:2021ehl,Xie:2022ght} with precision. Related investigations have also been extended from ultrarelativistic heavy-ion collisions to smaller collisions systems~\cite{Park:2016jap,Zhao:2020wcd,Liu:2021izt,Li:2021xbd} and collisions at lower center-of-mass energies~\cite{STAR:2017ieb,Wu:2022vbu}.

As jet partons plough through the QGP medium, they endure both elastic~\cite{Bjorken:1982tu,Braaten:1991we,Djordjevic:2006tw,Qin:2007rn} and inelastic~\cite{Baier:1996kr,Zakharov:1996fv,Gyulassy:1999zd,Wiedemann:2000za,Arnold:2002ja,Wang:2001ifa} scatterings with the medium. These interactions not only lead to nuclear modification of jets, but also energy-momentum deposition to the medium and depletion from the medium due to back-reaction. The latter two are collectively known as ``jet-induced medium excitation" or ``medium response", and are essential in understanding how the lost energy from jet partons are redistributed and thermalize. It has now been widely accepted that jet-induced medium excitation is responsible for some of the enhanced soft hadron production associated with jets~\cite{Chen:2017zte,Chen:2020tbl}, the increased jet energy within a given circular annulus at large angle with respect to the jet axis~\cite{Casalderrey-Solana:2016jvj,Tachibana:2017syd,KunnawalkamElayavalli:2017hxo,Luo:2018pto,Park:2018acg}, and recovering the momentum balance between the two hemispheres traversed by the leading and subleading jets in a dijet event~\cite{CMS:2011iwn,Casalderrey-Solana:2016jvj}. Unique signatures of medium response, such as the diffusion wake in the opposite direction to jet propagation~\cite{Yang:2021qtl,Yang:2022nei} and the enhanced baryon-to-meson ratio within and around the jet cone~\cite{Chen:2021rrp,Luo:2021voy,Sirimanna:2022zje} have been proposed and wait for confirmation from more precise experimental measurements. 

Measuring medium response is challenging because its signals related to soft hadrons are usually buried in the huge background from QGP hadronization. Despite delicate background subtraction techniques developed in heavy-ion experiments, jet constituents (shower partons including from medium-induced radiation) and medium response still inevitably overlap each other in reality. Therefore, seeking signatures of medium response relies on close comparisons between realistic model calculations and experimental data. Sophisticated event generators based on different model assumptions have been developed to simulate jet interactions with the QGP, such as the virtuality-ordered medium-modified splittings of highly virtual partons~\cite{Armesto:2009fj,Zapp:2011ya,Casalderrey-Solana:2014bpa,Cao:2017qpx}, the time-ordered transport descriptions of elastic and inelastic scatterings of on-shell partons through the QGP~\cite{Schenke:2009gb,Wang:2013cia}, and their combination according to the virtuality scale of each parton~\cite{JETSCAPE:2017eso,JETSCAPE:2022jer}. Different treatments of medium response are also implemented in these models, including perturbative approximation using recoil partons~\cite{Zapp:2011ya,He:2022evt}, hydrodynamic response to energy deposition~\cite{Tachibana:2020mtb}, and a concurrent simulation of jet and QGP evolution~\cite{Chen:2017zte}. 

The linear Boltzmann transport (LBT) model~\cite{He:2015pra,Cao:2017hhk} has been developed to describe jet propagation in QGP medium and to study the nuclear modification of jets and jet-induced medium excitation. It incorporates perturbative descriptions of both elastic and inelastic scatterings between hard and medium partons. The guiding principle of LBT is the treatment on the equal footing of the  offsprings of jet partons and the thermal partons scattered out of the QGP background which are known as ``recoil" partons. The production of these recoil partons is also accompanied by particles holes left behind inside the QGP, known as back-reaction or ``negative" partons. Recoil and ``negative" partons constitute the jet-induced medium excitation within the picture of linear response. This LBT model provides a satisfactory description of the nuclear modification of high $p_\mathrm{T}$ hadrons~\cite{Xing:2019xae}, single inclusive jets~\cite{He:2018xjv,He:2022evt} and $\gamma$-triggered jets~\cite{Luo:2018pto}, and has been embedded inside the JETSCAPE framework~\cite{Putschke:2019yrg} as a module of jet evolution at low virtuality. As a follow-up work of Ref.~\cite{He:2015pra},  we document a detailed description of the model structure and Monte Carlo implementations in LBT and their validation in this paper. In particular, we focus on how the medium-induced gluon bremsstrahlung process is modeled and how it affects the phase space structure of jet and thermal partons. Jet energy loss and momentum broadening are systematically studied in a static and uniform QGP medium. As a supplement to our earlier publications, additional jet observables in heavy-ion collisions with hydrodynamical  QGP background and event-by-event fluctuating initial condition will be presented and compared to data at RHIC and LHC.  These include
the nuclear modification of single inclusive jets, $\gamma$-triggered jets and dijet asymmetry, and their jet-cone size dependence.

The remainder of this paper is organized as follows.  In Sec.~\ref{sec:LBT}, we first present the structure and numerical implementations of elastic and inelastic scatterings between hard/recoil partons and the QGP within the LBT model. In Sec.~\ref{sec:inel}, we discuss in detail the inelastic scattering rate and the medium-induced gluon spectrum implemented in LBT. We then validate this model setup by numerically simulating the energy and angular distributions of hard partons after both single and multiple scatterings inside a static medium in Sec.~\ref{sec:partonInBrick}, including their dependences on the parton energy, medium temperature and energy loss mechanism. In Sec.~\ref{sec:response}, we analyze the spacetime evolution of jet-induced medium excitation in the static medium, such as the angular and momentum distributions of medium partons excited by jets. The energy loss of fully reconstructed jets, and the longitudinal and transverse momentum distributions of the jet constituents are then investigated within a static medium in Sec.~\ref{sec:jetInBrick}. We combine the LBT model with a hydrodynamic simulation of the QGP evolution in Sec.~\ref{sec:resultQGP}, and calculate the nuclear modification of both single inclusive jets, $\gamma$-triggered jets and dijets in realistic heavy-ion collisions at RHIC and LHC energies. The release of the LBT code package is presented in Sec.~\ref{sec:scode}. A summary is given in Sec.~\ref{sec:summary}.

\section{The linear Boltzmann transport model}
\label{sec:LBT}

The evolution of the phase space distribution of an incoming jet parton (denoted by ``$a$") inside a thermal medium is described in LBT according to the Boltzmann equation,
\begin{eqnarray}
  \label{eq:boltzmann1}
  p\cdot\partial f_a(x,p)=E\, \mathcal{C}_\mathrm{el+inel},
\end{eqnarray}
where the four-momentum of $a$ is given by $p^\mu=(E,\vec{p})$, and the $2 \rightarrow 2+n$ scattering processes denoted by the collision integral $\mathcal{C}_\mathrm{el+inel}$ can be decomposed into a sequence of $2 \rightarrow 2$ elastic scatterings and their induced gluon emissions. 

The elastic processes $ab\rightarrow cd$ can be simulated according to the scattering rate,
\begin{eqnarray}
 \label{eq:rate2}
 \Gamma_\mathrm{el}^{ab\rightarrow cd} &=&\frac{\gamma_b}{2E}\int \frac{d^3 p_b}{(2\pi)^3 2E_b}\int\frac{d^3 p_c}{(2\pi)^3 2E_c}\int\frac{d^3 p_d}{(2\pi)^3 2E_d}\nonumber\\
&\times& f_b(\vec{p}_b)\left[1\pm f_c(\vec{p}_c) \right]\left[1\pm f_d(\vec{p}_d)\right] S_2(\hat s,\hat t, \hat u)\nonumber\\
&\times& (2\pi)^4\delta^{(4)}(p+p_b-p_c-p_d)|\mathcal{M}_{ab\rightarrow cd}|^2,
\end{eqnarray}
where $b$ represents a thermal parton inside the medium, with a spin-color degeneracy $\gamma_b$ (6 for quarks and 16 for gluons), $c$ and $d$ are the final state particles of $a+b$ scattering. The distribution function of a given particle $i$ is denoted by $f_i(\vec{p}_i)$. The thermal distributions for $b$, $c$ and $d$ in Eq.~(\ref{eq:rate2}) are $f_i=1/(e^{p\cdot u/T}\mp 1)$, with ``$-$" sign for gluons and ``$+$" for quarks. The local temperature $T$ and the fluid velocity $u^\mu$ are provided by the hydrodynamic simulation of the QGP. 

In our current study, the initial-state-averaged and final-state-summed (over spin and color degeneracies) matrix elements $|\mathcal{M}_{ab\rightarrow cb }|^2$ are taken from perturbative QCD (pQCD) calculations at the leading order for all possible scattering channels~\cite{Auvinen:2009qm}. The corresponding elastic cross sections are $d\sigma_{ab\rightarrow cd}/d\hat t=|{\cal M}_{ab\rightarrow cd}|^2/(16\pi \hat s^2)$, with $s,t,u$ being the Mandelstam variables. Light flavor partons are assumed to be massless, and bare masses for heavy quarks are $m_c=1.27$~GeV for charm and $m_b=4.19$~GeV for beauty quarks. For scatterings between massless partons, $|\mathcal{M}_{ab\rightarrow cd}|^2$ diverges at small angle ($\hat u, \hat t\rightarrow 0$). This is regulated by a double-$\theta$ function 
\begin{eqnarray}
 \label{eq:S2}
 S_2(\hat s,\hat t, \hat u)=\theta(\hat s\ge2\mu_\mathrm{D}^2)\theta(-\hat s+\mu_\mathrm{D}^2\le \hat t\le -\mu_\mathrm{D}^2),
\end{eqnarray}
where  
\begin{equation}
    \mu_\mathrm{D}^2=(N_c+\frac{N_f}{2})\frac{g^2T^2}{3} 
\end{equation}
is the Debye screening mass, with $g^2=4\pi \alpha_\mathrm{s}$ being the strong coupling constant, $N_c=3$ and $N_f=3$ being the numbers of colors and flavors, respectively. 

The total elastic scattering rate for parton $a$ is
\begin{equation}
\Gamma^a_{\rm el}\equiv \sum_{b,(cd)} \Gamma_{\rm el}^{ab\rightarrow cd},
\label{eq-rate}
\end{equation}
where the summation is over all possible scattering channels $a+b\rightarrow c+d$.

\begin{figure}[!tbp]
\includegraphics[width=7.5cm]{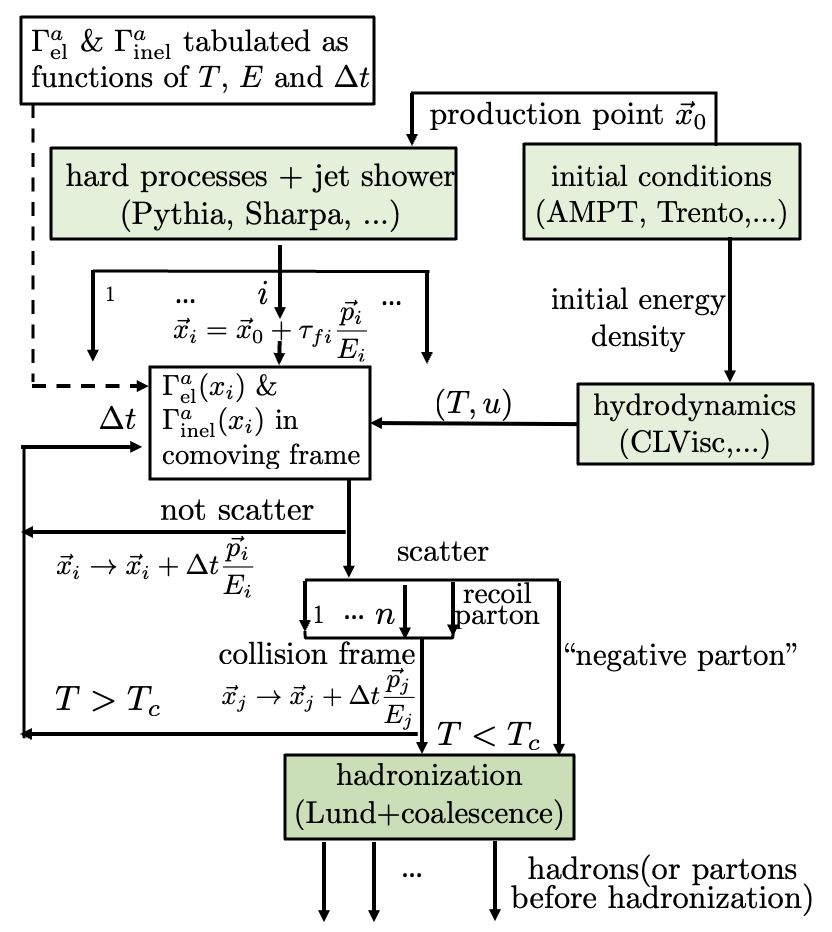}\\
\vspace{0.2cm}
\caption{(Color online) Flow chart of the LBT Monte Carlo model. The green shaded boxes represent modules that can be replaced by users with alternatives.  
}
\label{fig:chart}
\end{figure}

For inelastic scattering, we relate its rate to the medium-induced gluon spectrum as
\begin{equation}
\label{eq:rateInel}
\Gamma^a_\mathrm{inel}=\int dzdk_\perp^2 \frac{1}{1+\delta^{ag}}\frac{dN^a_g}{dzdk_\perp^2 dt},
\end{equation}
with the gluon spectrum taken from the higher-twist energy loss calculation~\cite{Guo:2000nz,Zhang:2003wk,Majumder:2009ge},
\begin{eqnarray}
\label{eq:gluondistribution}
\frac{dN_g^a}{dz dk_\perp^2 dt}=\frac{2\alpha_\mathrm{s} C_A \hat{q}_a(x) P_a(z)k_\perp^4}{\pi \left({k_\perp^2+z^2 m^2}\right)^4} \, {\sin}^2\left(\frac{t-t_i}{2\tau_f}\right),
\end{eqnarray}
where $z$ and $k_\perp$ are the fractional energy and the transverse momentum of the emitted gluon with respect to its parent jet parton, $\alpha_\mathrm{s}$ is the strong coupling constant, $C_A=N_c=3$, $m$ is the quark mass, $t_i$ is the production time of the jet parton or the last time when the parton has inelastic scattering with gluon radiation, and
\begin{equation}
    \tau_f=\frac{2Ez(1-z)}{k_\perp^2+z^2m^2}
    \label{eq:formationtime}
\end{equation}
is the formation time of the emitted gluon. The medium induced splitting functions $P_a(z)$ for $a+g({\rm medium}) \rightarrow a+g$ processes  are  \cite{Zhang:2003yn,Schafer:2007xh}, 
\begin{eqnarray}
P_q(z)&=&\frac{(1-z)[1+(1-z)^2]}{z},\\
P_g(z) &=&2\frac{(1-z+z^2)^3}{z(1-z)}.
\end{eqnarray}
To avoid divergence at $z\rightarrow 0$ in evaluating the inelastic scattering rate, a lower cut-off on the fractional energy $z_\mathrm{min}=\mu_\mathrm{D}/E$ is applied to the emitted gluon. In Eq.~(\ref{eq:rateInel}), a 1/2 factor is applied when we evaluate the scattering rate from the gluon spectrum if the parent jet parton is also a gluon. Note that in Eq.~(\ref{eq:gluondistribution}), the time (or path length) dependence of the gluon spectrum results from the interference between gluon radiation amplitudes from different scatterings. It was originally derived within the higher-twist formalism to take into account the interference between the radiation amplitude induced by the hard scattering (that generates the jet parton) and that induced by the secondary scattering~\cite{Guo:2000nz}. It is applied to multiple scatterings in the QGP medium in LBT and takes into account the inferences between gluon emissions induced by different scatterings. This will be referred to as the Landau–Pomeranchuk–Migdal (LPM) interference. Since not every scattering can induce a gluon emission, the radiation probability according to Eq.~(\ref{eq:gluondistribution}) for the current scattering depends on the distance since the last inelastic scattering. If induced gluon radiation does not happen for the current scattering, this distance for the propagating parton will be updated in the calculation of the radiation probability according to Eq.~(\ref{eq:gluondistribution}) for the next scattering as illustrated in the flow chart of LBT in Fig.~\ref{fig:chart}.

The jet quenching coefficient $\hat{q}_a(x)$ characterizes the transverse momentum broadening squared of the jet parton $a$ per unit length,  $\hat q_a =d\langle q^2_\perp\rangle_a/dt$, and can be evaluated using Eq.~(\ref{eq:rate2}) with a weight of $q_\perp^2=[\,\vec{p}_c-(\vec{p}_c\cdot\hat{\vec p})\hat{\vec p}\,]^2$ on the right-hand side. Here $\hat{\vec p}$ is the unit vector along parton $a$'s momentum before scattering. If the quantum statistics in the final state of $2\rightarrow 2$ scattering is neglected, the elastic scattering rate and jet transport coefficient can be expressed in terms of the elastic scattering cross sections,
\begin{eqnarray}
&&\Gamma_{\rm el}^{ab\rightarrow cd}(x) =\rho_b(x) \sigma_{ab\rightarrow cd},\\
&&\hat q_a(x)=\sum_{b,(cd)} \rho_b(x) \int d\hat t  q_\perp^2 \frac{d\sigma_{ab\rightarrow cd}}{d\hat t}.
\end{eqnarray}
Since pQCD $2\rightarrow 2$ processes are dominated by small angle $t$-channel processes, the elastic scattering rate, jet transport coefficient and the inelastic rate are all proportional to the 
quadratic Casimir $C_2(a)$ of the parton. They are a factor $C_A/C_F=9/4$ larger for a gluon than a quark. In the LBT model, we use Eq.~(2) to evaluate the elastic scattering rate and the jet transport coefficient $\hat q$ which can be approximated well by Eq.~(12) as shown in Ref.~\cite{He:2015pra}.

 The above elastic and inelastic scattering rates are the basic building blocks of the LBT model. They are computed and tabulated as functions of the local temperature $T$, parton energy $E$ and the propagation time $t-t_i$. This is done as a part of the initialization of the LBT model and provided as an input data file as a part of the LBT code. The structure of LBT built on top of these scattering rates are illustrated in Fig.~\ref{fig:chart}.  The default modules in green shaded boxes  can be replaced by alternatives supplied by users and the final hadronization module can also be customized.  Models such as the AMPT \cite{Lin:2004en} or Trento \cite{Moreland:2014oya} model provide the initial conditions for both the hydrodynamic evolution of the bulk medium and spatial probability distribution of hard binary collisions $\langle N_{\rm binary}\rangle(\vec x_{0 \perp })$ in the transverse plane. The default hydrodynamic model used in the past LBT and this study is the CLVisc model and can be replaced by users with other alternatives. The hydrodynamic model with the given initial condition provides the spatial and time evolution of the QGP medium: temperature $T$ and flow velocity $u$ as functions of spatial coordinates and time in the collision frame. Initial hard processes and jet showering are simulated with Monte Carlo models such as Pythia~\cite{Sjostrand:2007gs} and Sharpa~\cite{Gleisberg:2008ta} with their initial production points $\vec x_0$ sampled from the binary collision distribution $\langle N_{\rm binary}\rangle(\vec x_{0 \perp})$ in the transverse plane. The initial longitudinal position $x_{0 3}$ is set to 0 for hard processes in high energy heavy-ion collisions. 
 
 For each jet shower parton with momentum $\vec p_i$ and energy $E_i$ (the less energetic daughter from splitting) along the shower tree of the hard jet parton with initial momentum $\vec P$ and energy $E$, one can assign a formation time,
 \begin{equation}
 \tau_{fi}=\frac{2E}{P_{\rm T}^2}+\sum_{j=1}^i\frac{2E_j}{\left[\vec p_j - (\vec p_j\cdot \hat{\vec P}) \hat{\vec P}\right]^2},
 \label{eq:formation-time}
\end{equation}
where the summation is over all prior splittings along the branching tree. The formation time for the leading parton in the final splitting of the tree is assumed to be the same as the less energetic daughter. This parton is allowed to interact with the medium only after this formation time starting at the position $\vec x_i=\vec x_0+\tau_{fi}\vec p_i/E_i$.

The propagation and interaction of jet shower partons in the medium are simulated in small incremental time steps. For each time step $\Delta t$ in the collision frame, one first gets the elastic and inelastic scattering rates from the data table at the local temperature $T$ for parton's energy $E_i^\prime$ in the local comoving frame of the QGP. 
 
For small time step $\Delta t$ in the global computational frame, the average number of elastic and inelastic scatterings during the time interval are then given by 
\begin{eqnarray}
\langle \Delta N^a_\mathrm{el} \rangle(\vec x_i)&=&\Gamma^a_\mathrm{el}(\vec x_i)\Delta t',\\
\langle \Delta N^a_\mathrm{inel} \rangle(\vec x_i) &=&\Gamma^a_\mathrm{inel} (\vec x_i)\Delta t',
\end{eqnarray}
respectively, where $\Delta t'=\Delta t\,p_i\cdot u/E_i$ is the time interval in the local comoving frame. The probabilities for elastic and inelastic scattering in this time interval are
\begin{eqnarray}
P^a_\mathrm{el} &=&1-e^{-\langle \Delta N^a_\mathrm{el} \rangle(\vec x_i)},\\
P^a_\mathrm{inel} &=&1-e^{-\langle \Delta N^a_\mathrm{inel} \rangle(\vec x_i)}.
\end{eqnarray}
These scattering probabilities become $\Gamma_\mathrm{el}^a\Delta t'\ll 1$ and $\Gamma_\mathrm{inel}^a\Delta t'\ll 1$,  if $\Delta t'$ is  sufficiently small. 
To combine elastic and inelastic processes, the total scattering rate is given by $\Gamma_\mathrm{tot}^a(\vec x_i)=\Gamma_\mathrm{el}^a(\vec x_i)+\Gamma_\mathrm{inel}^a(\vec x_i)$, and the total number of scattering,  elastic and inelastic, is
\begin{equation}
\langle \Delta N^a_\mathrm{tot} \rangle(\vec x_i)=\Gamma_{\rm tot}^a(\vec x_i)\Delta t'.
\end{equation}
The total scattering probability is,
\begin{equation}
\label{eq:totP}
P_\mathrm{tot}^a=1-e^{-\langle \Delta N^a_\mathrm{tot} \rangle(\vec x_i)}=P^a_\mathrm{el} (1-P^a_\mathrm{inel})+P^a_\mathrm{inel},
\end{equation}
which can be understood as a sum of the probability of pure elastic scattering process without gluon emission (first term) and the probability of inelastic scattering with at lest one gluon emission (the second term). 

Using these probabilities within a given time step $\Delta t$, we first decide whether a given jet parton scatters with the medium thermal partons. If it does not scatter, it will propagate in the medium for the next time step $\vec x_i\rightarrow \vec x_i+\Delta t\, \vec p_i/E_i$ in the collision frame. The scattering rates at the new position will be used to determine the probabilities for scattering. Such propagation is repeated until a scattering occurs.

We then decide whether this scattering is elastic or inelastic.  If an elastic scattering occurs based on this probability, the branching ratio $\langle \Delta N_\mathrm{el}^{ab\rightarrow cd}\rangle/\langle \Delta N_\mathrm{el}^{a}\rangle$ is used to select a particular scattering channel, i.e., the species of $b$, $c$ and $d$. With a given channel, the differential scattering rate based on Eq.~(\ref{eq:rate2}) is then applied to determine the momenta of $b$, $c$ and $d$. For light flavor jets, we assign the parton between $c$ and $d$ with higher energy as the final state jet parton, while the other as the ``recoil parton" scattered out of the medium background. Meanwhile, $b$ represents the energy-momentum depletion from the medium, denoted as ``back-reaction” or ``negative parton” in LBT. Their propagation and interaction with the medium will be simulated in subsequent time steps according to the above procedure in the LBT model in order to provide a complete picture of jet-medium interaction and guarantee the energy-momentum conservation of the entire system.

If an inelastic scattering happens with a given number of time steps based on the given probability, we sample the number of emitted gluons using the Poisson distribution with a mean number of emissions  $\langle \Delta N^a_\mathrm{inel} \rangle$. We then assign the energy-momentum of each gluon using its differential spectrum given by Eq.~(\ref{eq:gluondistribution}). Note that the time $t_i$ in Eq.~(\ref{eq:gluondistribution}) is the time of the last inelastic scattering. Since the emitted gluons are induced by elastic scattering within our framework, we decompose each inelastic scattering ($2\rightarrow2+n$) into an elastic scattering ($2\rightarrow2$) plus a multiple gluon splitting ($1\rightarrow n$) process. In addition,  the final state jet parton directly sampled from Eq.~(\ref{eq:rate2}) is on-shell and is not able to split, we need to adjust the energy-momenta of the $2+n$ final state particles obtained from Eqs.~(\ref{eq:rate2}) and~(\ref{eq:gluondistribution}) together to ensure the energy-momentum conservation of the entire $2\rightarrow2+n$ process. Details on this adjustment are presented in Appendix~\ref{sec:appendix}. After each inelastic scattering the initial time $t_i$ in Eq.~(\ref{eq:gluondistribution}) is reset. Note that in each time step, we determine the momenta of final jet partons, radiated gluons, recoil and ``negative" partons in the local comoving frame first before boosting them back into the global collision frame. All these partons will continue to propagate and interact with the medium according to the above procedure until the local temperature reaches the QCD phase transition temperature $T<T_c$ according to the equation of state (EoS) used in the hydrodynamic model. In the current version of LBT model, parton-medium interaction below $T_\mathrm{c}$ is neglected. For the study of jets in heavy-ion collisions, we currently reconstruct jets with the final state partons using a jet finding algorithm. Since ``negative'' partons represent the depletion of energy and momentum in the medium due to the back-reaction, they should be subtracted from the jet energy within a given jet cone. In principle, the final jet shower partons, radiated gluons, recoil partons and  ``negative'' partons can be hadronized according to a hadronization model as done within the JETSCAPE framework~\cite{Putschke:2019yrg}. This is not included in the present work and needs to be implemented in our future effort.

In the following sections till the end of Sec.~VI, we will first validate the LBT model for the propagation of an on-shell single jet parton that enters a static medium at an initial time of $t_0=0$,  and develops into a jet that consists of recoil,  ``negative" partons and radiated gluons from elastic and inelastic scatterings. The formation time in Eq.~(13) does not apply to the initial single parton in these brick tests. Jet observables in realistic heavy-ion collisions will be discussed in Sec.~VII, in which initial jet shower partons are produced from Pythia simulations and start interacting with the QGP at their formation times

\section{The inelastic process}
\label{sec:inel}

\subsection{Inelastic scattering rate}
\label{subsec:rate}

The evolution of energetic partons in heavy-ion collisions is mainly governed by their scattering rates with the QGP. The elastic scattering rates for different scattering channels, including their energy and temperature dependences and the angular distribution of the recoil partons, were presented in  detail in an earlier publication~\cite{He:2015pra}. In this subsection, we concentrate on the medium-induced gluon spectra, which determine the rates for inelastic processes within LBT. 

\begin{figure}[htbp]
    \centering
    \includegraphics[width=7.5cm]{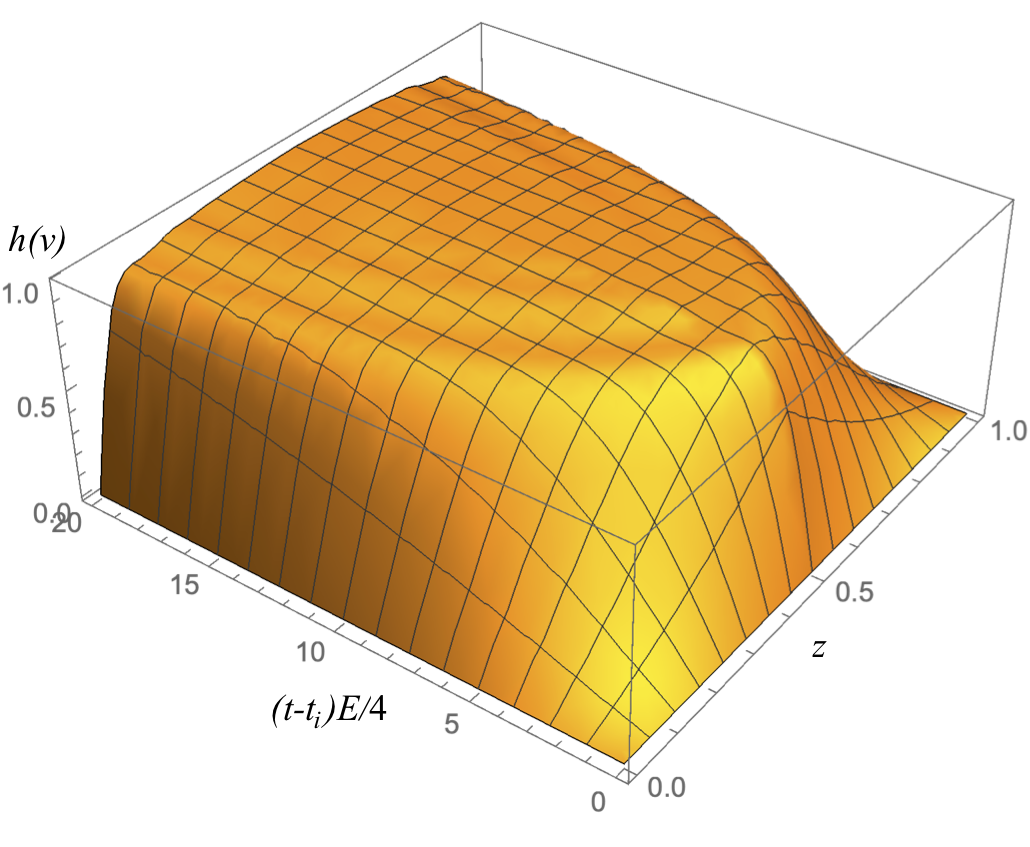}
    \caption{(Color online)  $h(v)$ defined in Eq.~(\ref{eq:funh}) as a function of fractional momentum $z$ and $(t-t_i)E$.}
    \label{fig:funh3D}
\end{figure}

\begin{figure}[htbp]
    \centering
    \includegraphics[width=7.5cm]{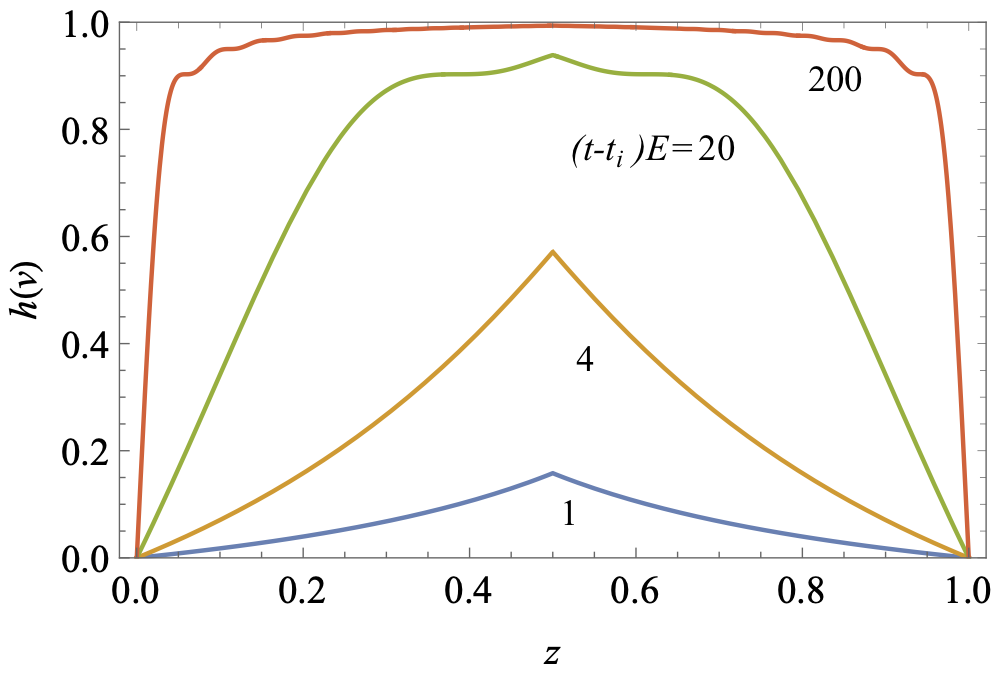}
    \caption{(Color online)  $h(v)$ defined in Eq.~(\ref{eq:funh}) as a function of fractional momentum $z$ for four different values of $(t-t_i)E$.}
    \label{fig:funh}
\end{figure}

For massless partons (light quarks and gluons), one can complete the integration over the transverse momentum of the radiated gluon in Eq.~(\ref{eq:gluondistribution}) and obtain the differential inelastic rate,
\begin{equation}
\frac{d\Gamma^a_{\rm inel}}{dz}=\frac{\alpha_\mathrm{s} C_A \hat{q}_a}{1+\delta^{ag}} \frac{(t-t_i)}{4 E}\frac{P_a(z)}{z(1-z)} h(v),
\label{eq:dgdz}
\end{equation}
where the kinetic limit for $k_\perp$ is set as $k_\perp\le \min(z,1-z)E$ and
\begin{eqnarray}
\label{eq:funh}
h(v)&=&\frac{2}{\pi}\int_0^v dx \frac{\sin^2 x}{x^2},\\ \nonumber
v&=&\frac{(t-t_i)E}{4z(1-z)}\min(z^2, (1-z)^2),
\end{eqnarray}
which is normalized with the limit $h(\infty)=1$.  Since
\begin{equation}
\left[\frac{P_a(z)}{z(1-z)}h(v) \right]_{z\rightarrow 0}=\frac{2}{\pi}\frac{(t-t_i)E}{2z},
\label{eq:splitt}
\end{equation}
the above differential inelastic rate has a logarithmic infrared divergence. We therefore impose an infrared cut-off $z\ge \mu_D/E$ when we calculate the integrated inelastic scattering rate $\Gamma^a_{\rm inel}$. Shown in Figs.~\ref{fig:funh3D} and \ref{fig:funh} are $h(v)$ as a function of $z$ and $(t-t_i)E/4$ and  $h(v)$ as a function of $z$ for four different values of $(t-t_i)E$, respectively. One can see that it approaches to its asymptotic value $h(\infty)=1$ when $(t-t_i)E$ is very large and the effect of the kinetic restriction on the transverse momentum $k_\perp\le z E$ is negligible. 

\begin{figure}[htbp]
    \centering
     \includegraphics[width=7.5cm]{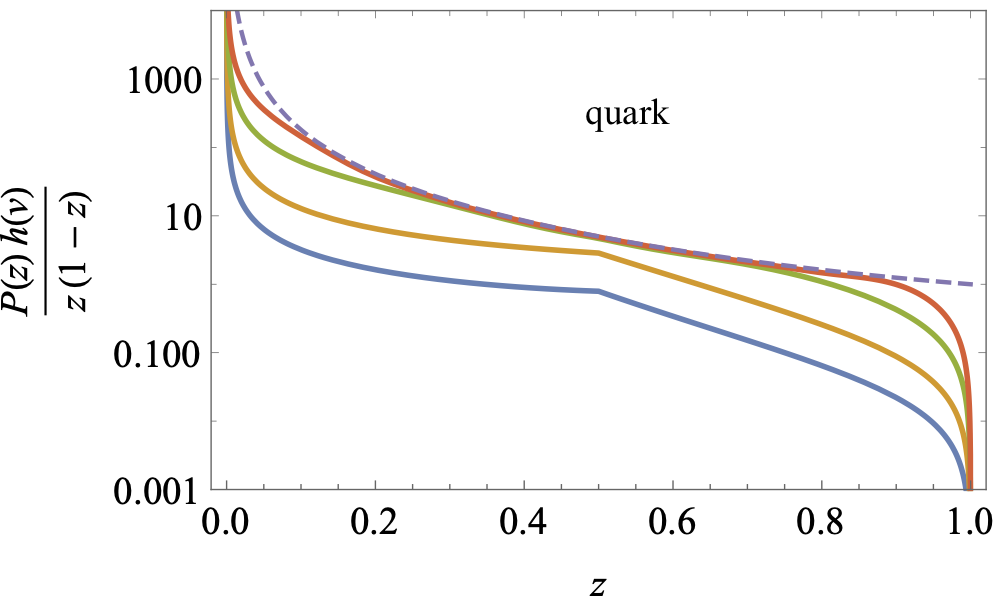}\\
     \vspace{-0.80cm}
    \includegraphics[width=7.5cm]{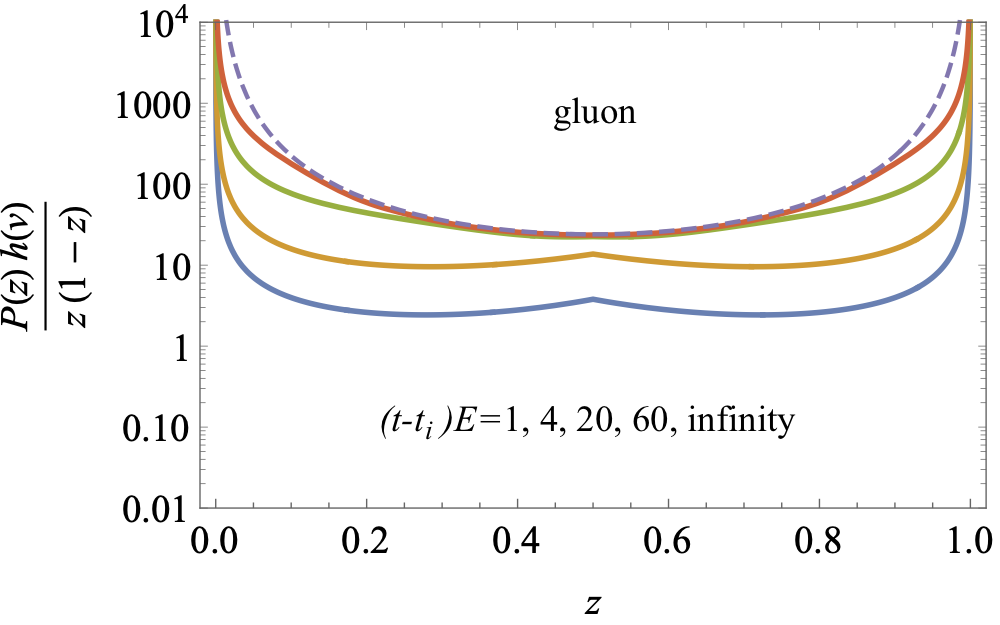}\\
    \caption{(Color online)  The effective medium-induced parton splitting function $P_a(z)h(v)/[z(1-z)]$ for a quark (upper panel) and a gluon (lower panel)  for four different values of $(t-t_i)E=1,4, 20, 60$ (from bottom to top) as compared to the asymptotic limit (dashed line) when $(t-t_i)E\rightarrow \infty$.}
    \label{fig:partonrate}
\end{figure}

The effective medium-induced parton splitting functions $P_a(z)h(v)/[z(1-z)]$ are shown in Fig.~\ref{fig:partonrate} for four different values of $(t-t_i)E$, which approaches to the asymptotic form (dashed line) when $(t-t_i)E$ is large. 

\begin{figure}[htbp]
    \centering
    \vspace{-0.4cm}
    \includegraphics[width=7.5cm]{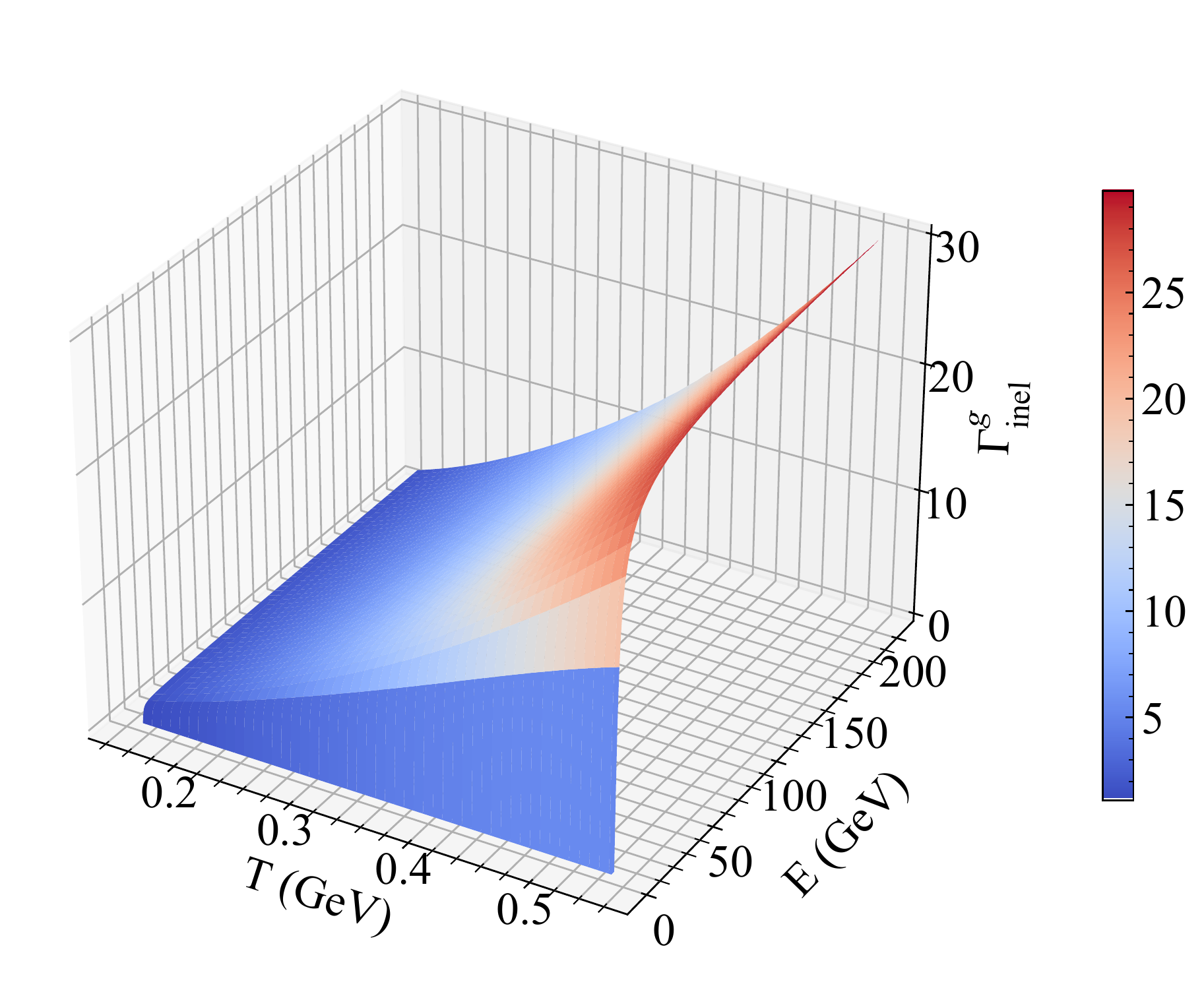}
    \caption{(Color online)  The rate of gluon emission from a gluon jet parton at $t-t_i=1$~fm/$c$ as a function of temperature $T$ and initial parton energy $E$. The strong coupling constant $\alpha_{\rm s}$ is set to 0.3.}
    \label{fig:quark-totrate}
\end{figure}

Integrating over the momentum fraction, we can obtain the total inelastic rate or the number of radiated gluons per unit time as shown in Fig.~\ref{fig:quark-totrate} for a gluon at $t=1$~fm/$c$ as a function of the local temperature $T$ and parton energy $E$. This rate has only a logarithmic energy dependence at large $E$, which is contributed by $\hat{q}_a$. Since the effective parton splitting functions have an asymptotic limit $P_a(z)/[z(1-z)]$ for large $(t-t_i)E>4$, one can obtain the integrated inelastic scattering rate analytically,
\begin{equation}
\Gamma^a_{\rm inel}= (t-t_i)\frac{\alpha_\mathrm{s} C_A \hat{q}_a }{2\mu_{\rm D}}\left[1+\frac{\mu_{\rm D}}{E}\ln\frac{\mu_{\rm D}}{E} +{\cal O}(\frac{\mu_{\rm D}}{E})\right].
\end{equation}
For $(t-t_i)E<4$, the integrated inelastic rate is instead
\begin{equation}
    \Gamma^a_{\rm inel}\approx (t-t_i)^2 \frac{\alpha_\mathrm{s} C_A\hat{q}_a}{4\pi}\ln\frac{E}{\mu_{\rm D}}.
\end{equation}
Using small angle approximation for the parton scattering cross section, the jet transport coefficient is \cite{He:2015pra}
\begin{equation}
\hat q_a=C_2(a)\frac{42\zeta(3)}{\pi}\alpha_{\rm s}^2T^3 \ln\frac{c_aET}{\mu_{\rm D}^2},
\end{equation}
where $\zeta(3)\approx 1.202$ is the Ap\'{e}ry's constant, $C_2(q)=4/3$, $C_2(g)=3$, $c_q=5.8$ and $c_g=5.6$. The energy and temperature dependence of $\Gamma^a_{\rm inel}$ at large $(t-t_i)E$ is therefore determined mostly by that of $\hat q_a$ for a fixed value of the strong coupling constant $\alpha_{\rm s}$.

\begin{figure}[!tbp]
\vspace{-2.0cm}
\includegraphics[width=7.5cm,bb=15 150 585 687]{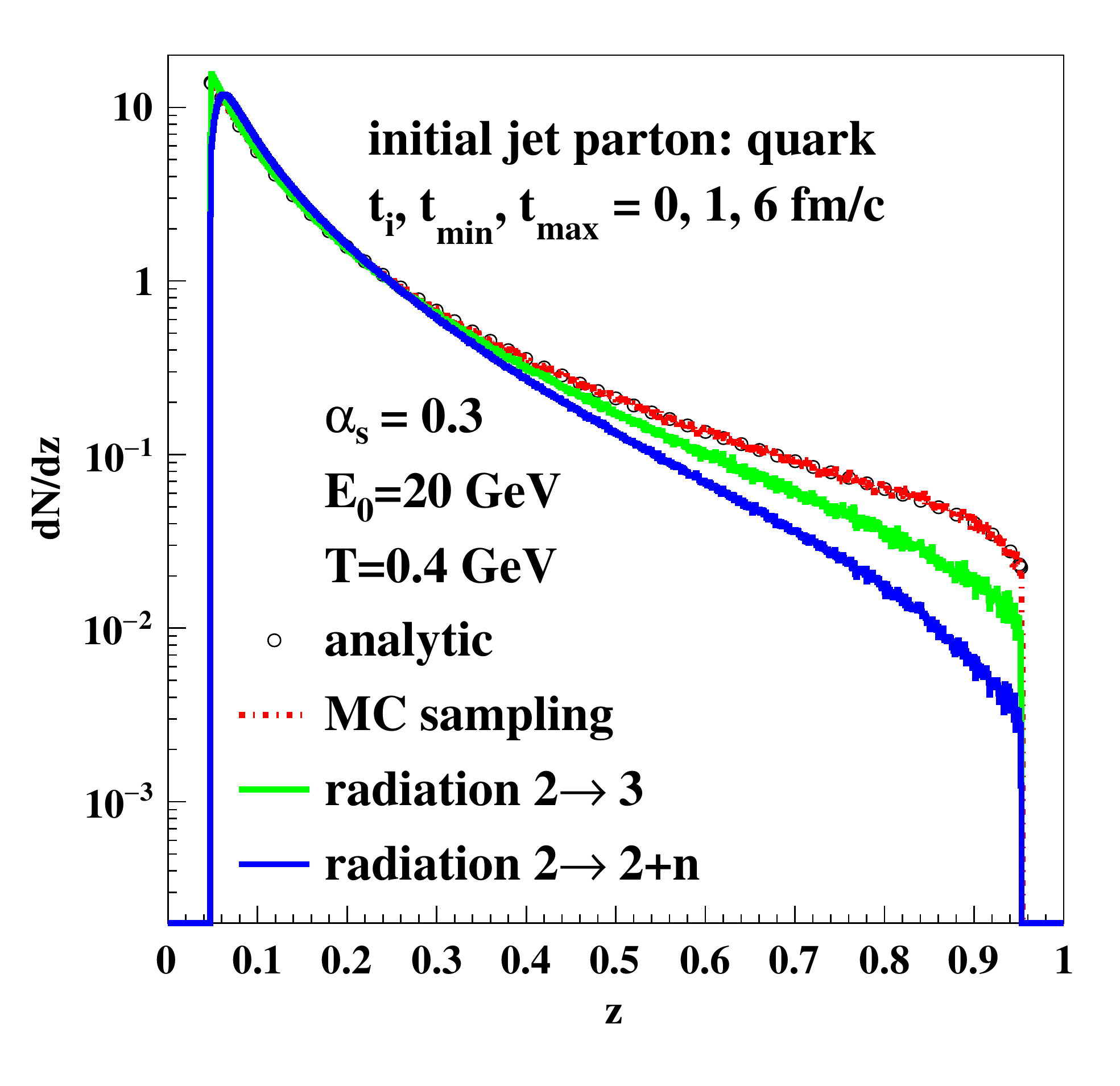}\\
\vspace{-0.86cm}
\includegraphics[width=7.5cm,bb=15 150 585 687]{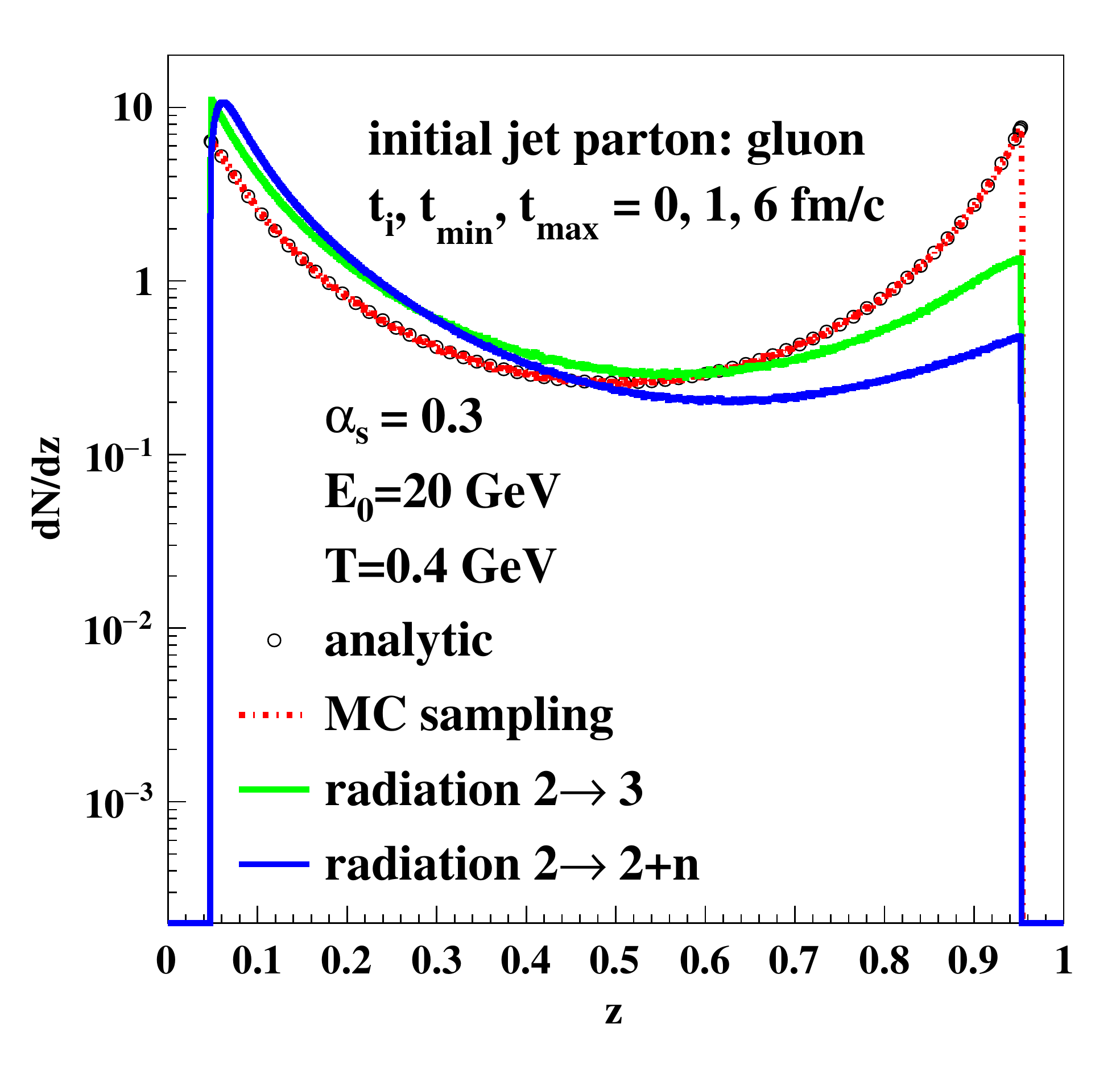}\\
\vspace{1.8cm}
\caption{(Color online) Fractional energy distributions of the radiated gluons in $q\rightarrow q+g$ (upper) and $g\rightarrow g+g$ (lower) processes with an initial parton energy $E_{0}=20$~GeV in a static medium at temperature $T=400$~MeV from semi-analytical calculations (black open circles), direct MC sampling (red dashed lines) and LBT simulations for $2\rightarrow 3$ (green solid lines) and $2\rightarrow 2+n$ $(n\ge 1)$ (blue solid lines) .
}
\label{fig:dngde}
\end{figure}

\subsection{Medium-induced gluon spectra}

For each inelastic scattering, we use Eq.~(\ref{eq:gluondistribution}) to sample the momentum fraction $z$ and the transverse momentum $\vec k_\perp$ of the radiated gluon. Since in each inelastic process in LBT simulations, the medium-induced gluon emissions are always accompanied by a $2\rightarrow 2$ elastic scattering process and the jet parton will also experience transverse momentum transfer (or broadening) and elastic energy loss. The energy-momentum conservation in this inelastic process  with one or multiple gluon emissions should be considered together. We adopt a procedure as described in Appendix \ref{sec:appendix} to maintain the global energy-momentum conservation in $2\rightarrow 2+n(g)$ processes. 

Shown in Fig.~\ref{fig:dngde} are the normalized fractional energy distributions of gluons, 
\begin{equation}
    \frac{dN}{dz}=\int_{t_{\rm min}}^{t_{\rm max}} dt \frac{d\Gamma^a_{\rm inel}}{dz}\left/ \int_{t_{\rm min}}^{t_{\rm max}} dt \Gamma^a_{\rm inel}\right. ,
\end{equation} 
emitted from a quark (upper panel) or gluon (lower panel) jet parton with an initial energy of $E_0=20$~GeV in a single inelastic process with different treatments of global energy-momentum conservation. In order to compare to results from semi-analytical calculations, we consider a static and uniform medium at a constant temperature $T=0.4$~GeV and set the strong coupling constant at $\alpha_{\rm s}=0.3$. In the figure, the black open circles represent results from numerical integration of Eq.~(\ref{eq:gluondistribution}) over $k_\perp^2$ and time between $t_{\rm min}=1$~fm/$c$ and $t_{\rm max}=6$~fm/$c$ with $t_i=0$.
The red dashed lines represent results from direct Monte-Carlo sampling according to Eq.~(\ref{eq:gluondistribution}), which are in exact agreement with the circles as expected. The gluon spectra become softer and deviate from the spectrum given by Eq.~(\ref{eq:gluondistribution}) when energy-momentum conservation is considered for $2\rightarrow 3$ process as shown by the green curves. This softening is caused by the elastic energy loss of the jet parton when considered together with the induced gluon radiation.
When multiple gluon emissions are taken into account in $2\rightarrow 2+n$ processes, the emitted gluon spectra (the blue lines) are further shifted towards lower energies because of the successive energy loss of the jet parton when it emits multiple gluons. Comparing the upper and the lower panel, one observes that the energy spectrum of emitted gluons peaks at low energy for a quark jet, but peaks at both small and large energy for a gluon jet. This can be understood with the $\sim 1/z$ splitting function for the $q\rightarrow qg$ process, while the splitting is symmetric $\sim 1/[z(1-z)]$ between the emitted gluon and the jet gluon for $g\rightarrow gg$. The symmetry around $z=0.5$ can be seen in the semi-analytical result, but is broken when the realistic $2\rightarrow 2+n$ process is taken into account.

\begin{figure}[tbp!]
    \centering
    \includegraphics[width=8.5cm]{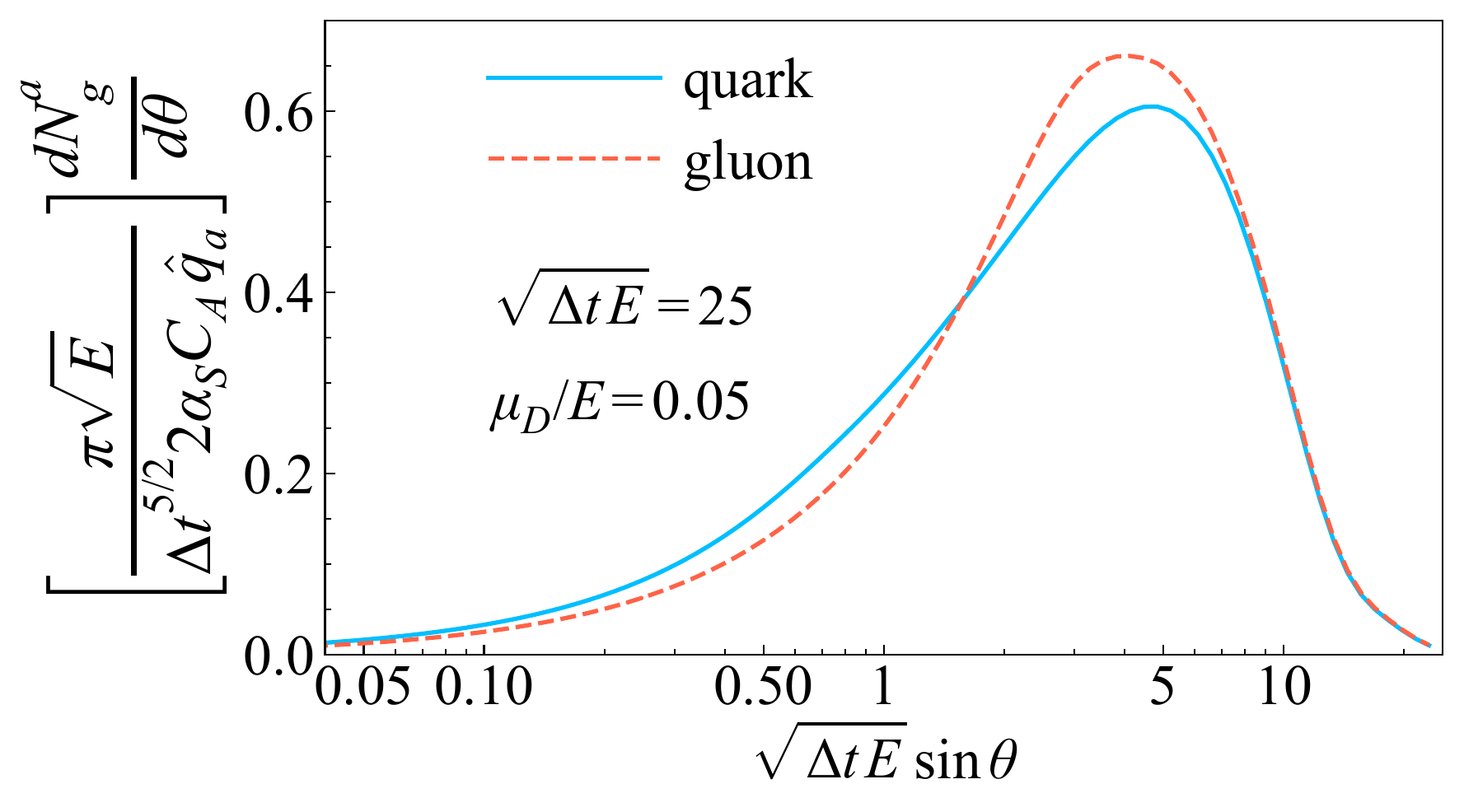}
    \caption{(Color online) The angular distribution of medium-induced radiated gluon for a typical parton energy and length $\sqrt{\Delta t E}$ and Debye screening mass $\mu_D/E$.}
    \label{fig:scale-angle}
\end{figure}

In a static and uniform medium, one can complete the time integral in Eq.~(\ref{eq:gluondistribution}) and obtain the angular distribution of the radiated gluon,
\begin{eqnarray}
    \frac{dN_g^a}{d\theta}&=& \frac{2\alpha_\mathrm{s} C_A \hat{q}_a }{\pi \sqrt{E}} \frac{\Delta t^{5/2}\cos\theta}{(\sqrt{\Delta t E}\sin\theta)^3}\int_{\mu_D/E}^1 dz \frac{P_a(z)}{z^2} \nonumber \\
    &\times& \left\{1 -\frac{\sin\left(\Delta t E z \sin^2\theta/[2(1-z)]\right)}{\Delta t E z \sin^2\theta/[2(1-z)]}\right\},
\end{eqnarray}
where the polar angle is defined as $\sin\theta=k_\perp/(zE)$. Note that this polar angle is different from the relative angle between the final two partons after the splitting. For a gluon jet, the angle is defined for the less energetic gluon in the final state $\sin\theta=k_\perp/(\min[z,1-z]E)$ and the result is twice of the integration over $z\le 1/2$.

Unlike gluon radiation from a virtual parton in vacuum that has a collinear divergence $dN_g/d\theta \sim 1/\theta$,  medium-induced gluons tend to be emitted with a finite angle $\theta \sim 1/\sqrt{\Delta t E}$ with respect to the direction of the incoming jet parton. Small angle emissions whose formation times are longer than the medium length are suppressed due to the LPM interference as shown in Fig.~\ref{fig:scale-angle}. The angular distribution vanishes as $dN_g\sim \theta $ when $\theta\ll 1/\sqrt{\Delta t E}$.  At large angles, gluon emission becomes incoherent without LPM interference for large values of $\sqrt{\Delta t E}$, and the angular distribution becomes $dN_g/d\theta \sim \cos\theta/\sin^3\theta$, corresponding to the transverse momentum spectrum $dN_g/dk_\perp^2 \sim 1/k_\perp^4$.

\begin{figure}[!tbp]
\vspace{-2.2cm}
\includegraphics[width=7.5cm,bb=15 150 585 687]{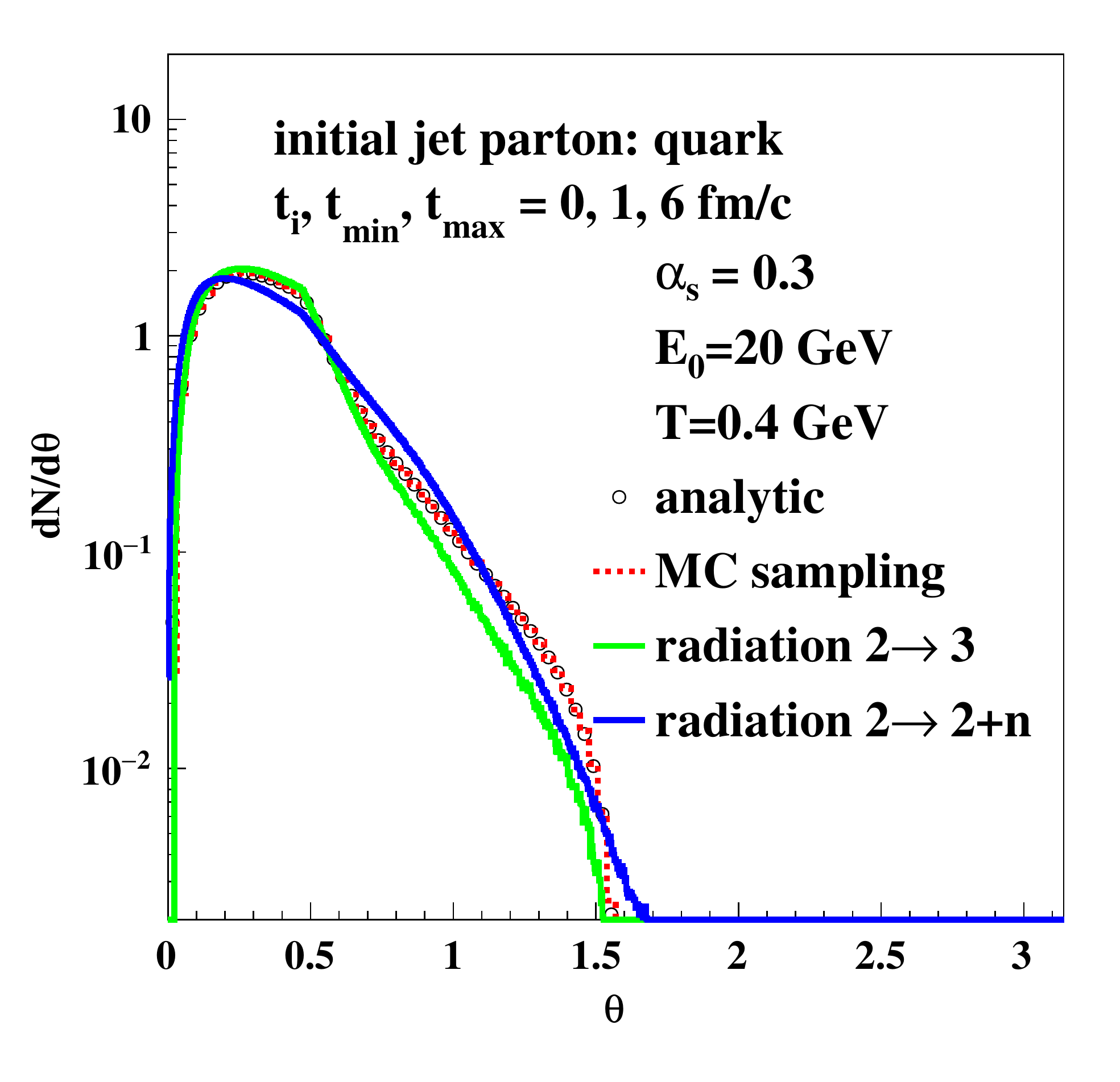}\\
\vspace{1.5cm}
\caption{(Color online) Polar angle distributions of the radiated gluons in $q\rightarrow q+g$ process with an initial parton energy $E_{0}=20$~GeV in a static medium at temperature $T = 400$~MeV from semi-analytical calculations (black open circles), direct MC sampling (red dashed line), LBT simulations for $2\rightarrow 3$ (green solid line) and $2\rightarrow 2+n$ $(n\ge 1)$ (blue solid line) processes.
}
\label{fig:dngdtheta}
\end{figure}

In Fig.~\ref{fig:dngdtheta}, we show the normalized polar angle distribution of the emitted gluon with respect to a parent quark from LBT model. Distributions from a parent gluon is very similar. Our Monte Carlo sampling faithfully reproduce the semi-analytic results for medium-induced gluon emissions. When elastic scatterings are considered together with the gluon radiation in the $2\rightarrow 3$ process, the global energy-momentum conservation that suppresses the gluon emission with large momentum fraction (as shown in Fig.~\ref{fig:dngde}) also enhances the polar angle distribution at small angle slightly. 
When multiple gluon emissions are allowed in LBT simulations (the blue curves), the average polar angles become larger compared to the case where only single gluon emission is allowed (the green curves) due to radiative broadening.

\section{Parton energy loss and transverse momentum broadening}
\label{sec:partonInBrick}

\subsection{Parton energy loss}

\begin{figure}[!tbp]
\vspace{-2.2cm}
\includegraphics[width=7.5cm,bb=15 150 585 687]{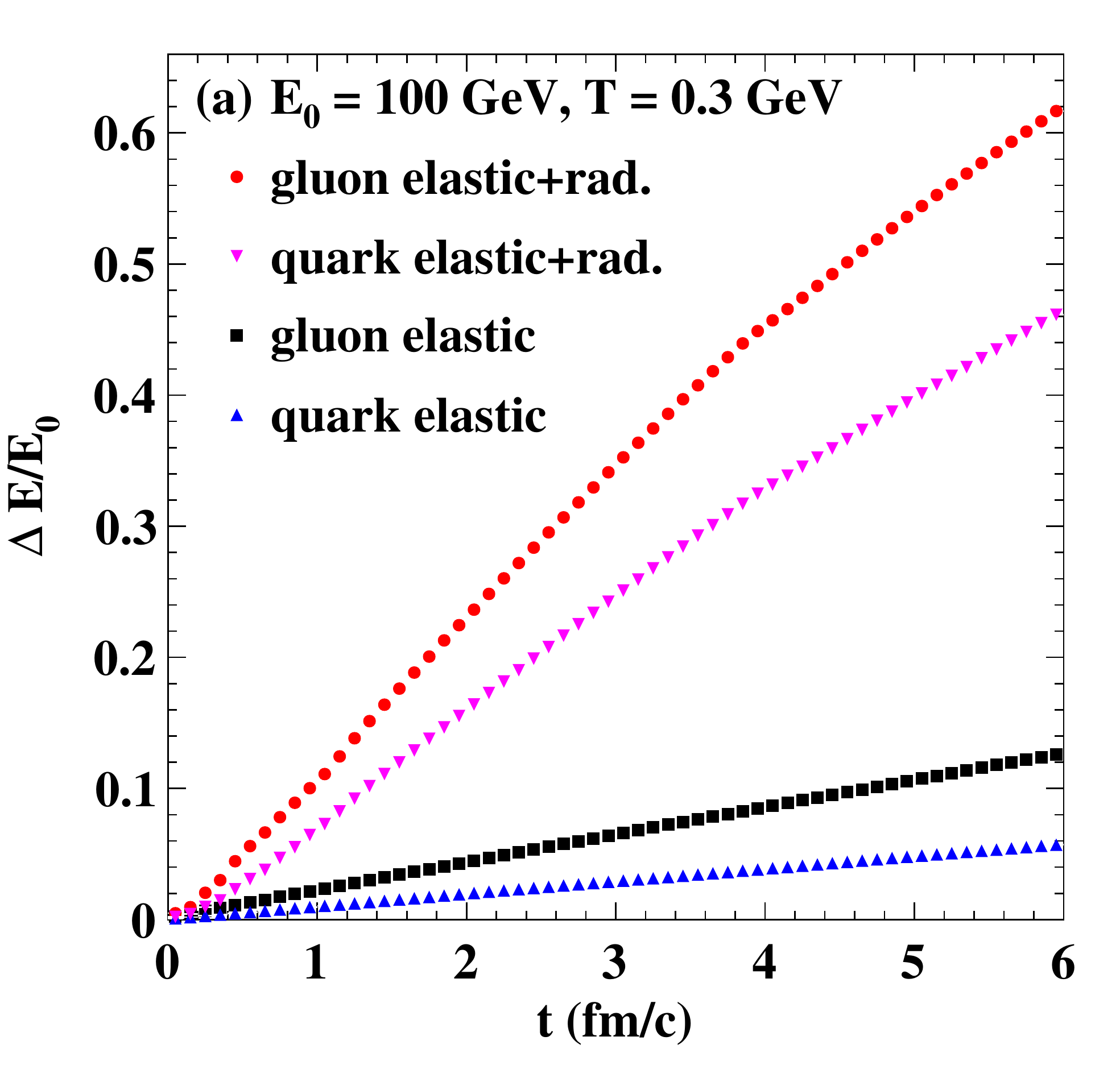}\\
\vspace{-0.85cm}
\includegraphics[width=7.5cm,bb=15 150 585 687]{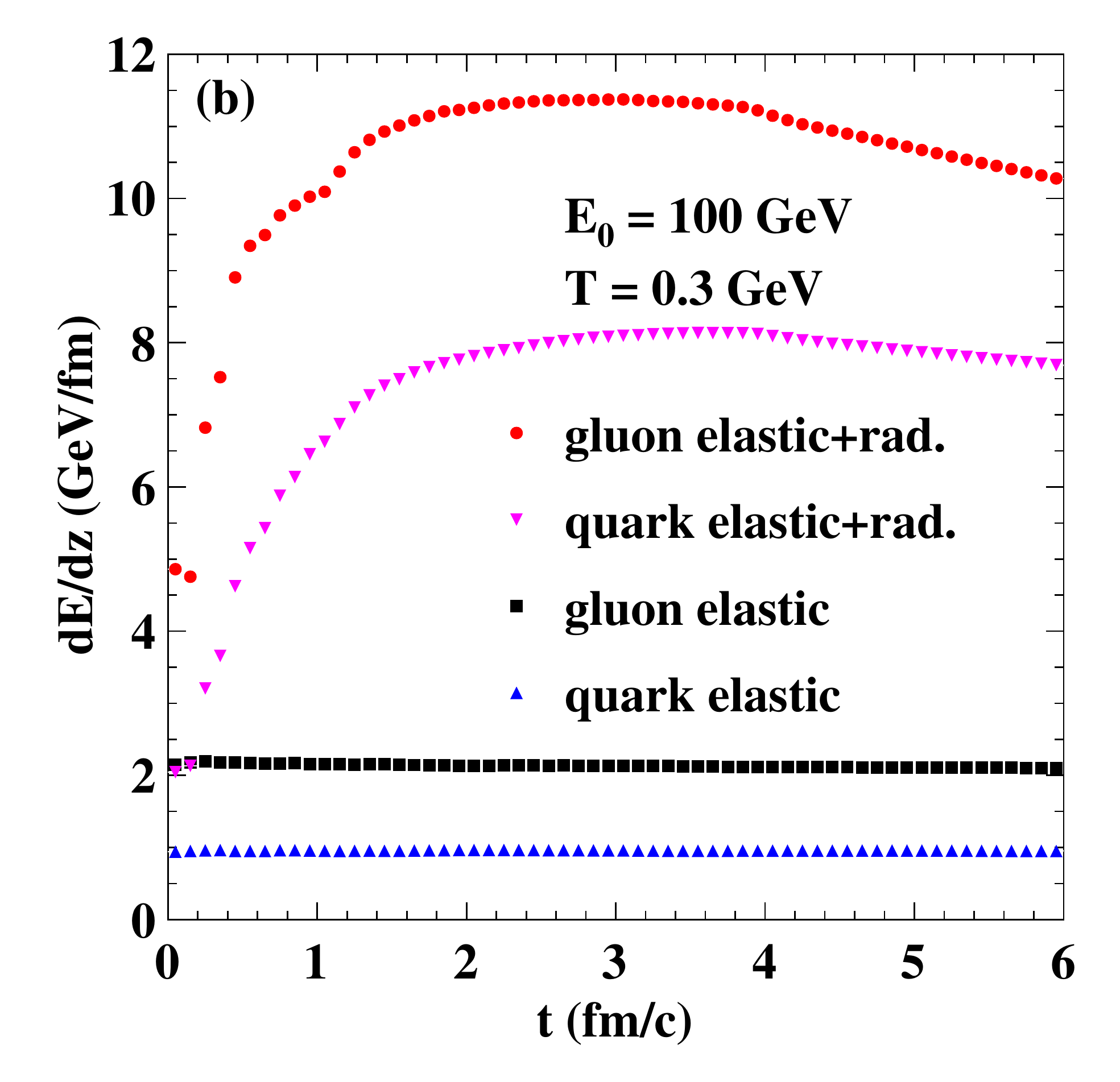}\\
\vspace{1.8cm}
\caption{(Color online) The time dependence of the fractional energy loss $\Delta E/E_{0}$ (upper panel) and energy loss per unit length $dE/dz$ (lower panel) of a quark (up and down triangle) or a gluon (circle and square) with initial energy $E_{0}=100$~GeV through a static medium with temperature $T = 300$~MeV, with only elastic (black and blue) and elastic + gluon radiation processes (red and magenta).}
\label{fig:elossparton}
\end{figure}

\begin{figure}[!tbp]
\vspace{-2.2cm}
\includegraphics[width=7.5cm,bb=15 150 585 687]{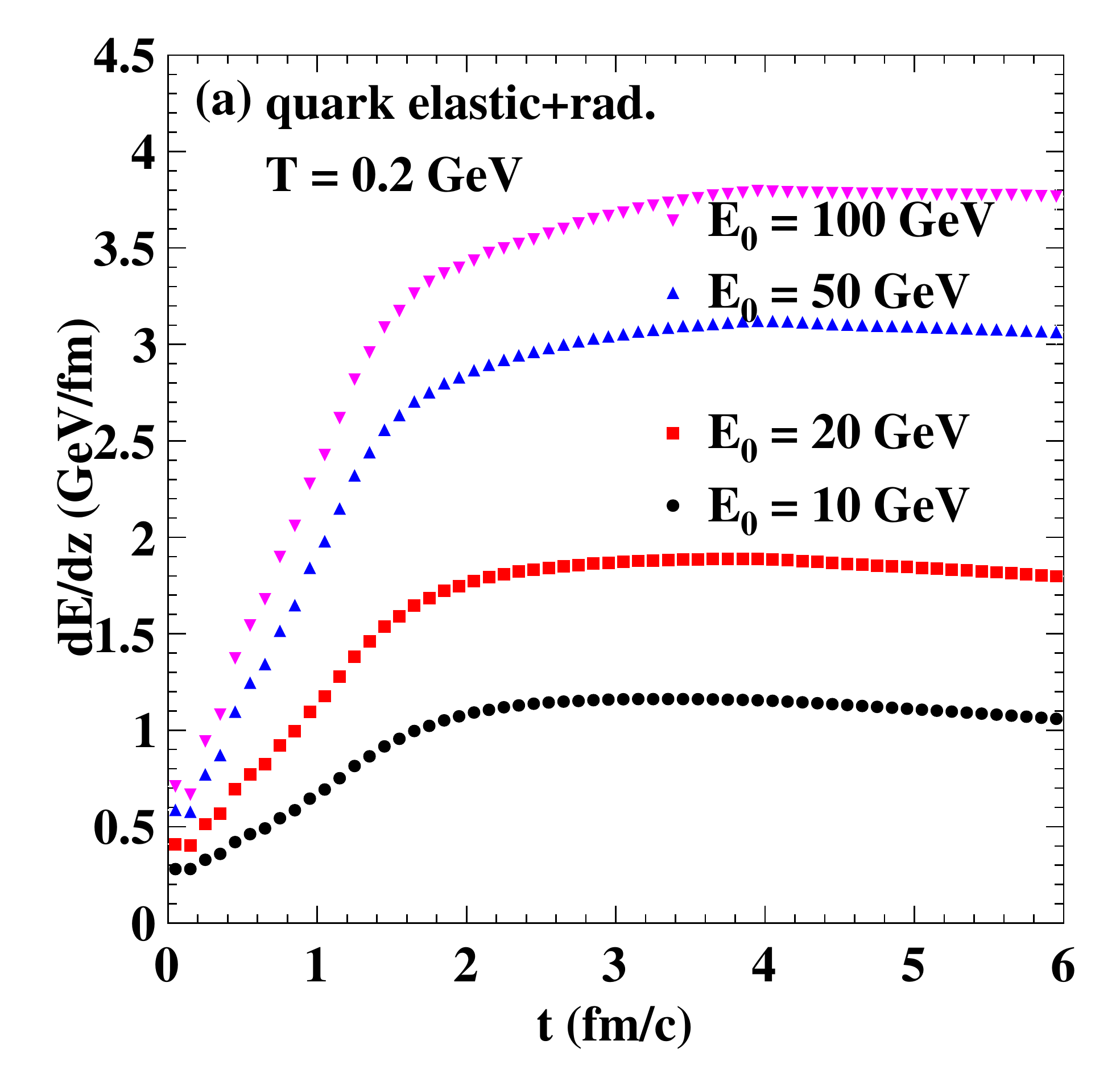}\\
\vspace{-0.86cm}
\includegraphics[width=7.5cm,bb=15 150 585 687]{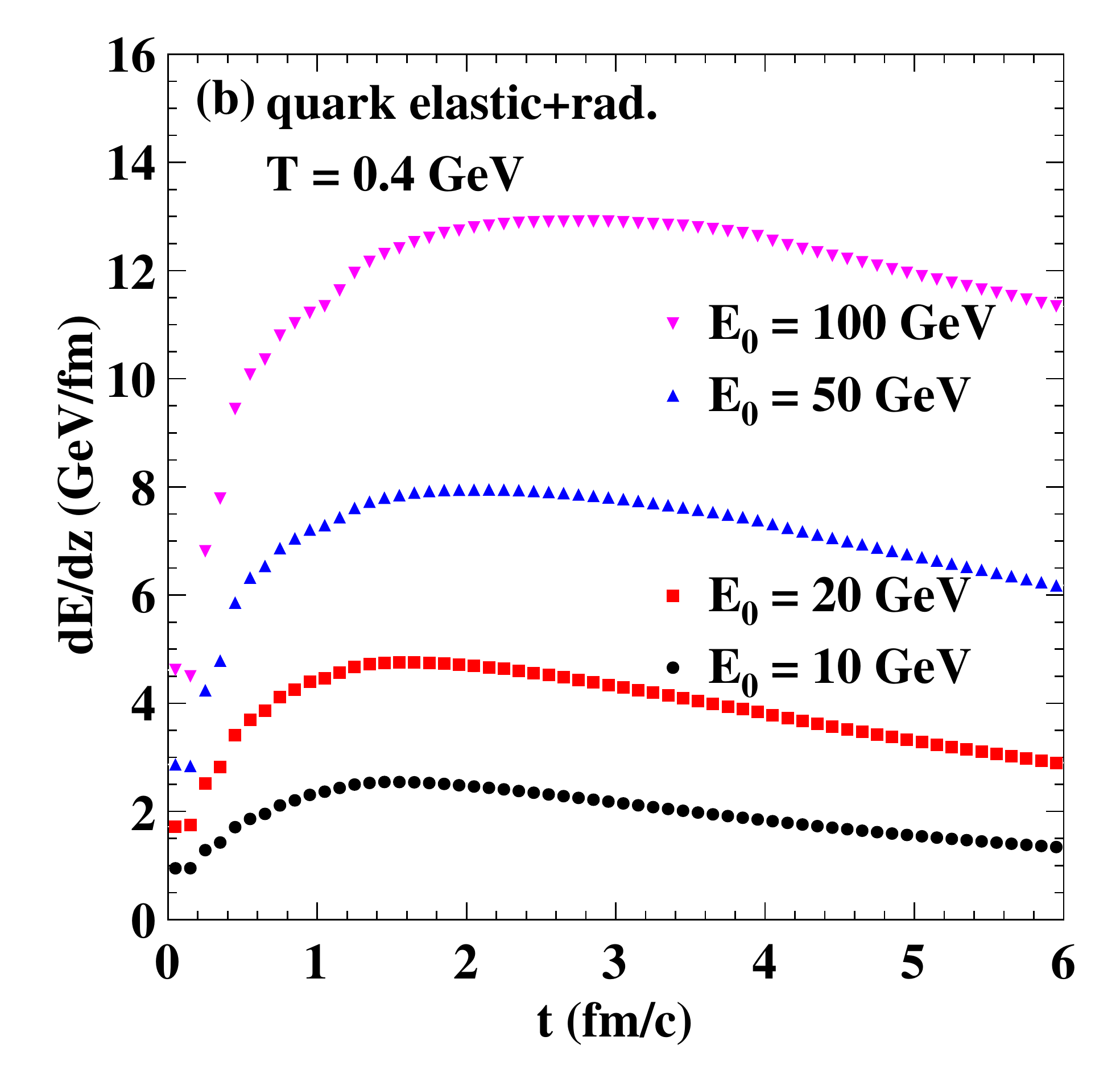}\\
\vspace{1.8cm}
\caption{(Color online) Quark energy loss per unit length as a function of time in a static medium with temperatures $T = 200$~MeV (upper panel) and $T = 400$~MeV (lower panel), for different initial energies $E_0$. Both elastic and gluon radiation processes are included.}
\label{fig:elosspartonQ}
\end{figure}

\begin{figure}[!htb]
\vspace{-2.2cm}
\includegraphics[width=7.5cm,bb=15 150 585 687]{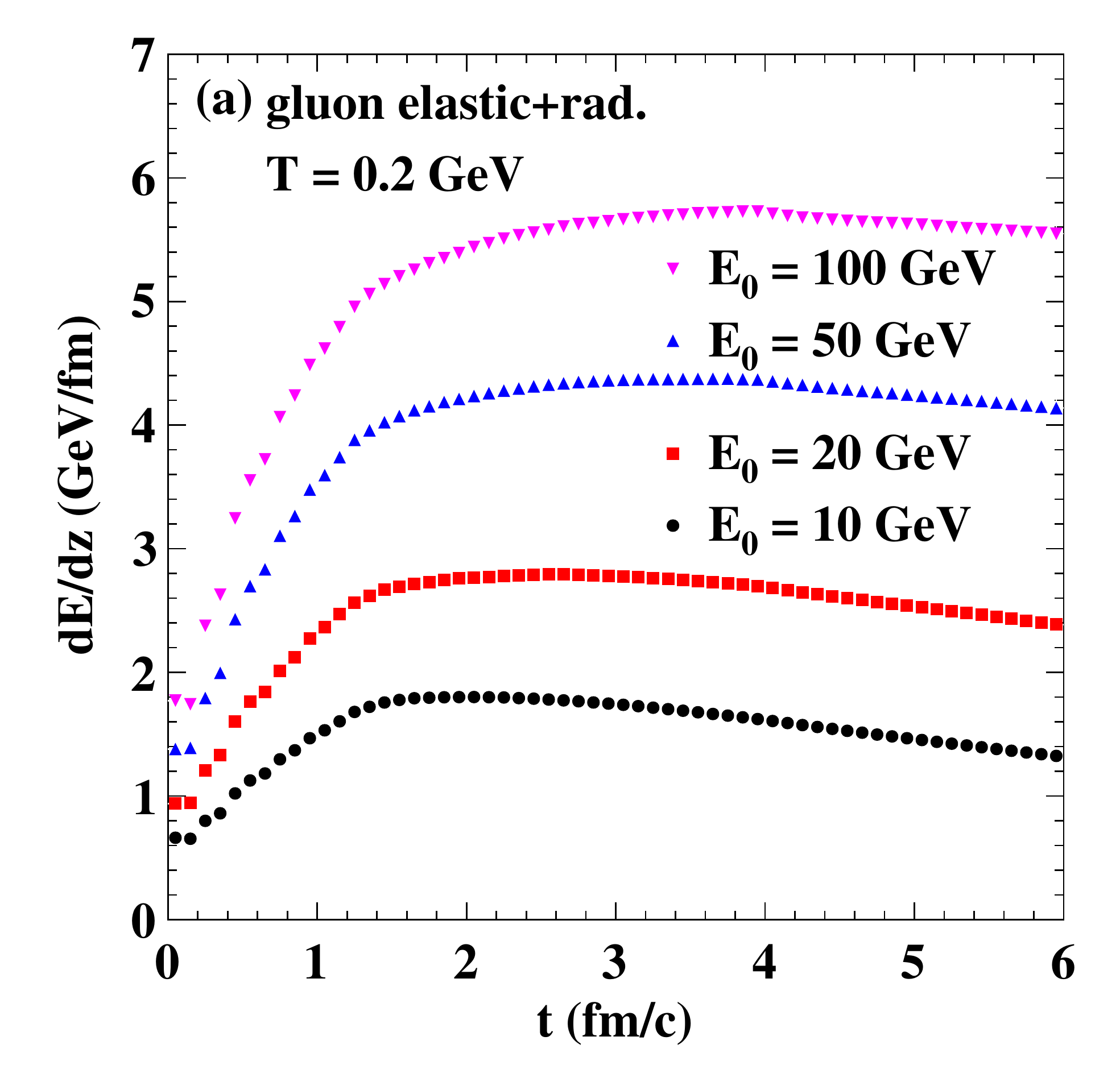}\\
\vspace{-0.86cm}
\includegraphics[width=7.5cm,bb=15 150 585 687]{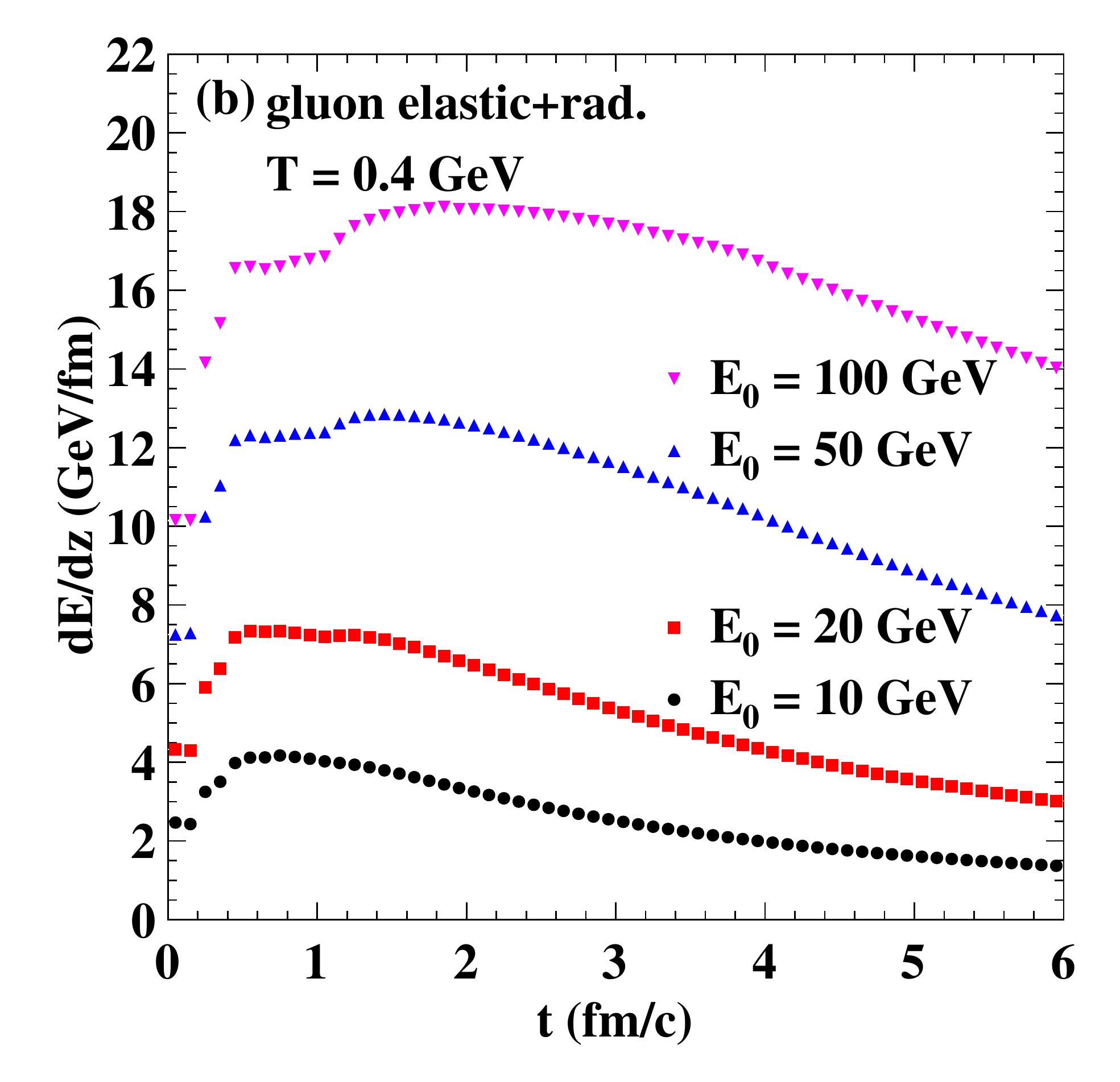}\\
\vspace{1.8cm}
\caption{(Color online) The same as Fig.~\ref{fig:elosspartonQ} except for a gluon.
}
\label{fig:elosspartonG}
\end{figure}

To investigate the energy loss and transverse momentum broadening of a propagating parton within the LBT model, we again consider a static and uniform medium with no flow at a constant temperature in this section. We start with the energy loss of a single parton in Fig.~\ref{fig:elossparton} as a function of its propagation time (or distance) in a static medium with $T=300$ MeV. The parton is initialized with 100~GeV energy along the longitudinal ($z$) direction. In the upper panel, we present the accumulated fractional energy loss of the parton, compared between a quark and a gluon jet parton and different scattering processes. One can clearly observe a stronger energy loss of a gluon than a quark due to the larger color factor of the former. Compared to the energy loss from only elastic scatterings, the parton energy loss significantly increases when inelastic scattering, or medium-induced gluon radiation, is included. This indicates the dominating contribution from gluon radiation to parton energy loss. In the lower panel of Fig.~\ref{fig:elossparton}, we show the parton energy loss per unit length $dE/dz$. For elastic scattering, a constant energy loss rate $dE/dz$ is reached after the traversed length of the parton surpasses its mean free path, resulting in a linear increase of the accumulated parton energy loss as shown in the upper panel. In contrast, when induced gluon radiation process is incorporated, the energy loss rate $dE/dz$ increases with the path length almost linearly for a rather long time until the mean formation time of single gluon emission is reached. It becomes less than linear in the distance dependence afterwards. As a result, the corresponding accumulated parton energy loss shown in the upper panel increases almost quadratically with respect to the path length at early times and returns to a slower length dependence at later times.  This change of the energy loss rates with time is caused by the energy dependence of the energy loss. Since energy loss rate increases with parton energy, it should slow down as the parton loses more energy. In addition, such an energy dependence of the energy loss also leads to the change of the ratio between the energy loss of a gluon and a quark. The energy loss rate of a gluon is larger than that of a quark by a factor of $C_A/C_F=9/4$ at early times as seen in Fig.~\ref{fig:elossparton}. At later times, the energy loss of these hard partons, which is more for a gluon than a quark, leads to the reduction of further energy loss and narrows the difference of energy loss between a quark and a gluon. Note that if one does not reset $t_i$ in Eq.~(\ref{eq:gluondistribution}) to zero after a radiative gluon is formed and ignores the variation of the parton energy as it propagates through the medium, the accumulated energy taken by emitted gluons then keeps increasing quadratically with respect to the path length and the relative difference between gluon and quark energy loss remains at $C_A/C_F$~\cite{Cao:2017hhk}.

The dependence of parton energy loss on the parton energy and the medium temperature is further illustrated in Fig.~\ref{fig:elosspartonQ} for a quark and Fig.~\ref{fig:elosspartonG} for a gluon. In each figure, the upper plot is for a static medium at $T=200$~MeV and the lower plot is for $T=400$~MeV. In each plot, results from different initial parton energies are compared. Contributions from both elastic scattering and gluon radiation are included here. Similar to the findings previously presented in Fig.~\ref{fig:elossparton}, the increase of parton energy loss per unit length $dE/dz$ is rapid at early times and then significantly slows down and even starts to decrease with time as multiple gluon emissions commence and the hard parton's energy becomes smaller. This variation of the energy loss rate is more prominent for lower energy partons and higher medium temperature, considering the larger fractional energy loss in these scenarios. For the same reason, it is more prominent for gluons than for quarks.

\subsection{Transverse momentum broadening}

\begin{figure}[!tbp]
\vspace{-2.4cm}
\includegraphics[width=7.5cm,bb=15 150 585 687]{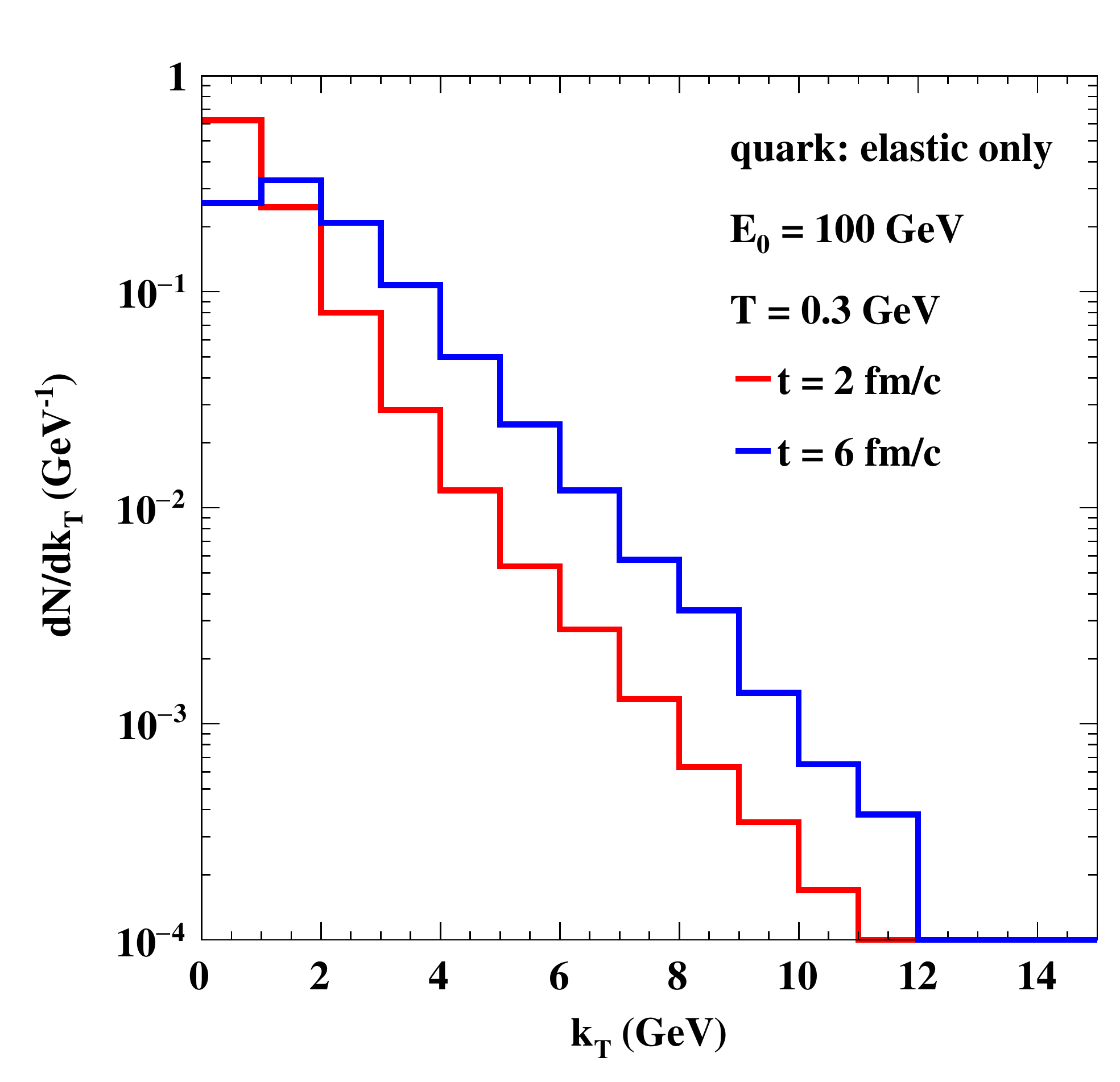}\\
\vspace{-0.72cm}
\includegraphics[width=7.5cm,bb=15 150 585 687]{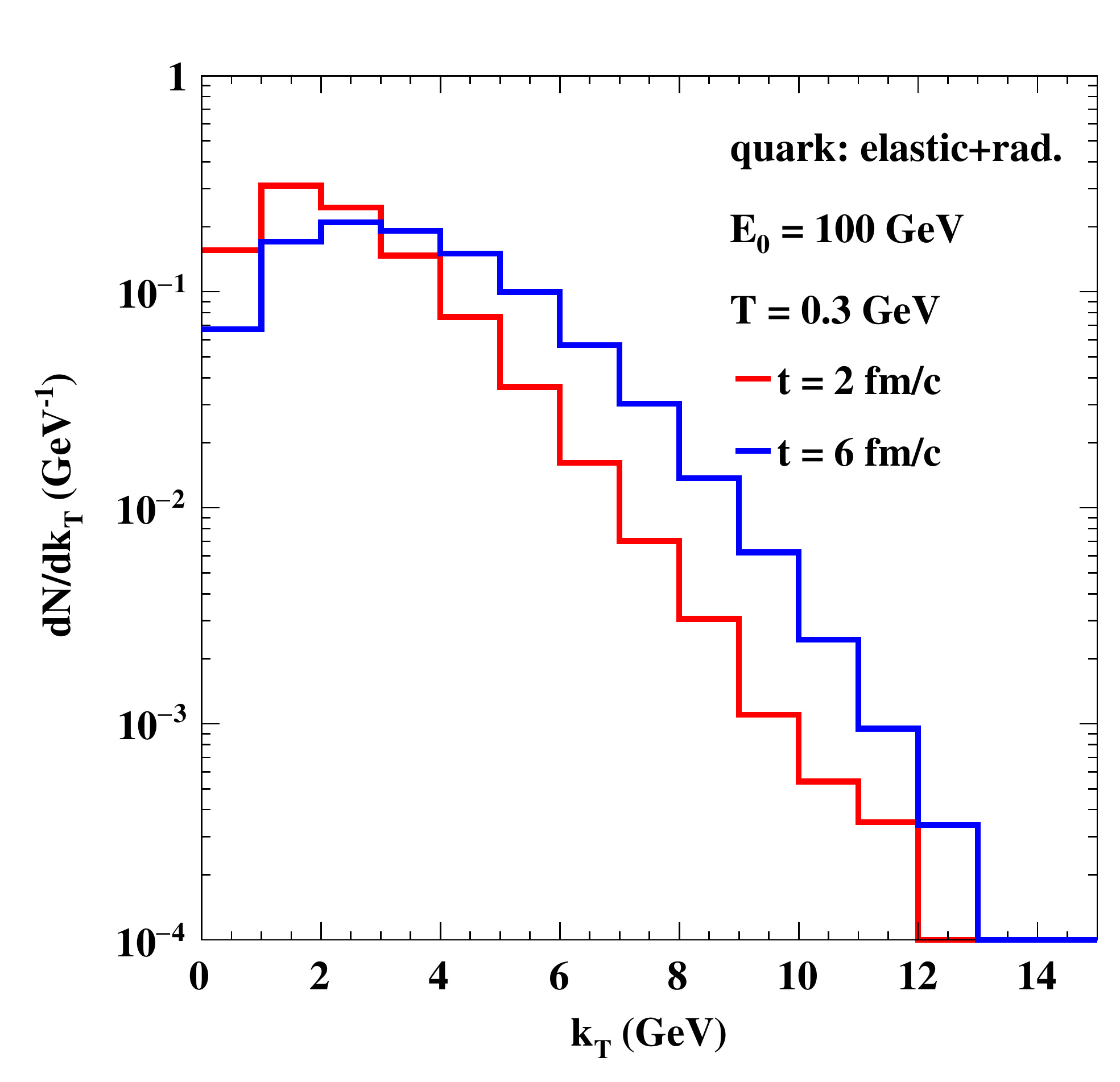}\\
\vspace{1.8cm}
\caption{(Color online) Transverse momentum distribution of a quark with an initial energy $E_0=100$~GeV at different evolution times through a static medium at temperature $T=300$~MeV. The upper panel includes only elastic scatterings, while the lower panel includes both elastic scatterings and gluon emissions.}
\label{fig:ptbroadening_q}
\end{figure}

\begin{figure}[!tbp]
\vspace{-2.4cm}
\includegraphics[width=7.5cm,bb=15 150 585 687]{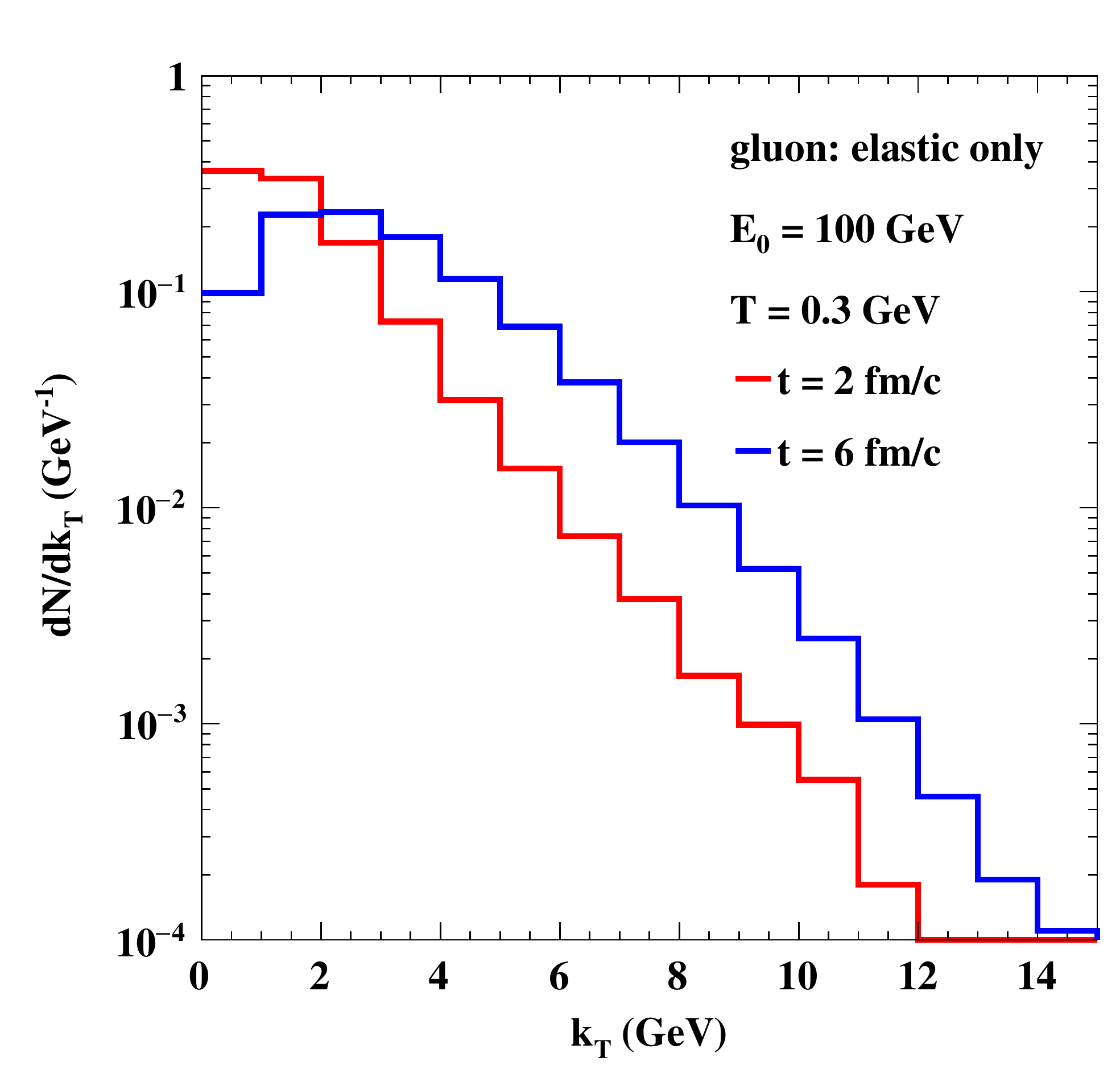}\\
\vspace{-0.72cm}
\includegraphics[width=7.5cm,bb=15 150 585 687]{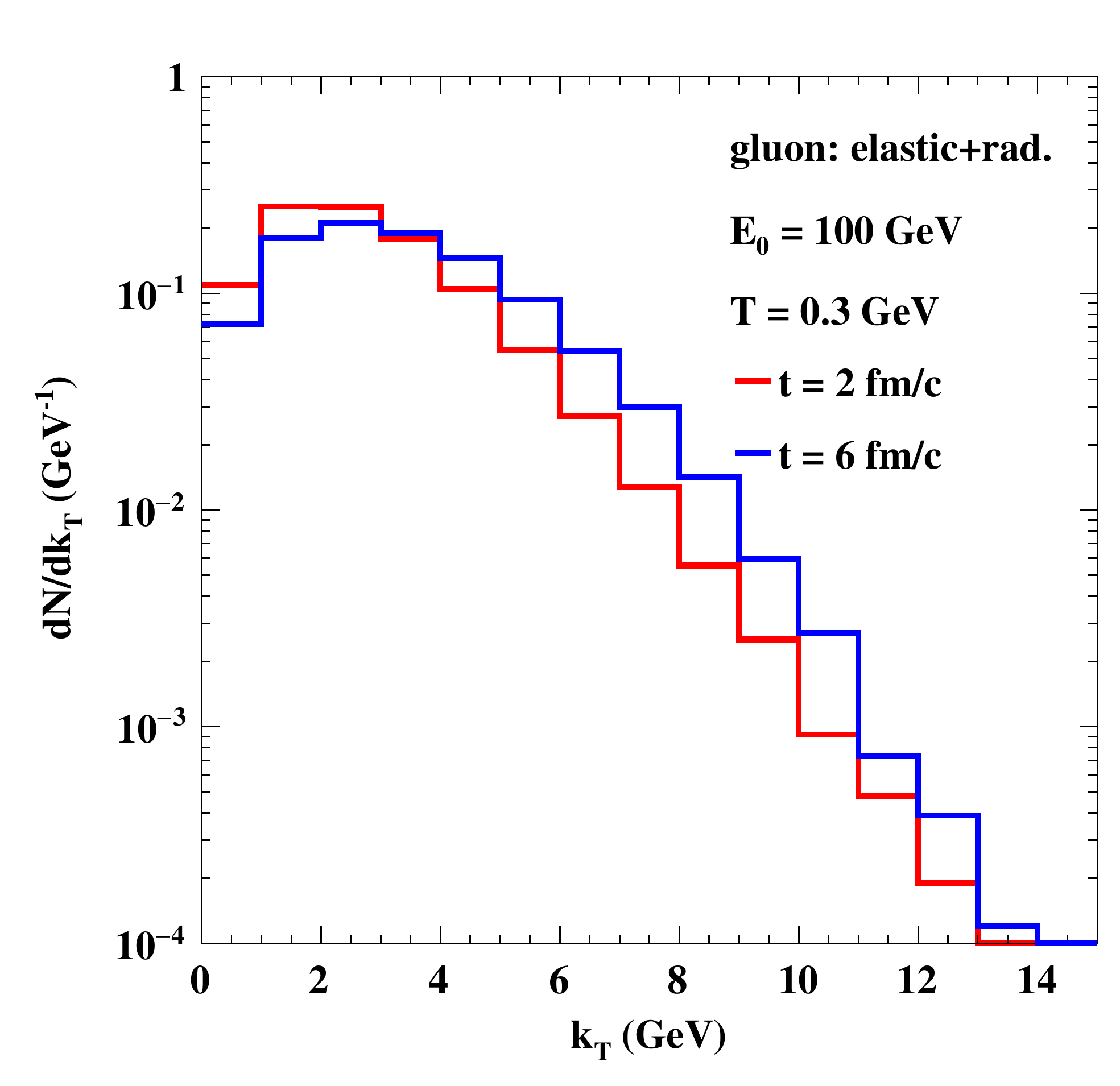}\\
\vspace{1.8cm}
\caption{(Color online) The same as Fig.~\ref{fig:ptbroadening_q} except for a gluon.}
\label{fig:ptbroadening_g}
\end{figure}

As discussed in Sec.~\ref{sec:LBT}, the transverse momentum broadening of a hard parton, or the $\hat{q}$ parameter, is a key quantity that governs the inelastic scattering rate of a parton. To investigate this broadening of transverse momentum ($k_\mathrm{T}$) of a hard parton traversing a QGP medium, we present the $k_\mathrm{T}$ distributions of the leading parton at different times through a static medium in Fig.~\ref{fig:ptbroadening_q} for an initial quark and Fig.~\ref{fig:ptbroadening_g} for an initial gluon. In each figure, the upper panel shows the $k_{\rm T}$ distributions due to only elastic scatterings, while the lower panel shows the distributions due to both elastic and inelastic scatterings. Here, the hard parton is initialized with an energy $E_0=100$~GeV along the $z$ direction, and the medium temperature is set as $T=300$~MeV. Therefore, starting as a $\delta$-function at $k_\mathrm{T}=0$, the transverse momentum distribution becomes broader as time evolves. Comparing the upper and lower panels, one observes a faster broadening of this $k_\mathrm{T}$ after the gluon radiation process is included. The non-exponential $k_{\rm T}$ distributions due to inelastic scatterings is caused by the restriction on small $k_{\rm T}$ gluon radiations that have long formation times due to the LPM interference.   Comparing Fig.~\ref{fig:ptbroadening_q} and Fig.~\ref{fig:ptbroadening_g}, one observes a broader $k_\mathrm{T}$ distribution for a gluon than for a quark due to the difference of their jet transport coefficients.

\section{Jet induced medium excitation}
\label{sec:response}

After the discussion of energy loss and transverse momentum broadening of a propagating parton in the last section, we will study the evolution of the jet shower in the phase space from an initial hard parton propagating in a static and uniform QGP medium. The partons in the jet shower include contributions from the medium-modified leading parton, gluons from medium-induced radiations, recoil partons and ``negative" partons. As previously discussed, the recoil and ``negative" partons collectively constitute the jet-induced medium excitation in the LBT model.

\subsection{Conic structures in jet-induced medium excitation}

\begin{figure}[!tbp]
\vspace{-2.2cm}
\includegraphics[width=7.5cm,bb=15 150 585 687]{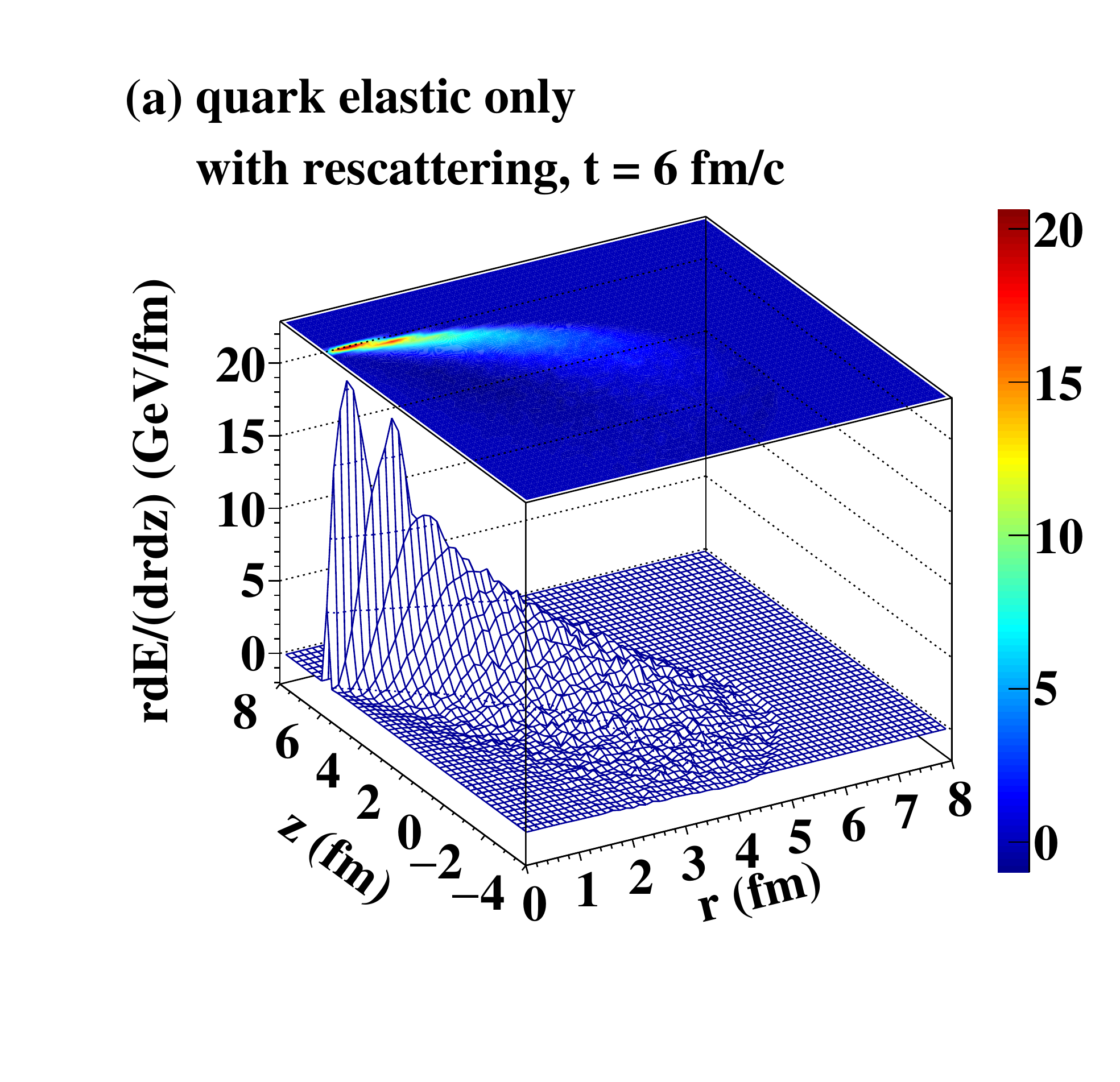}\\
\vspace{-0.8cm}
\includegraphics[width=7.5cm,bb=15 150 585 687]{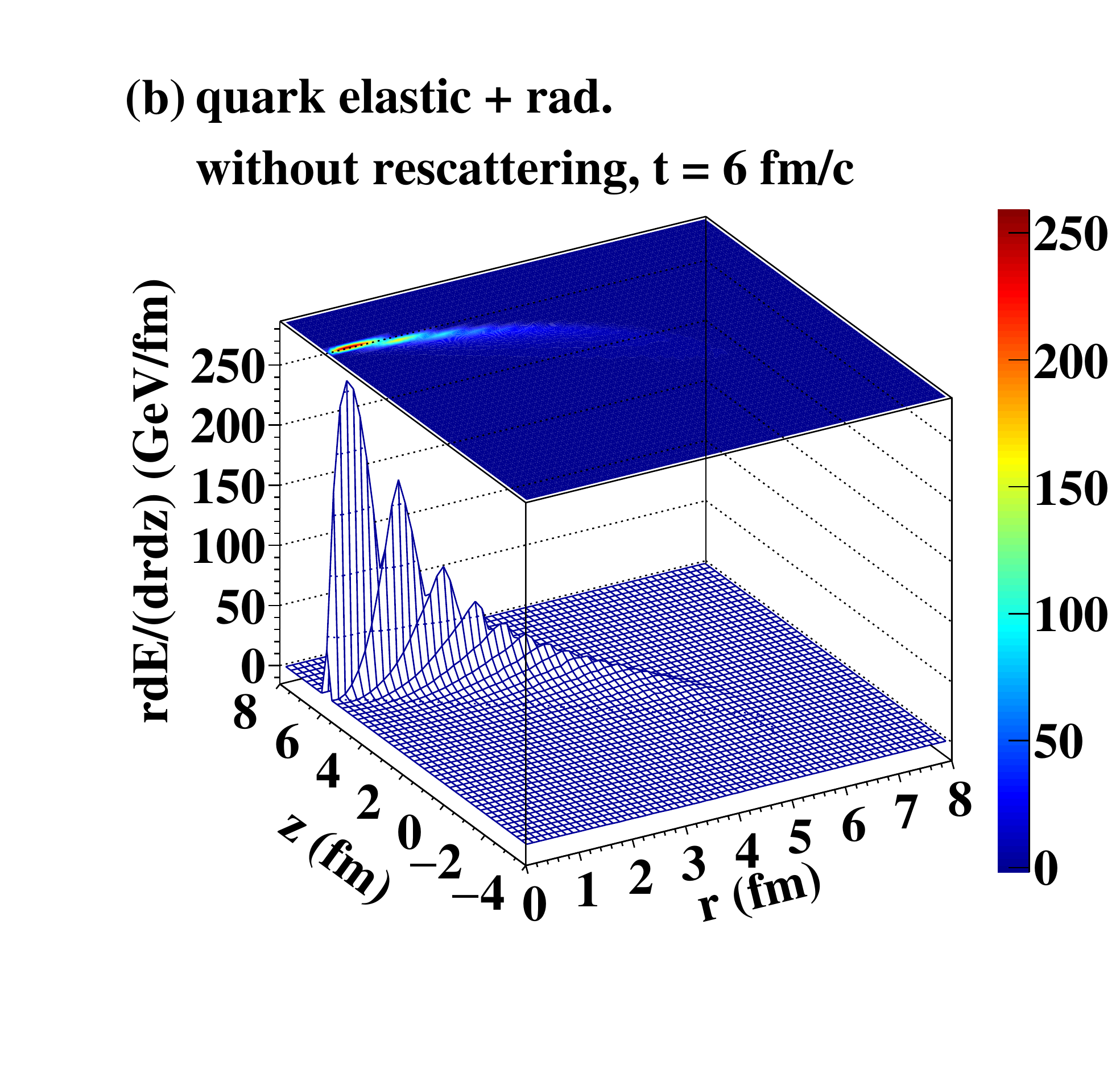}\\
\vspace{-0.8cm}
\includegraphics[width=7.5cm,bb=15 150 585 687]{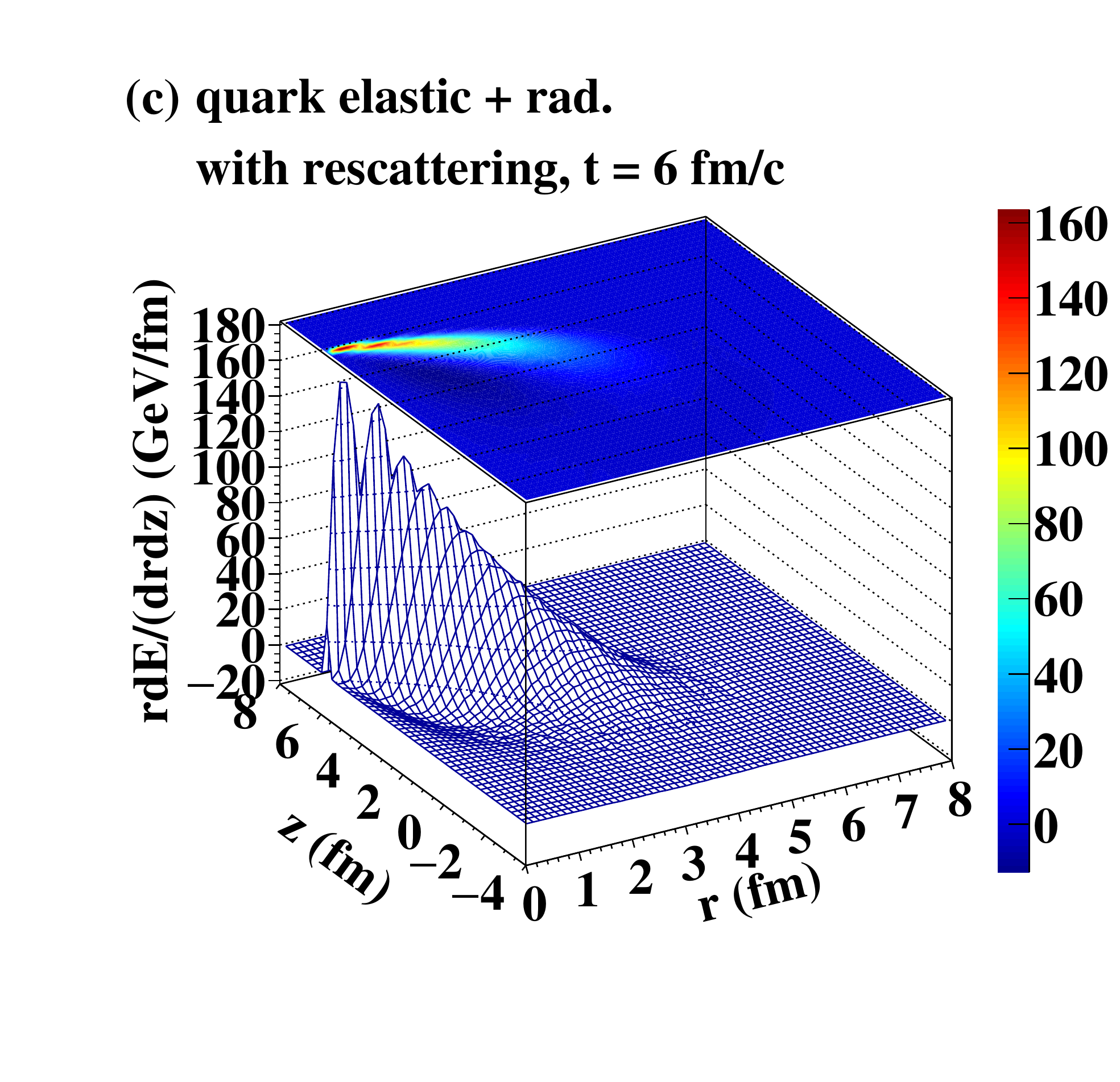}\\
\vspace{1.6cm}
\caption{(Color online) Energy density distribution of a jet shower originating from a 100~GeV quark at $x=y=z=0$ and propagating along the $z$ direction in a static medium at $T = 300$~MeV. Results are shown at $t=6$~fm/$c$ with contributions from only elastic scatterings with rescatterings (upper panel), elastic + inelastic scatterings without rescatterings (middle panel), and elastic + inelastic scatterings with rescatterings (lower panel).}
\label{fig:rdedrdzq}
\end{figure}

\begin{figure}[!tbp]
\vspace{-2.2cm}
\includegraphics[width=7.5cm,bb=15 150 585 687]{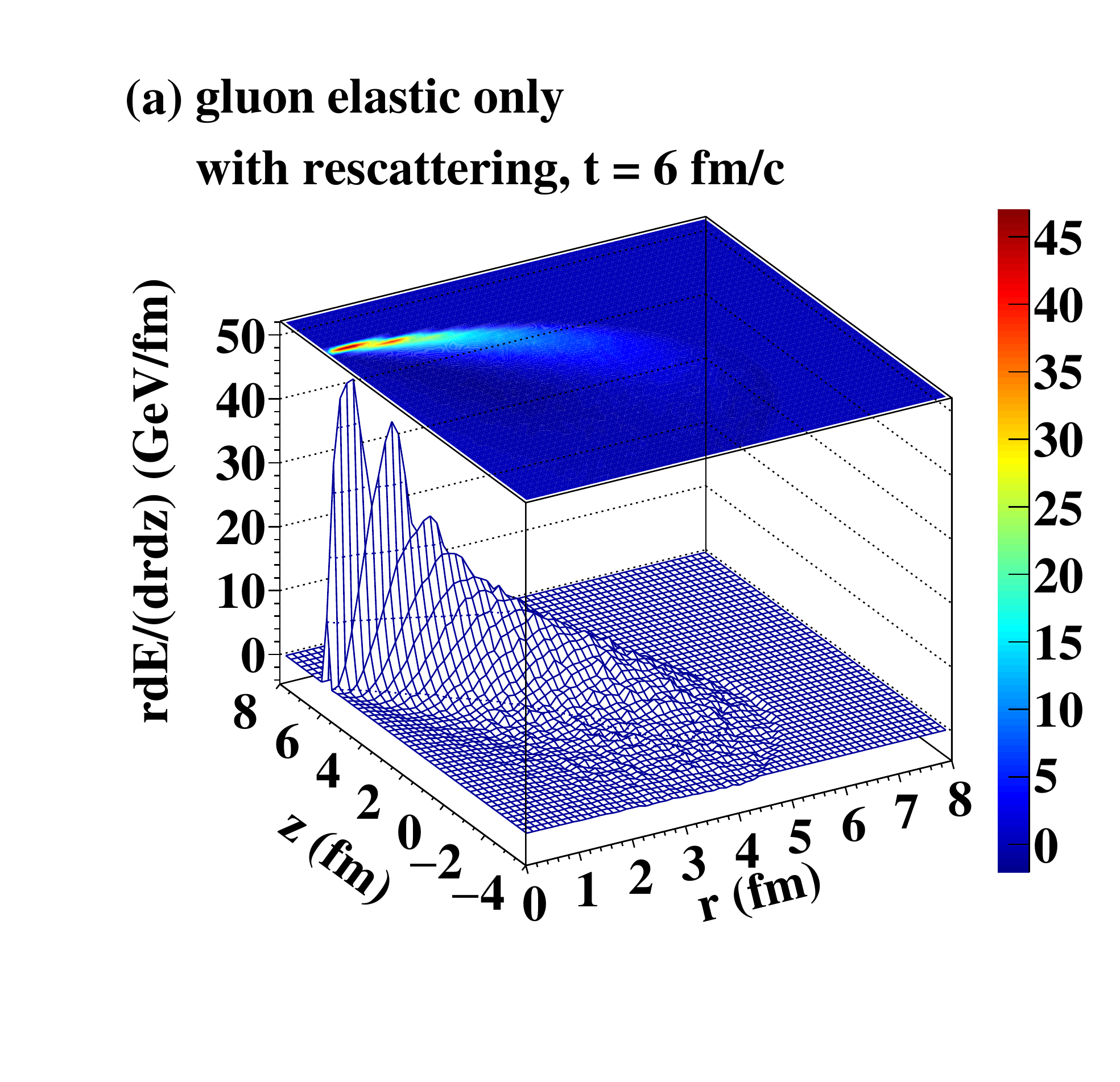}\\
\vspace{-0.8cm}
\includegraphics[width=7.5cm,bb=15 150 585 687]{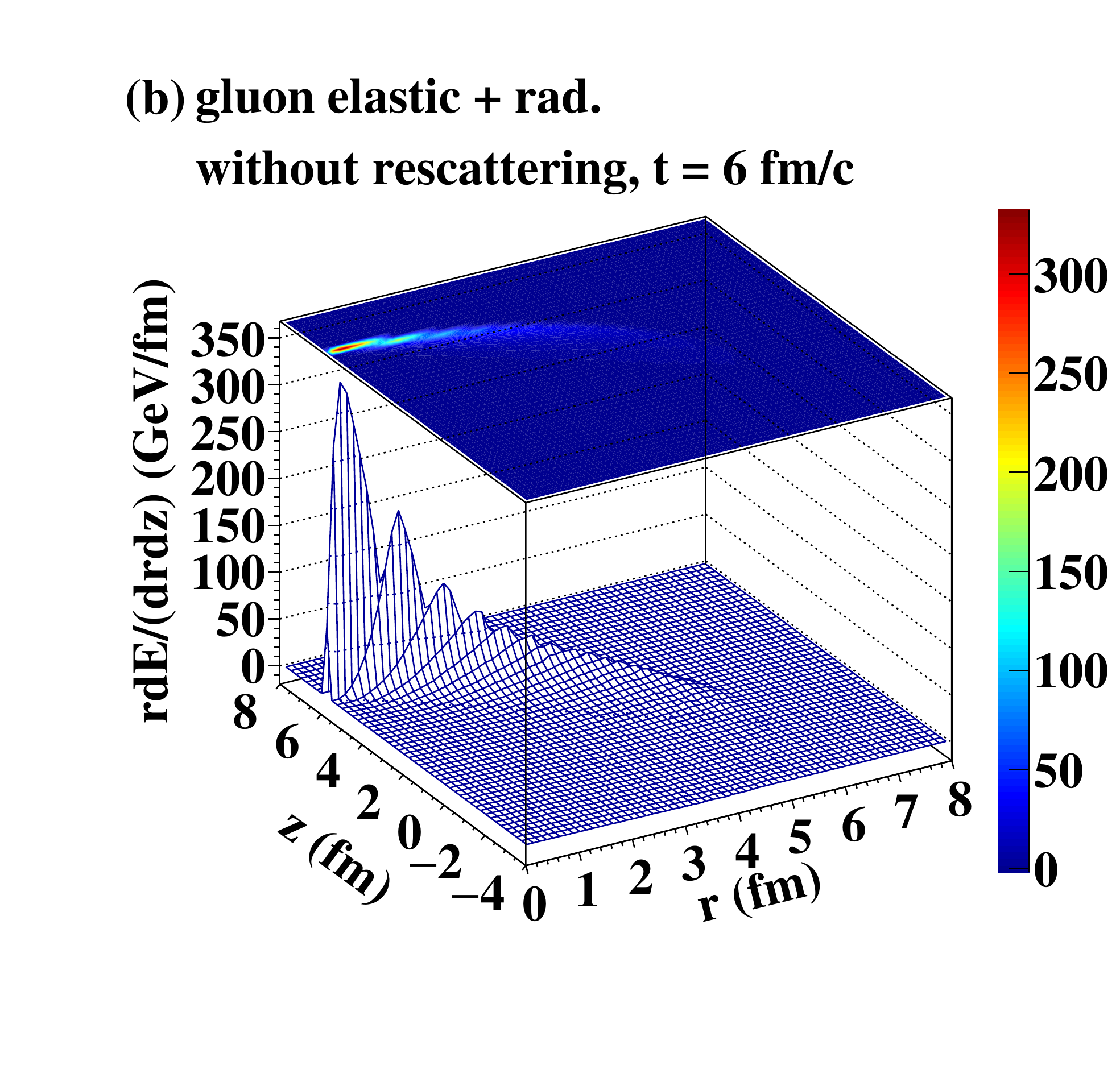}\\
\vspace{-0.8cm}
\includegraphics[width=7.5cm,bb=15 150 585 687]{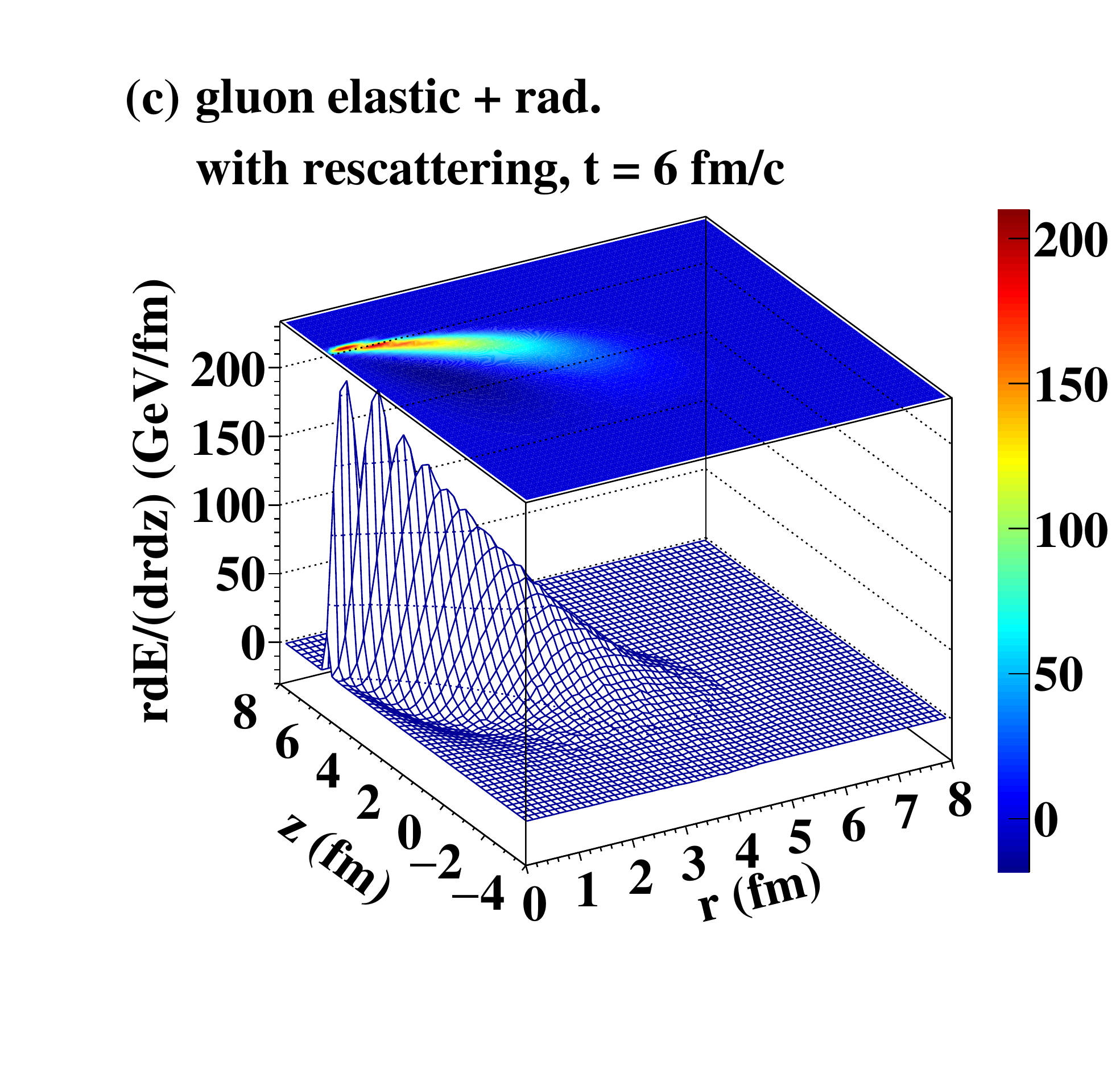}\\
\vspace{1.6cm}
\caption{(Color online) The same as Fig.~\ref{fig:rdedrdzq} except for a gluon.
}
\label{fig:rdedrdzg}
\end{figure}

\begin{figure}[!tbp]
\centering
\includegraphics[width=8.5cm]{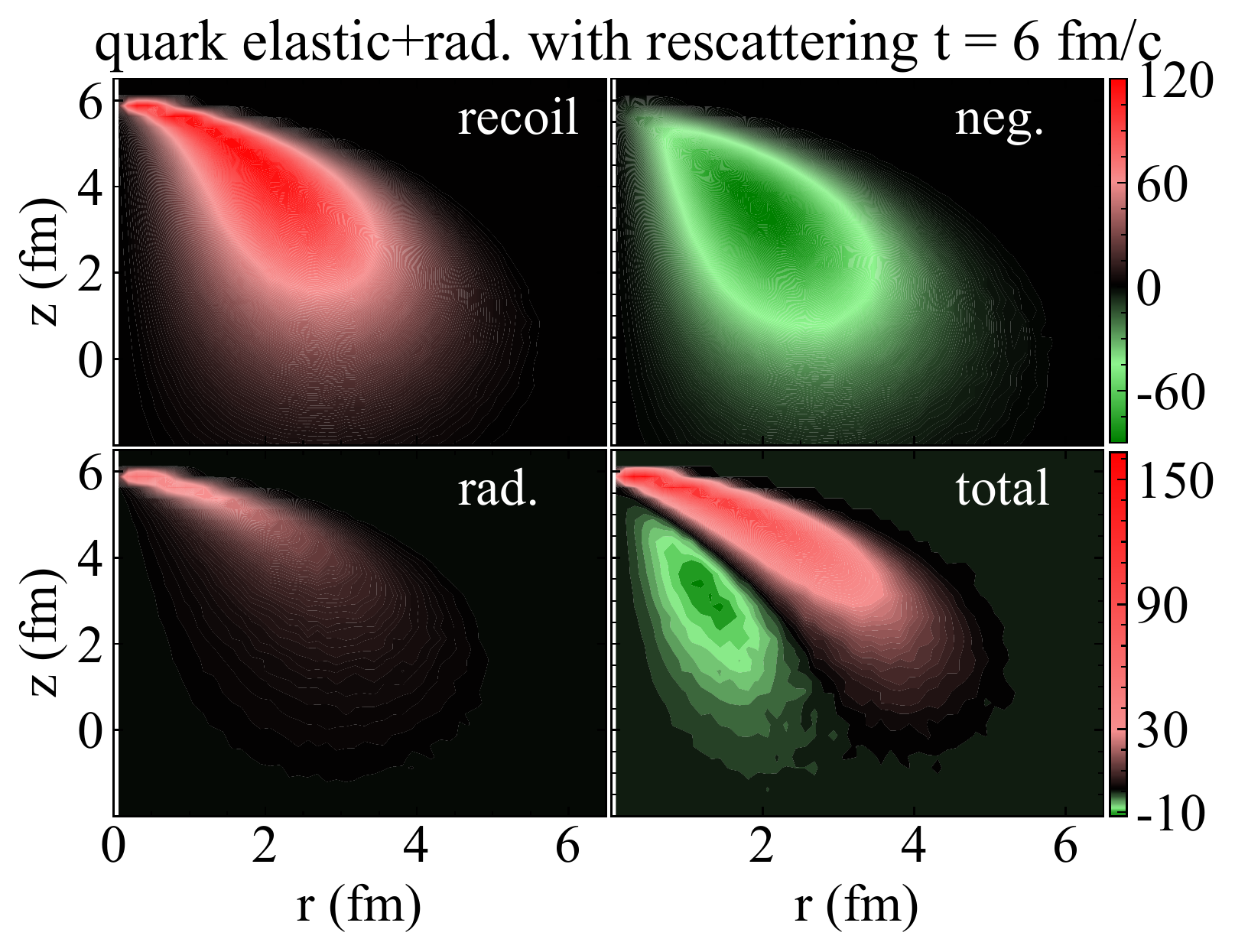}\\
\caption{(Color online) 
(Color online)  The same as the contour plot in Fig.~\ref{fig:rdedrdzq} (c) except for the energy density carried by recoil partons, “negative” partons, radiated gluons and their sum. Note the color scales in upper and lower panels are different for clarity.
}
\label{fig:wake}
\end{figure}

In Fig.~\ref{fig:rdedrdzq}, we first present the energy density distribution of a jet shower, containing radiated gluons from parton-medium interaction, recoil and ``negative'' partons but excluding the leading parton, initiated by a hard quark with an initial energy of $E_0=100$~GeV when it propagates through a static medium at a temperature of $T=300$~MeV. The hard quark starts its propagation at the origin ($x=y=z=0$) with its momentum along the longitudinal ($z$) direction. The energy density is calculated at the time of $t=6$~fm/$c$ in the coordinate space ($z$-$r$ plane) where $r$ denotes the transverse distance $r=\sqrt{x^2+y^2}$. The contribution from ``negative" partons is subtracted from the total energy of radiated gluons and recoil partons. The energy density distributions are averaged over many events.  In the figure, the upper panel is generated with the LBT simulations that only include the elastic scattering processes, in which rescatterings of recoil partons with the medium are allowed; the middle panel includes both elastic and inelastic processes, with rescatterings of radiated gluons and recoil partons switched off; while the lower panel represents the full LBT simulations that include both elastic and inelastic scatterings and allow rescatterings of radiated gluons and recoil partons. 

In the panels that include rescatterings, conic structures of the energy density distribution can be observed along the path of the jet parton.
In the upper panel that only includes elastic scatterings, this energy density distribution can approximate the Mach-cone structure of the medium response induced by the jet parton, where the wave front is produced by the propagation and medium interaction of recoil partons, corresponding to the hydrodynamic evolution of the energy deposited by the jet parton. The negative net distribution in the direction opposite to the jet propagation, known as the ``diffusion wake", is caused by the energy depletion when thermal partons are scattered out of their original phase space by the jet parton. Comparing the upper and lower panel, we observe significant enhancement of both the magnitude and the width of shock wave in the energy density distribution when inelastic scatterings are introduced, because the medium-induced gluons can further scatter with the medium and make the jet-induced medium excitation stronger and broader. 
Comparing the middle and the lower panels, we observe the wave front of the energy distribution is close to an arc shape when rescatterings are not included, a conic structure occurs and the shock wave is more diffused when rescatterings of the radiated gluons and recoil partons with the medium are allowed. These rescatterings with the medium slow down the average speed of parton propagation, which is equivalent to an effective velocity of sound in a hydrodynamic description of Mach-cone evolution that is smaller than the velocity of light (for massless partons) when interaction is considered in the calculation of the equation of state (EoS). Rescatterings also excite additional partons out of the medium and further broaden the shock wave induced by the jet parton much like the effect of viscosity in the hydrodynamic description
Similar features can also be seen in Fig.~\ref{fig:rdedrdzg} for a jet shower initiated by a gluon. Comparing Fig.~\ref{fig:rdedrdzq} and Fig.~\ref{fig:rdedrdzg}, we see a stronger medium excitation induced by a gluon jet than by a  quark jet due to the stronger interactions between a gluon and the medium. Note that the conic structures obtained from LBT simulations here are not rigorously Mach cones. Mach cones in a rigorous sense are only present in the hydrodynamic response to jet energy loss.

To better illustrate the energy density deposited into and depleted from the medium, we present in Fig.~\ref{fig:wake} the energy density contour plots from recoil partons, ``negative" partons, radiated gluons and their sum. One observes that the distribution of ``negative" partons overlaps with that of recoil partons and radiated gluons. They reduce the total energy density carried by recoil partons and radiated gluons around the forward direction of jet propagation. They become dominant in the backward region where the net total energy density becomes negative. This flow of negative net energy density is often referred to as the diffusion wake trailing the propagating jet parton.

\begin{figure}[!tbp]
\vspace{-2.0cm}
\includegraphics[width=7.5cm,bb=15 150 585 687]{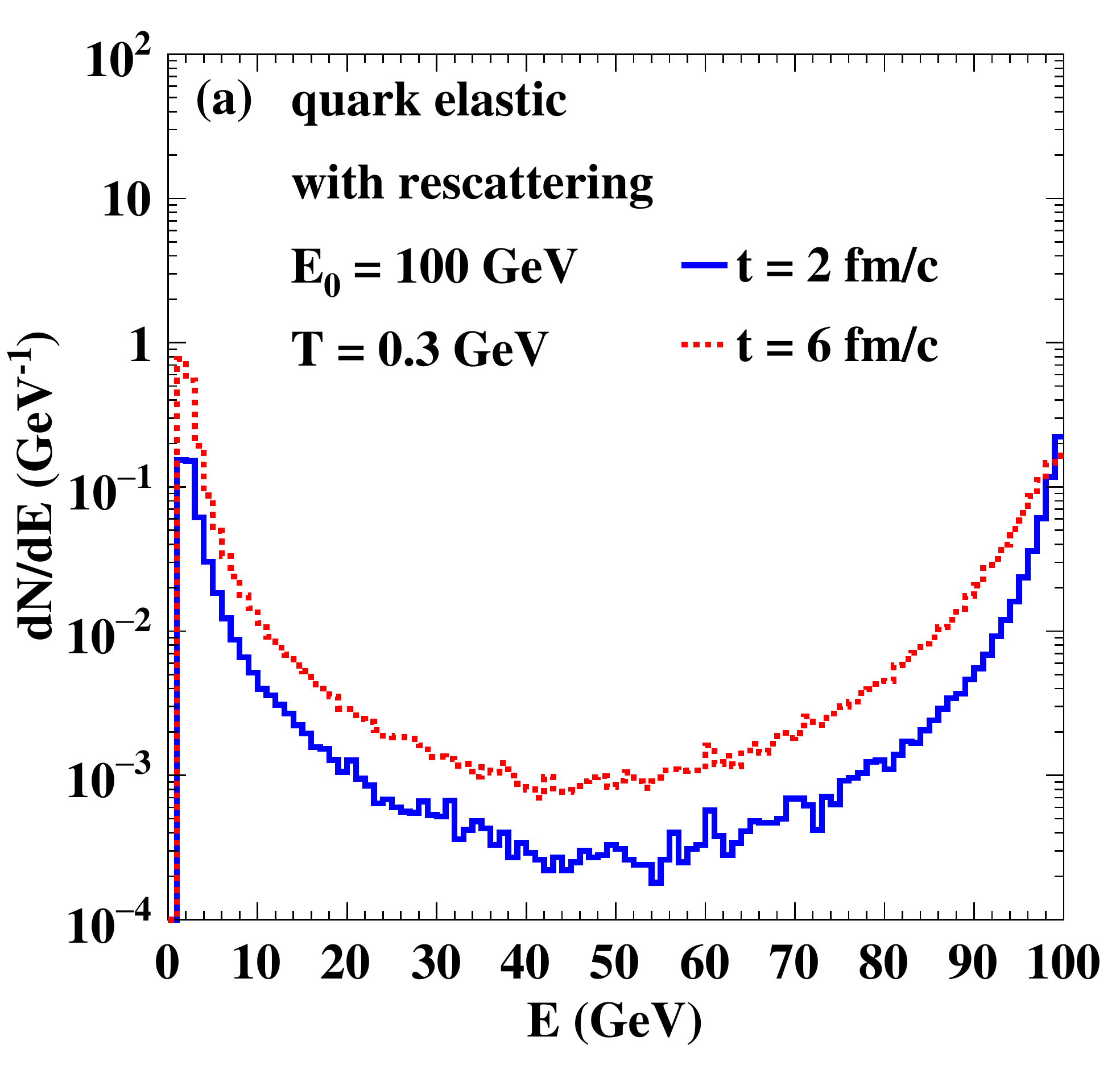}\\
\vspace{-0.82cm}
\includegraphics[width=7.5cm,bb=15 150 585 687]{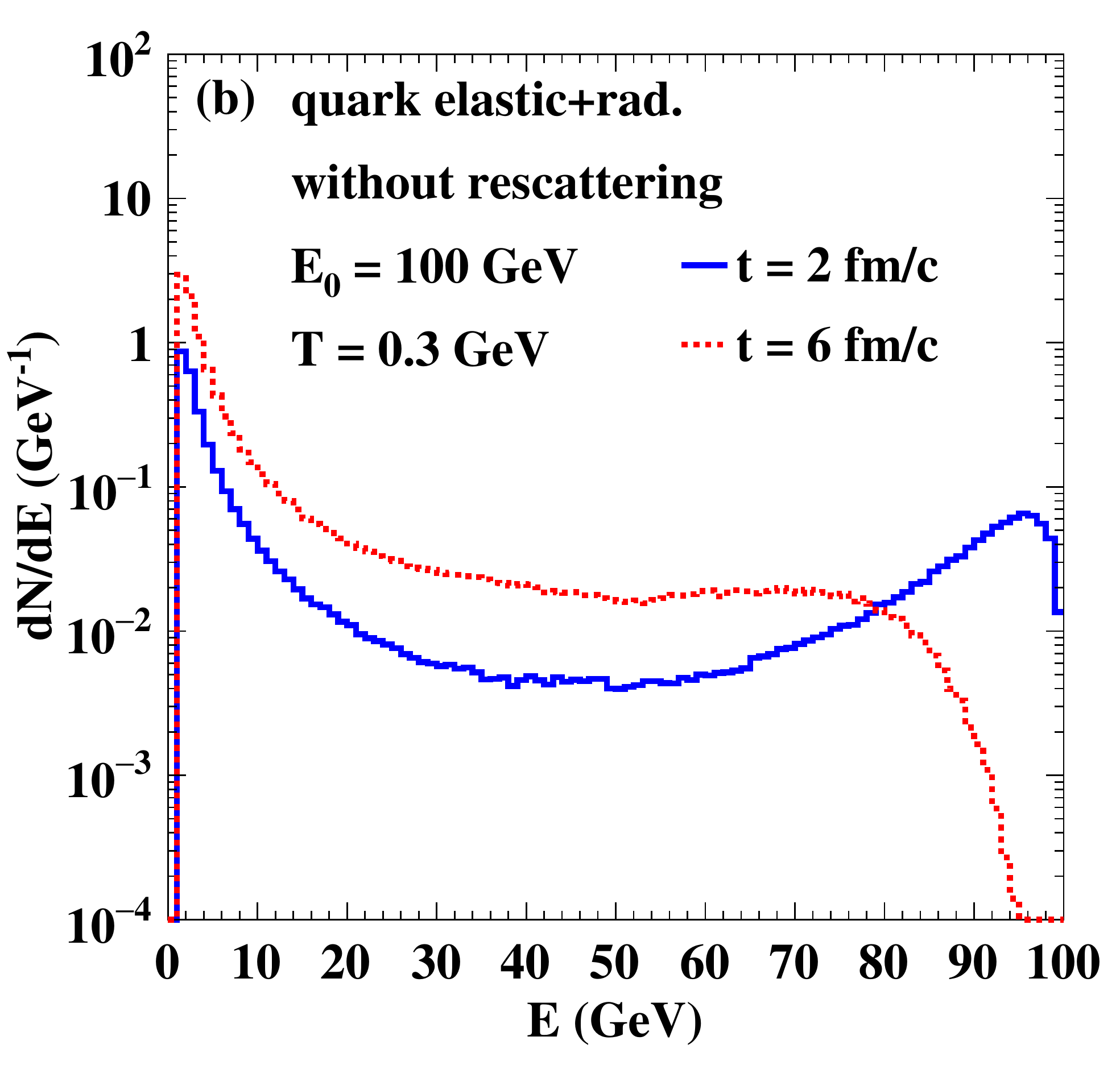}\\
\vspace{-0.82cm}
\includegraphics[width=7.5cm,bb=15 150 585 687]{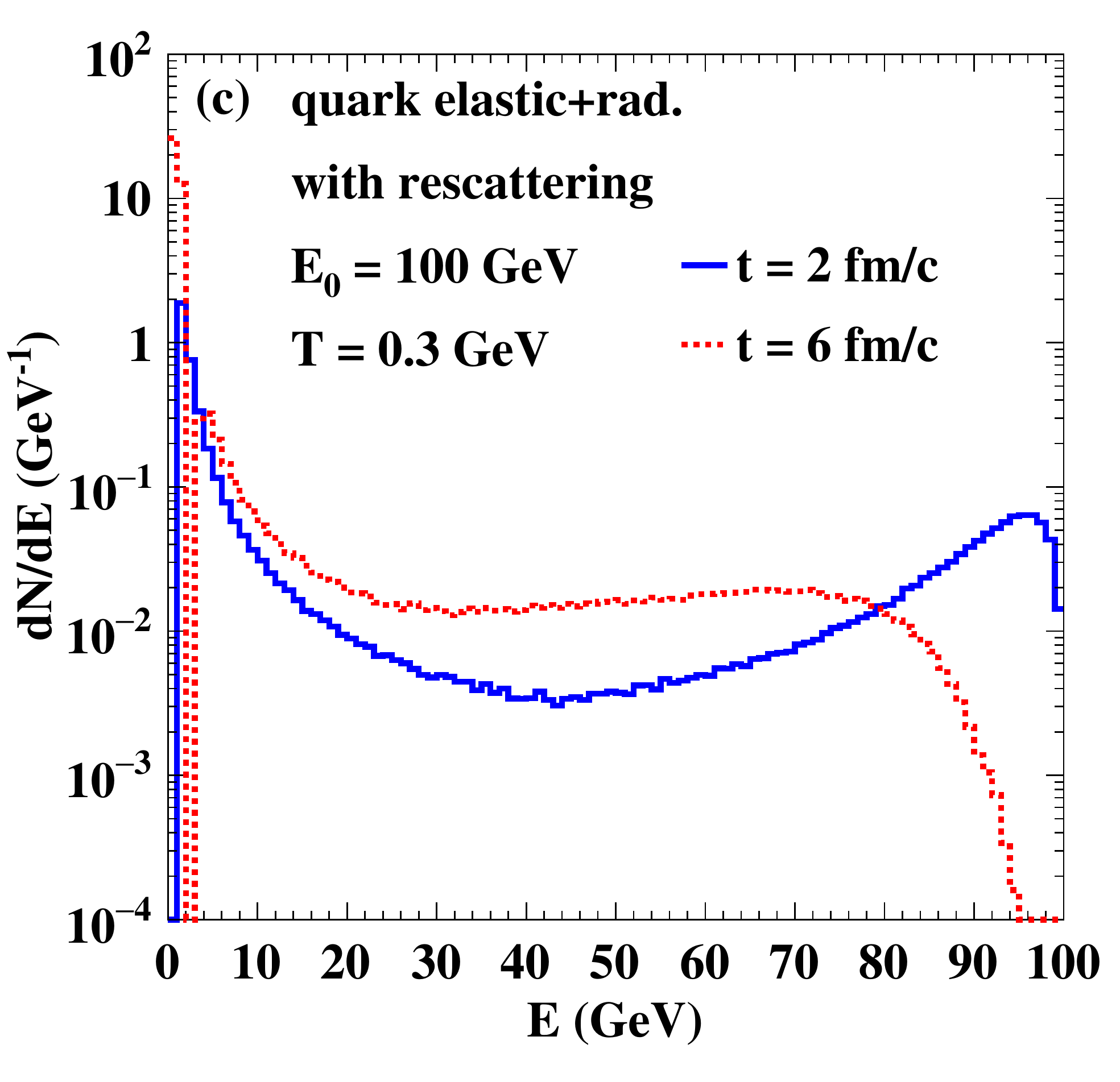}\\
\vspace{1.8cm}
\caption{(Color online) Energy distribution of partons inside a jet shower at different times developed from a 100~GeV quark in a static medium at $T = 300$~MeV, with only elastic scatterings with rescatterings (upper panel), elastic + inelastic scatterings without rescatterings (middle panel), and elastic + inelastic scatterings with rescatterings (lower panel).}
\label{fig:dndeq}
\end{figure}

\begin{figure}[!tbp]
\vspace{-2.0cm}
\includegraphics[width=7.5cm,bb=15 150 585 687]{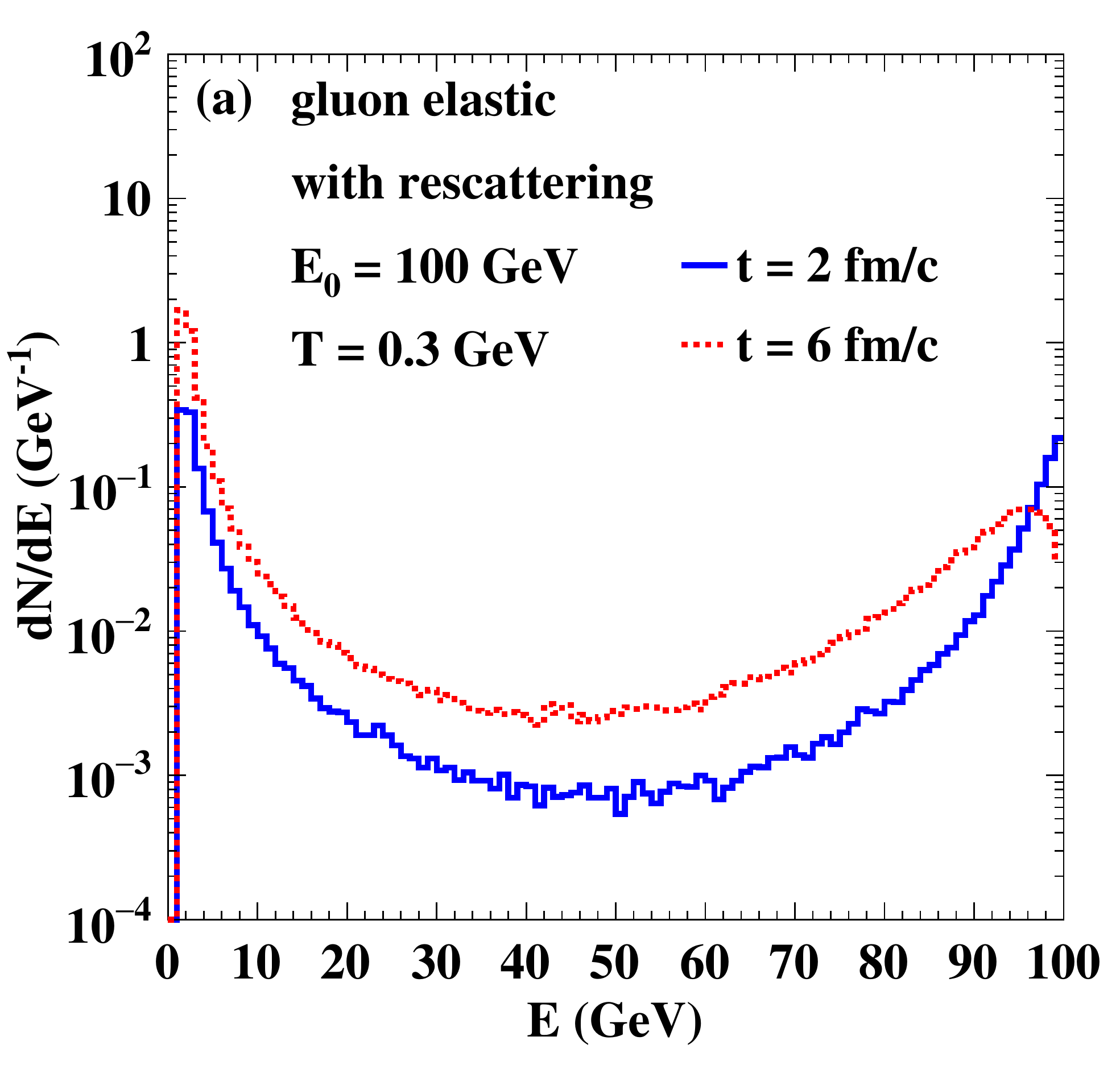}\\
\vspace{-0.82cm}
\includegraphics[width=7.5cm,bb=15 150 585 687]{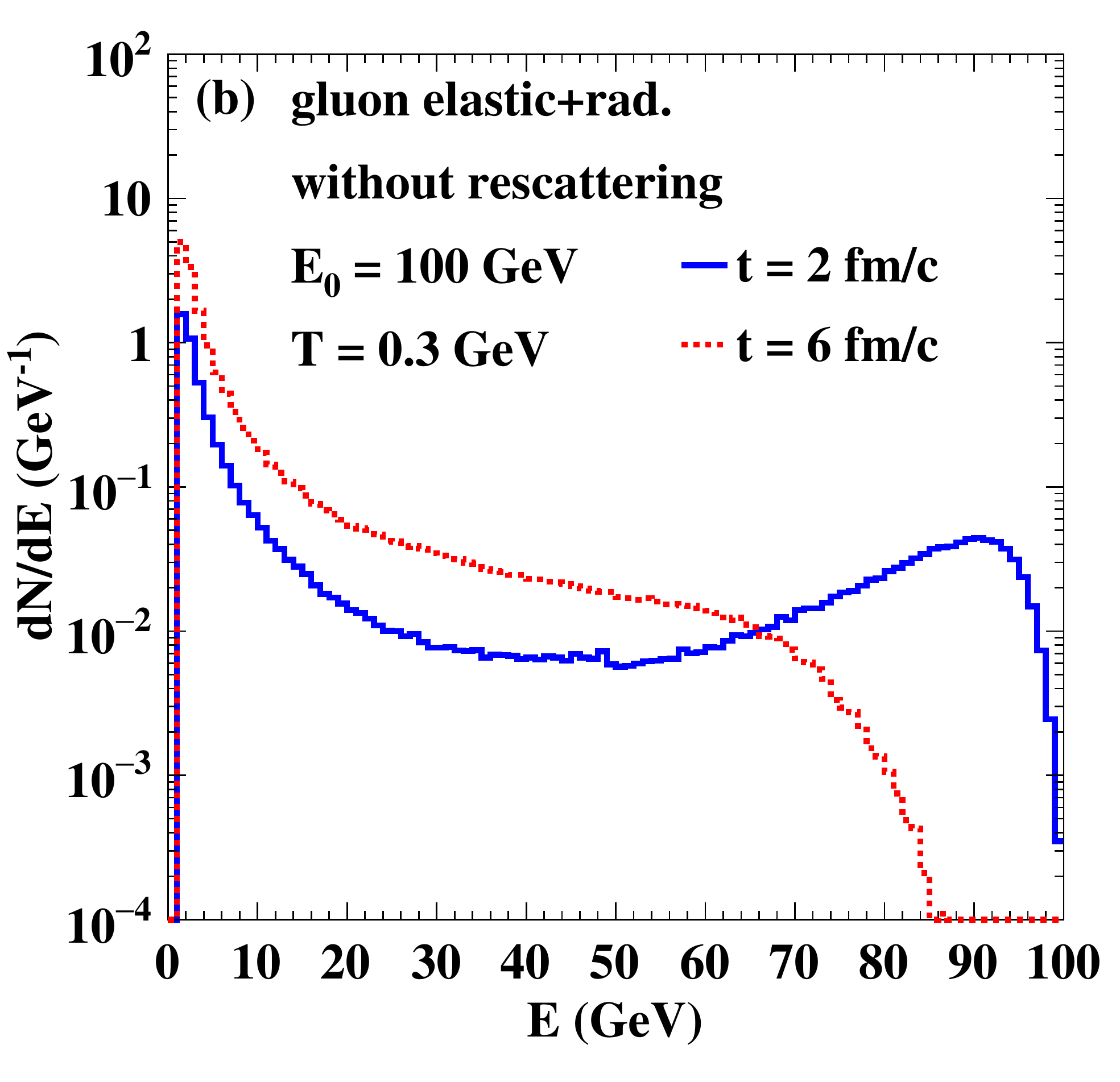}\\
\vspace{-0.82cm}
\includegraphics[width=7.5cm,bb=15 150 585 687]{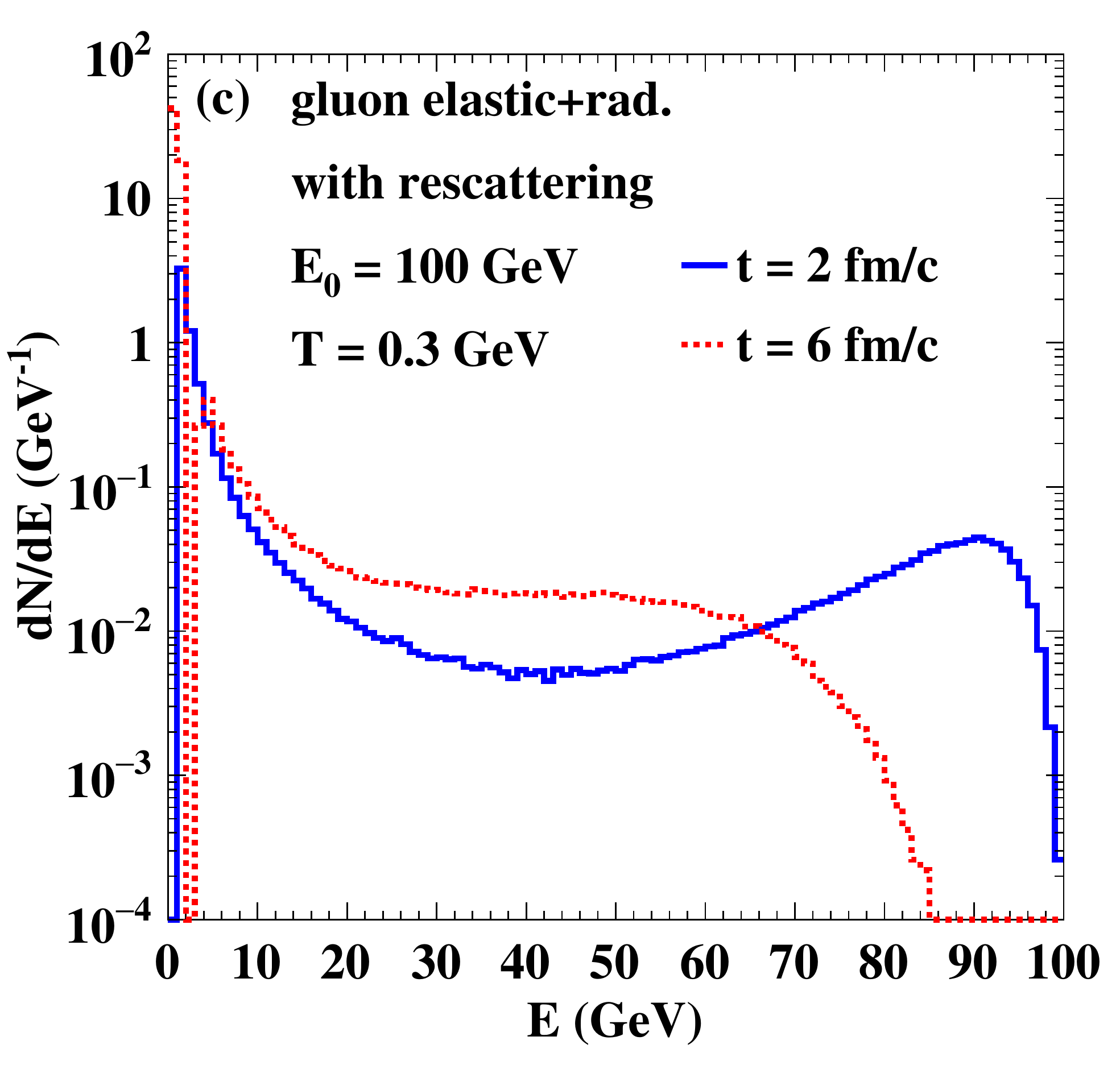}\\
\vspace{1.8cm}
\caption{(Color online) The same as Fig.~\ref{fig:dndeq} except for an initial gluon.
}
\label{fig:dndeg}
\end{figure}

\subsection{Energy distribution}

In the LBT simulations, by subtracting the energy-momentum of ``negative" partons from that of radiated gluons and recoil partons, we can rigorously respect the energy-momentum conservation in each scattering between a hard parton and a thermal medium parton. Apart from the energy taken by the medium-induced gluon radiations, the energy lost from the leading parton is transferred to the jet-induced medium excitation, i.e., the energy of recoil partons subtracted by that of ``negative" partons. To illustrate how the energy is transferred, we present in Fig.~\ref{fig:dndeq} the energy distribution of all partons within a jet shower as time evolves. Here, the jet shower is initiated by a 100~GeV quark through a static medium with a temperature of 300~MeV. From the top to the bottom panel, we compare simulations including only elastic scatterings with rescatterings of recoil partons, both elastic and inelastic scatterings without rescatterings of recoil partons and radiated gluons, and both elastic and inelastic scatterings with rescatterings of recoil partons and radiated gluons in the full LBT model. At early times, the energy distribution peaks at both the high energy and low energy ends, with the former contributed by the leading parton close to its initial energy and the latter contributed by medium response around the temperature scale. As time evolves, further energy loss of the leading parton suppresses the peak at the high energy end and broadens the high-energy peak towards the lower energy region. Meanwhile, successively induced medium excitation enhances the peak at the low energy end. Comparing the upper and lower panel, we observe the gluon emission process significantly accelerates the reduction and broadening of the high energy peak, while enhances the low energy peak from the medium response. Comparing the middle and lower panel, we observe rescatterings of the radiated gluons and recoil partons also enhance the effect of medium response at the low energy end, though have no impact on the leading parton distribution at the high energy end. In Fig.~\ref{fig:dndeg}, we present similar energy distribution of partons inside a jet shower initiated by a hard gluon. Due to the stronger interaction of a gluon with the medium than a quark, the energy spectrum of partons inside a gluon jet is broadened faster than inside a quark jet.

\begin{figure}
\includegraphics[width=6.5cm]{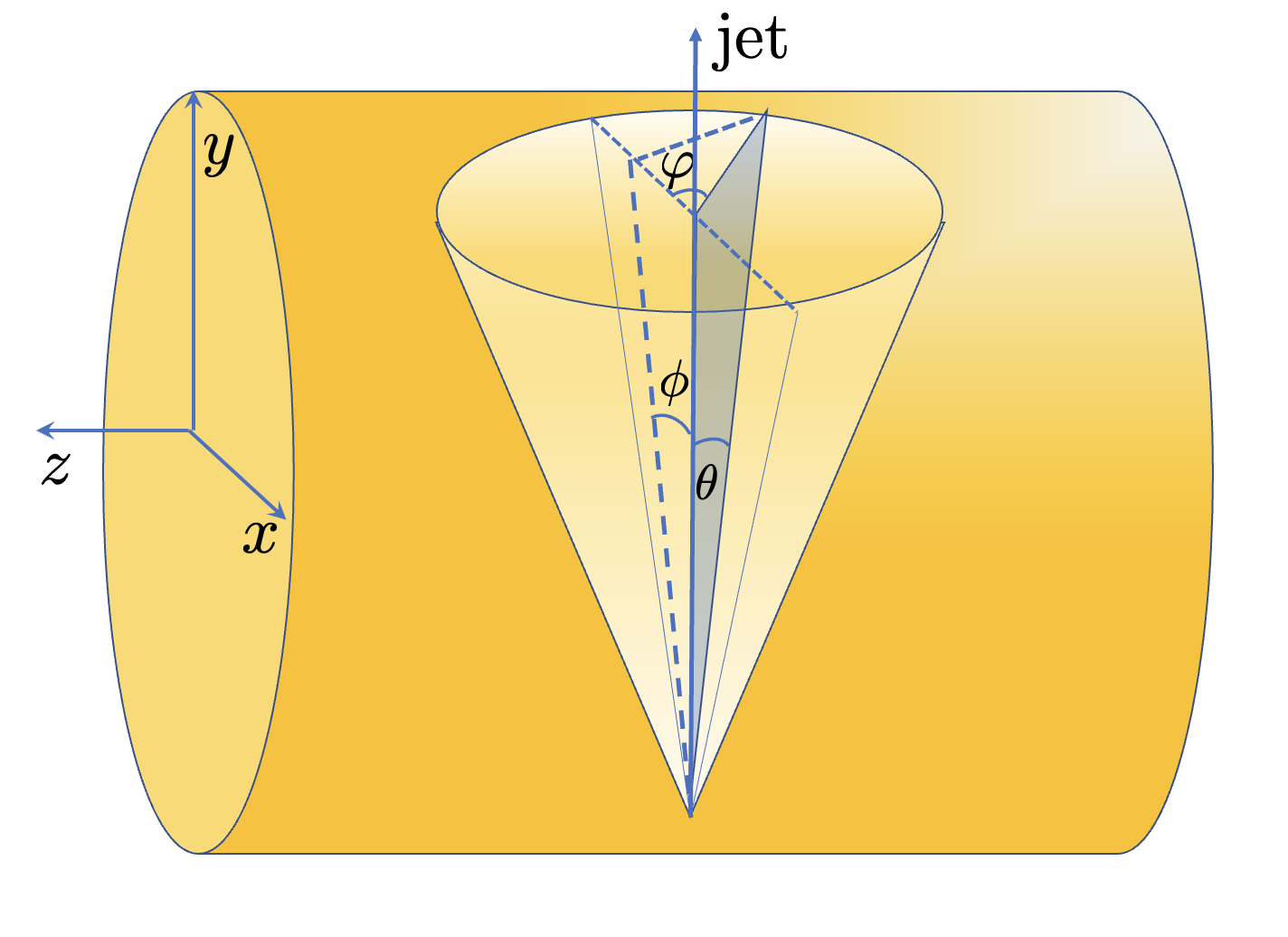}
\caption{(Color online) Projection from azimuthal angle $\varphi$ relative to the jet parton direction $y$ to the azimuthal angle $\phi$ in the transverse plane of the collision frame.}
\label{fig:dNdphi}
\end{figure}

\subsection{Angular distribution}

To have a more detailed illustration of how the energy is transferred from the leading parton to soft components inside a jet, and how a jet gets broadened as it evolves through the medium, we look at the azimuthal angle distributions of partons inside a medium-induced jet shower. In heavy-ion collisions, the most considered case of jet propagation is in the transverse direction with jet's rapidity $\eta$ set to 0. To imitate such propagation and the corresponding azimuthal angle distribution of the produced partons, one can imagine the longitudinal direction in a static and uniform medium is along such a transverse direction ($y$) in real heavy-ion collisions. One then can project a symmetric azimuthal angle ($\varphi$) distribution with respect to the initial jet parton direction to the angular ($\phi$) distribution in the transverse plane in the cylindrical frame of the heavy-ion collisions as illustrated in Fig.~\ref{fig:dNdphi}. This project can be written as
\begin{equation}
\frac{dN}{d\phi}=\int d\theta d\varphi \frac{dN}{d\theta d\varphi} \delta\left( \phi - \arctan[\tan \theta \cos\varphi]\right),
\end{equation}
where $\theta$ and $\varphi$ are the polar and azimuthal angles, respectively, relative to the initial jet parton direction $y$.  We refer to the angle $\phi$ in the transverse plane of the collisions as the projected azimuthal angle and the distribution as the projected azimuthal distribution.

\begin{figure}[!tbp]
\vspace{-2.0cm}
\includegraphics[width=7.5cm,bb=15 150 585 687]{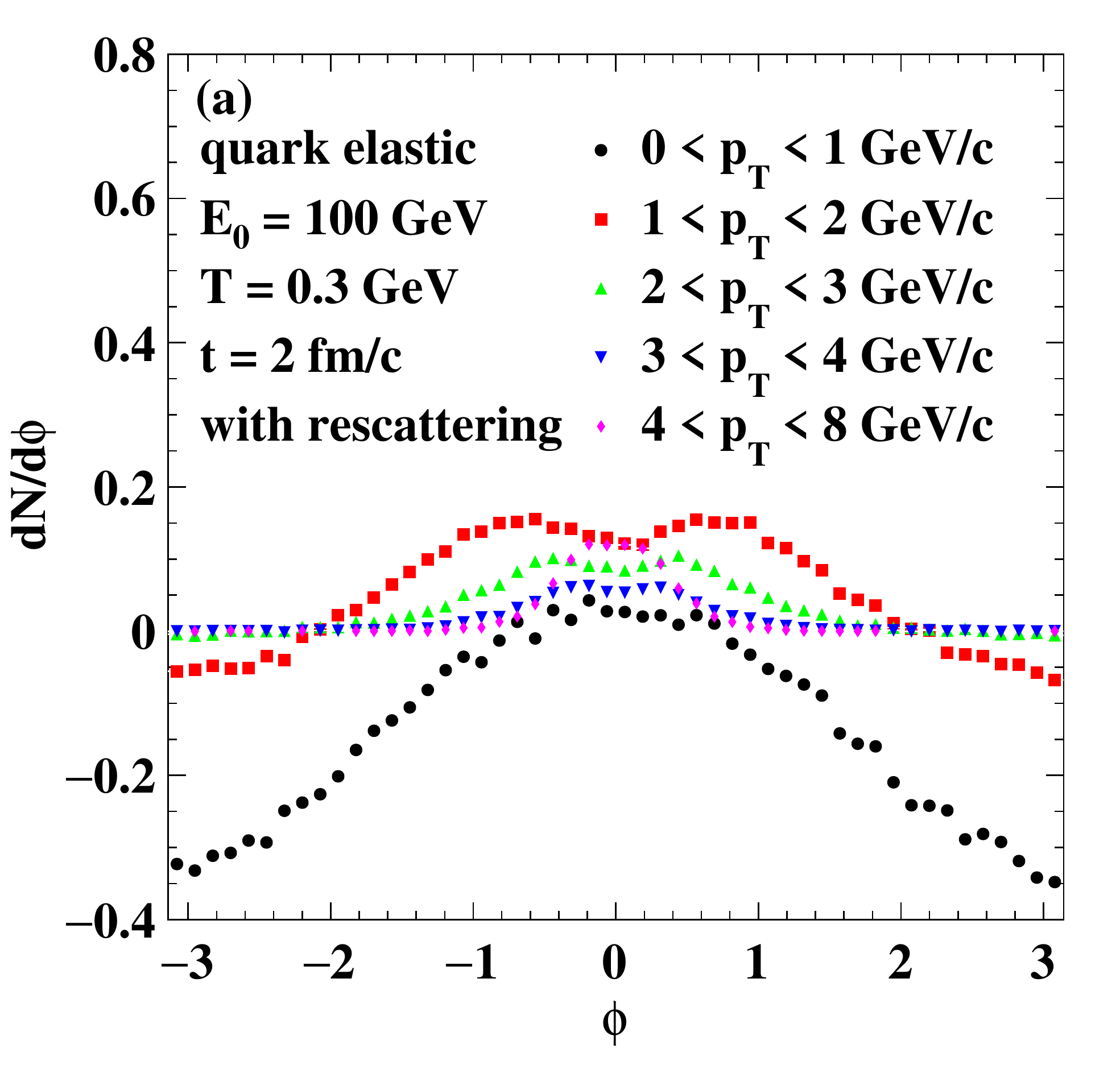}\\
\vspace{-0.84cm}
\includegraphics[width=7.5cm,bb=15 150 585 687]{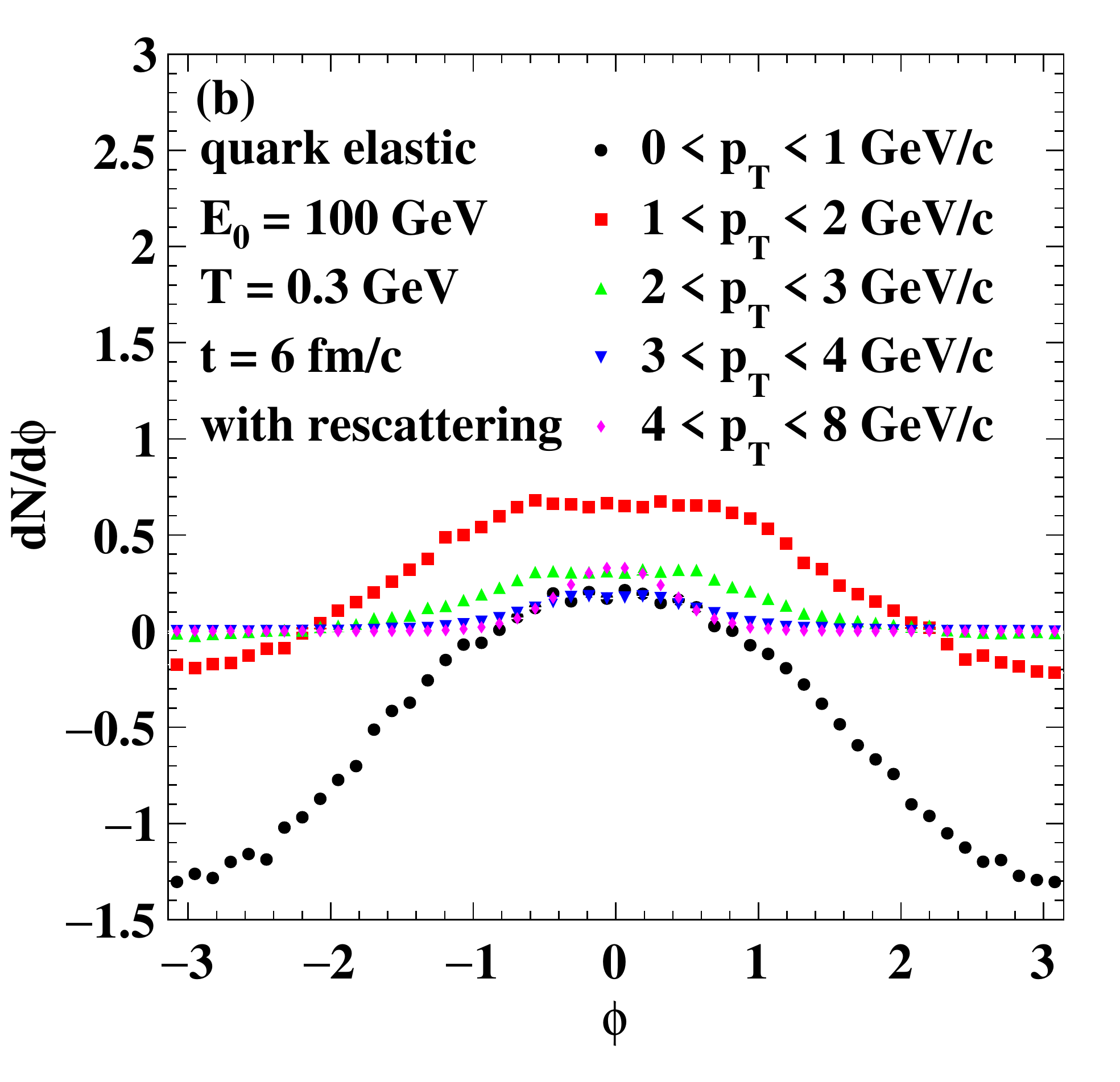}\\
\vspace{1.8cm}
\caption{(Color online)  Projected azimuthal angle distribution of partons within different energy intervals inside a jet shower developed from a 100~GeV gluon in a static medium at $T =300$~MeV. The upper panel shows results at time $t=2$~fm/$c$, and the lower panel at $t=6$~fm/$c$. Only elastic scatterings are included, with rescatterings of recoil partons.}
\label{fig:dngdphi_go}
\end{figure}

\begin{figure}[!tbp]
\vspace{-2.0cm}
\includegraphics[width=7.5cm,bb=15 150 585 687]{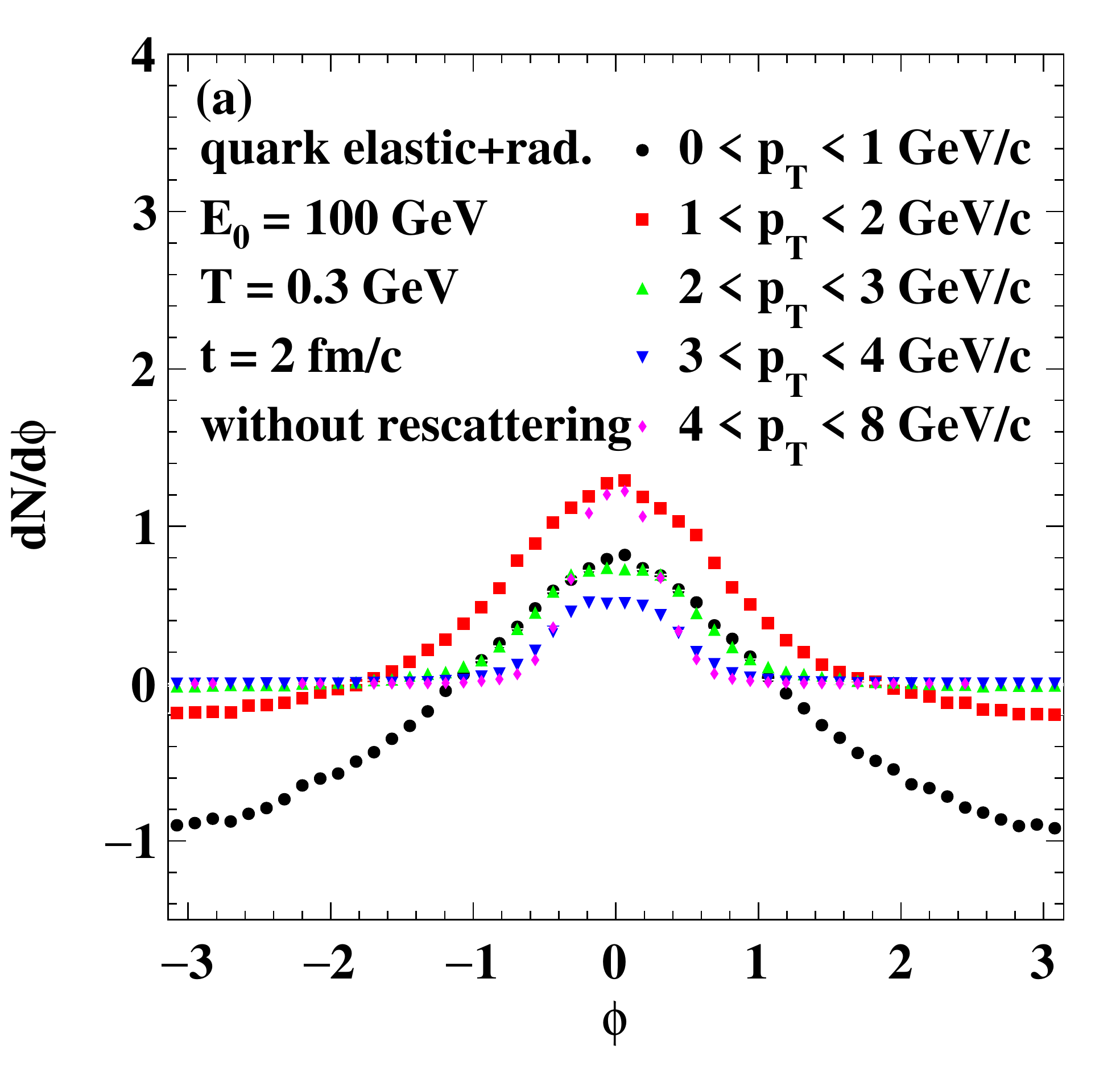}\\
\vspace{-0.84cm}
\includegraphics[width=7.5cm,bb=15 150 585 687]{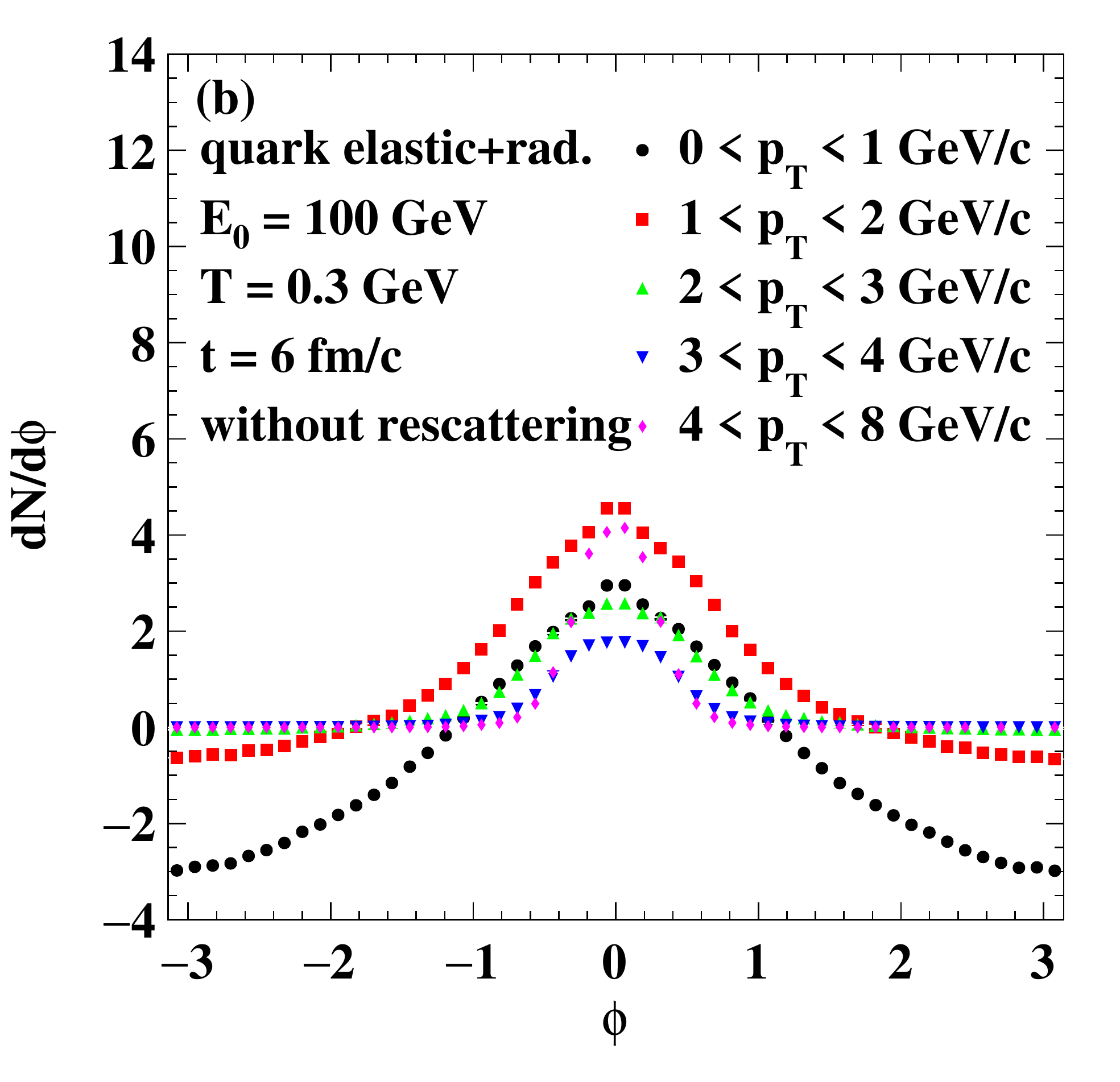}\\
\vspace{1.8cm}
\caption{(Color online) The same as Fig.~\ref{fig:dngdphi_go} except that both elastic and inelastic scatterings are included
without rescatterings of radiated gluons and recoil partons.
}
\label{fig:dngdphi_gr}
\end{figure}

\begin{figure}[!tbp]
\vspace{-1.8cm}
\includegraphics[width=7.5cm,bb=15 150 585 687]{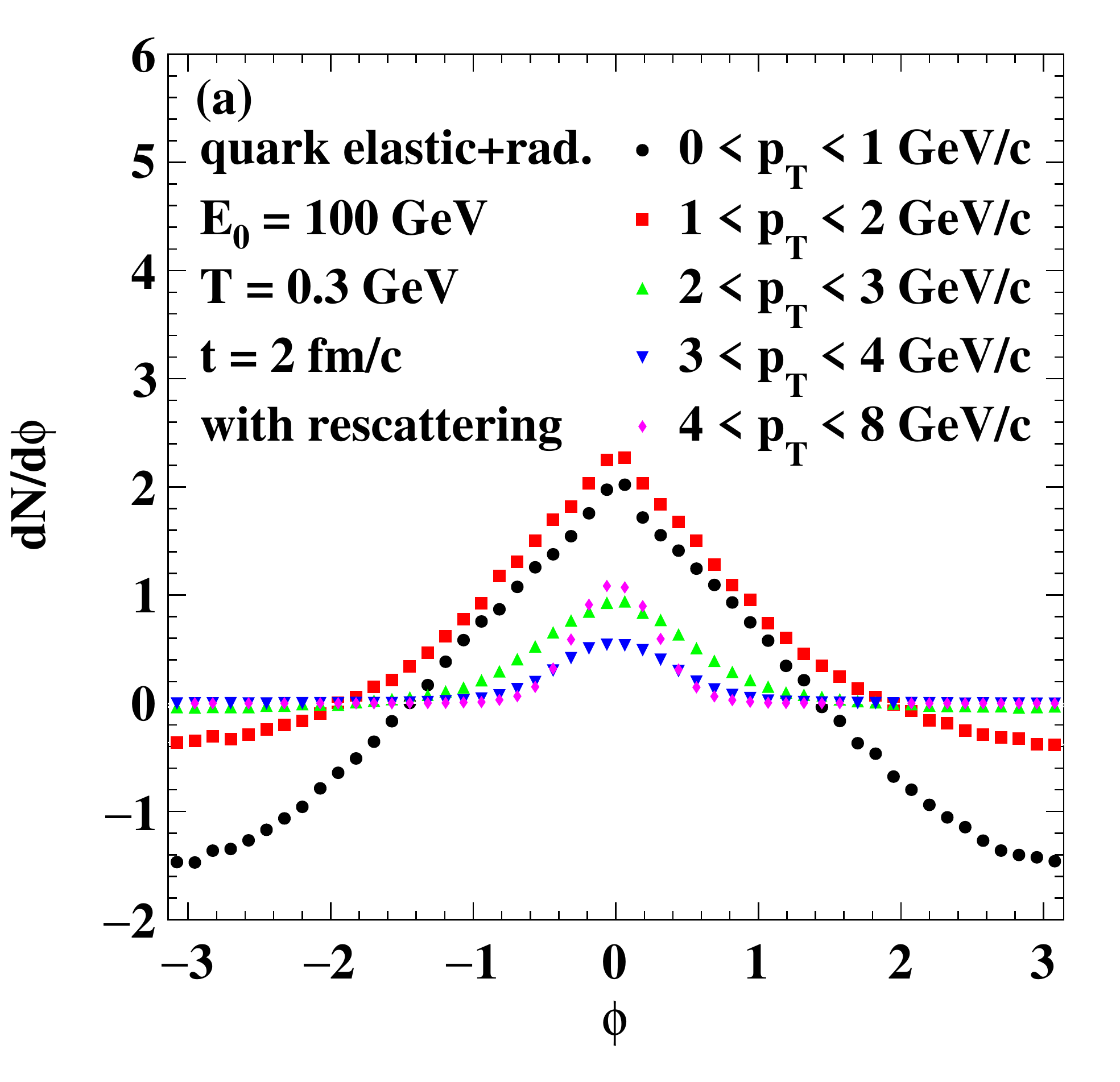}\\
\vspace{-0.84cm}
\includegraphics[width=7.5cm,bb=15 150 585 687]{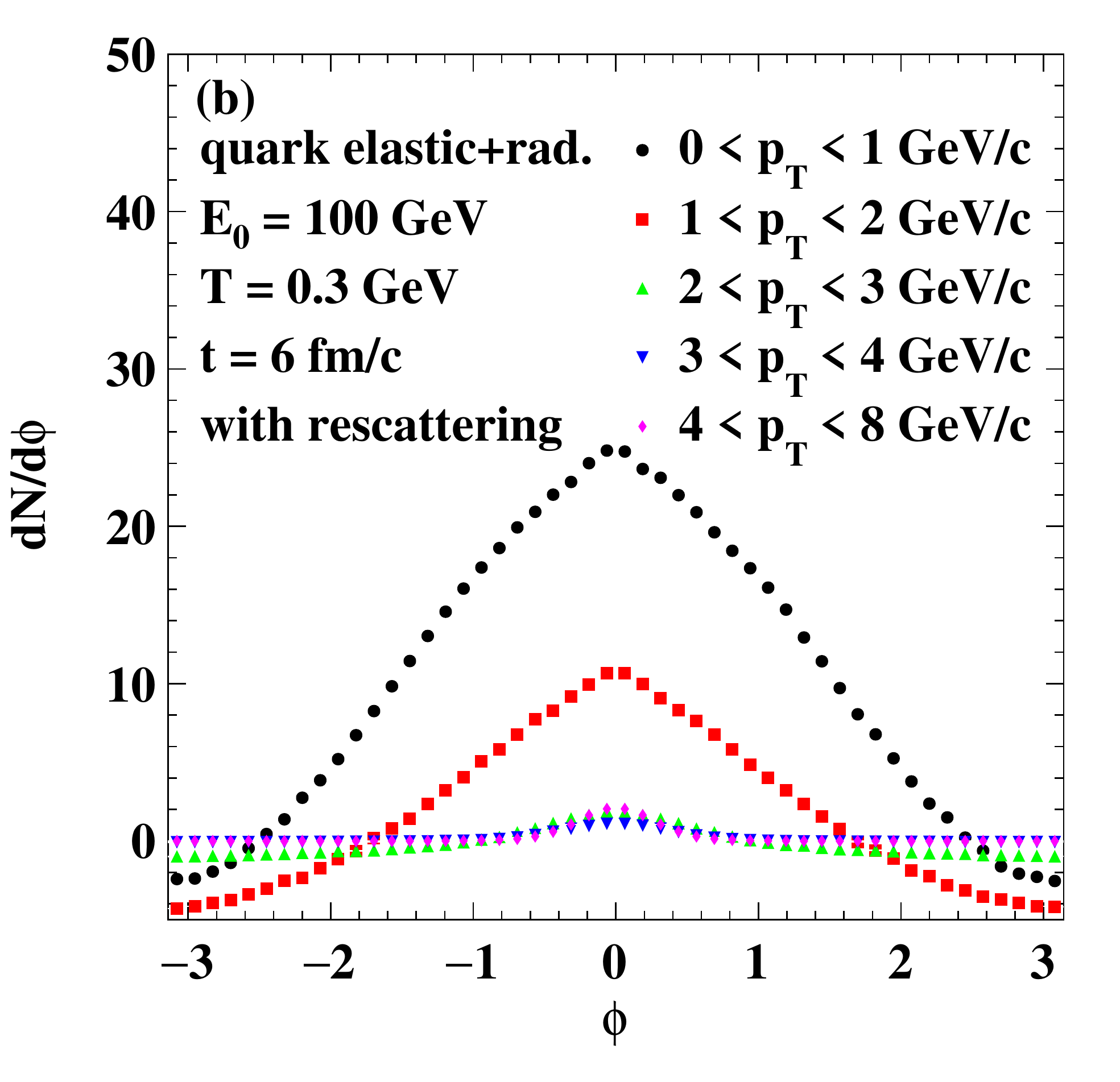}\\
\vspace{1.8cm}
\caption{(Color online) The same as Fig.~\ref{fig:dngdphi_go} except that both elastic and inelastic scatterings are included
with rescatterings of radiated gluons and recoil partons.
}
\label{fig:dngdphi_gw}
\end{figure}

Shown in Figs.~\ref{fig:dngdphi_go} -~\ref{fig:dngdphi_gw} are the projected azimuthal angle distributions of jet-induced medium partons (radiated gluons and recoil partons with ``negative'' partons subtracted) within different ranges of the parton energy $p_{\rm T}=\sqrt{p_x^2+p_y^2}$. The jet parton originates from a 100~GeV quark and evolves inside a static medium with a constant temperature of $T=300$~MeV. In each figure, the upper panel shows the angular distributions at $t=2$~fm/$c$ and the lower panel at $t=6$~fm/$c$. 

In Fig.~\ref{fig:dngdphi_go}, we first study the angular distributions of soft partons produced by pure elastic scatterings. One can observe that the net distribution of very low energy partons ($p_{\rm T}$=0 \textendash ~1~GeV) is mostly negative, because this range is dominated by the ``negative" partons left behind the jet parton inside the medium when medium partons are scattered out into recoil partons. In the momentum space, these ``negative" partons center around the direction opposite to the jet. In contrast, the distribution of recoil partons peaks around the jet parton direction in the range (1 \textendash ~2~GeV) that is slightly higher than the thermal energy scale considering the energy they gain from scatterings with the leading parton. In a small angle scattering approximation, the average energy transfer per scattering is \cite{He:2015pra},
\begin{equation}
    \langle\delta E_{\rm el}\rangle =\frac{1}{\Gamma^a_{\rm el}} \left\langle \Gamma^a_{\rm el}\frac{q_\perp^2}{2\omega} \right \rangle \approx \frac{18\pi^3}{84\zeta(3)}\alpha_{\rm s}T 
    \ln\frac{2.6ET}{\mu_{\rm D}^2}.
\end{equation}
This energy scale for recoil partons should increase logarithmically with the jet parton's energy. For $\alpha_{\rm s}=0.3$, $E=100$ GeV and and $T=300$ MeV, $\delta E_{\rm el}\approx 2.5$ GeV, consistent with what we see in Fig.~\ref{fig:dngdphi_go}.

This distribution of recoil partons exhibits a double peak structure in the projected azimuthal angle at early times, which results from the typical transverse momentum transfer $\mu_{\rm D}$ between the hard parton and the medium constituent in the LBT model simulations. As time evolves, more recoil and ``negative" partons are generated. Meanwhile, the two peaks in the projected azimuthal angle distribution of recoil partons are smeared and in the end merge due to the successive rescatterings of both the jet parton and recoil partons with the medium. Although the impact of $\mu_\mathrm{D}$ on the particle distribution here can be easily diminished by other factors of jet-medium interactions, it worths further investigation considering that $\mu_\mathrm{D}$ is a fundamental quantity characterizing the medium properties, as recently explored in Ref.~\cite{Yang:2023dwc}.

When the gluon radiation process is included in the simulations, as shown in Figs.~\ref{fig:dngdphi_gr} and~\ref{fig:dngdphi_gw}, significantly more soft particles are produced through medium-induced radiation as the jet evolves through the medium. Clearly, the energy deposition (positive value of the distribution function) can be observed in the forward direction ($\phi\approx 0$) even for the softest energy range (0 \textendash ~1~GeV), though energy depletion can still be seen in the backward direction ($\phi\approx \pi$). Since gluons from medium-induced radiation are dominantly produced in the collinear direction with respect to the hard parton, the angular distributions of partons within all energy ranges peak at $\phi=0$. Since gluons from medium-induced radiation are dominantly produced in the collinear direction with respect to the hard parton, the angular distributions of partons within all energy ranges peak near $\phi = 0$. Note that the LPM suppression previously shown in Fig.~\ref{fig:scale-angle} at very small angle ($\theta<1/\sqrt{\Delta t E} \approx 1/25$ for a 100 GeV parton traveling through a 1 \textendash ~2~fm medium) cannot be observed in these angular distributions due to contributions from recoil partons, multiple gluon emissions and rescatterings they experience. Recent studies show, however, a new subjet observable known as energy-energy correlator can resolve the angular structure caused by the LPM interference of gluon emissions induced by multiple scattering in the QGP medium~\cite{Andres:2022ovj,Yang:2023dwc}. Including rescatterings of radiated gluons and recoil partons (Fig.~\ref{fig:dngdphi_gw}) generates much broader distributions of the soft partons than not including these rescatterings (Fig.~\ref{fig:dngdphi_gr}).

\section{Jet modification in a static medium}
\label{sec:jetInBrick}

In this section, we study properties of fully reconstructed jets originating from a single hard jet parton in a static and uniform medium. Constituents of these reconstructed jets include the leading jet parton, medium-induced gluons, recoil partons, and ``negative'' partons whose energies are subtracted from the jet energy within the jet cone.  
A jet in high-energy collisions is defined as a cluster of collimated hadrons or partons reconstructed within a cone,
\begin{equation}
\sqrt{(\eta-\eta_\mathrm{jet})^2+(\phi-\phi_\mathrm{jet})^2} \le R,
\end{equation}
based on a given jet finding algorithm~\cite{Cacciari:2011ma} and a particular background subtraction scheme. In the expression above, $\eta$ ($\eta_\mathrm{jet}$) and $\phi$ ($\phi_\mathrm{jet}$) are the pseudorapidity and azimuthal angle of the final state parton/hadron (jet), respectively, and $R$ is a pre-defined jet cone size.
In our present study, we assume the effect of hadronization on the reconstructed jet energy is small and calculate reconstructed jet energy using the final state partons in LBT simulations. 
For other observables such as the jet fragmentation functions one  should include hadronization in the future. 

We use a modified version of the FASTJET package~\cite{Cacciari:2011ma} to reconstruct jets with the anti-$k_\mathrm{T}$ algorithm, in which $\eta$ and $\phi$ of ``negative" partons are calculated in the same way as for regular partons, while their energy-momentum is subtracted from that of the regular partons in each iteration of the jet finding algorithm. All jet partons from the LBT simulations, including the leading parton, gluons from medium-induced radiations, recoil partons and ``negative" partons, are fed to the FASTJET package. This corresponds to a perfect subtraction of the unperturbed portion of the QGP background.

\begin{figure}[!tbp]
\vspace{-1.8cm}
\includegraphics[width=7.5cm,bb=15 150 585 687]{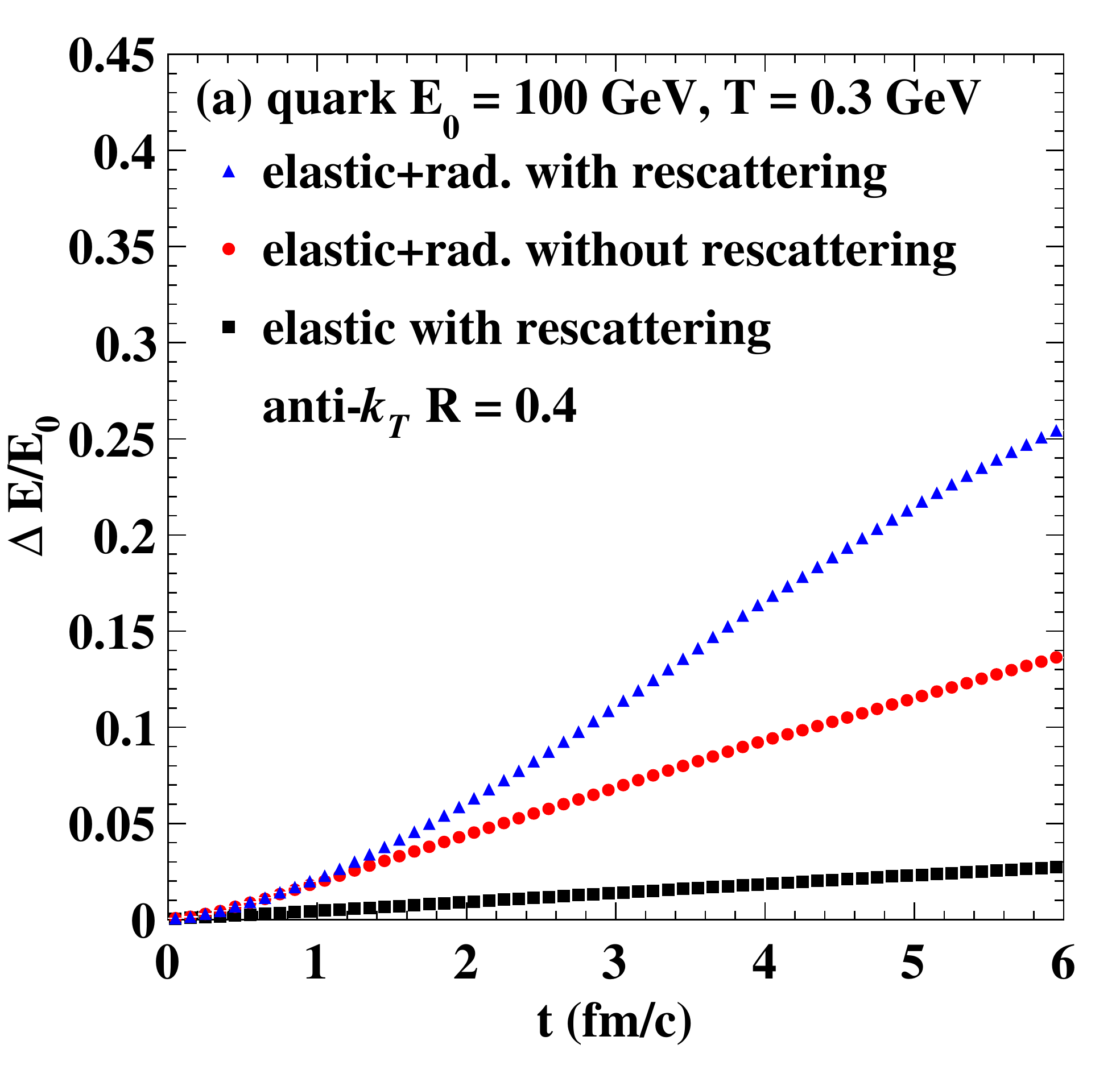}\\
\vspace{-0.84cm}
\includegraphics[width=7.5cm,bb=15 150 585 687]{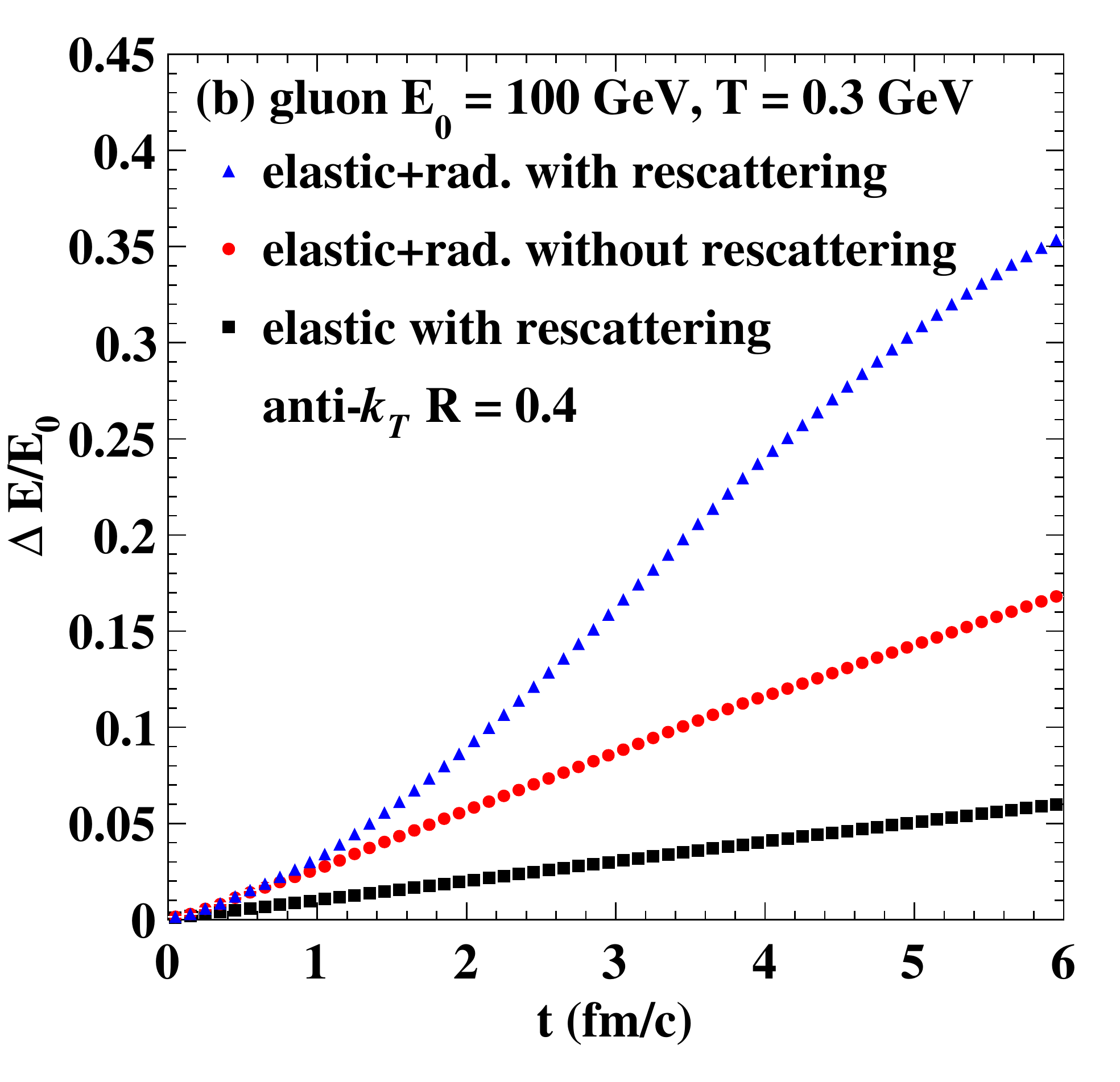}\\
\vspace{1.8cm}
\caption{(Color online) Accumulated fractional energy loss $\Delta E/E_{0}$ of a jet with cone size $R=0.4$ as a function of the evolution time including only elastic scatterings with rescatterings (solid), elastic + inelastic scatterings without rescatterings (dotted), and elastic + inelastic scatterings with rescatterings (dot-dashed). The jet is initiated with a 100~GeV quark (upper panel) or gluon (lower panel), propagating through a static medium at $T = 300$~MeV.}
\label{fig:elossjetf}
\end{figure}

\subsection{Jet energy loss}

During the propagation of a fast parton, it loses energy and experiences transverse momentum broadening through multiple scatterings with the medium partons. The lost energy is carried by radiated gluons and jet-induced medium partons which continue propagating through the medium and interacting with the medium partons. While some fraction of the lost energy from the leading parton will still remain inside the jet as long as its carriers are inside the jet cone, some fraction will be transported outside the jet cone via large-angle scatterings, large-angle emissions and successive diffusion of the jet-induced medium partons. The latter results in a reduction of the energy of a reconstructed jet. Shown in Fig.~\ref{fig:elossjetf} is the accumulated energy loss of a reconstructed jet with a cone size of $R=0.4$ as a function of the evolution time. The jet is initiated by either a quark (upper panel) or a gluon (lower panel) with an initial energy of 100~GeV and travels through a static medium at a temperature of 300~MeV. In each panel, one may observe the jet energy loss is significantly enhanced after the processes of medium-induced gluon radiation are included and a significant fraction of the radiated gluons are emitted outside the jet cone as the LPM interference suppresses collinear gluon radiations. The jet energy loss also becomes larger when rescatterings of radiated gluons and recoil partons are allowed, since these rescatterings further transport partons from small to large angles with respect to the jet axis. For each of the three model settings in each panel, the jet energy loss appears to increase approximately linearly with time (or distance). This is different from the feature of the energy loss for a single parton previously shown in Figs.~\ref{fig:elosspartonQ} and~\ref{fig:elosspartonG} due to the different energy loss mechanisms for a single parton and a reconstructed jet. As expected, the energy loss of a gluon jet (lower panel) is larger than that of a quark jet (upper panel). Effects of different components of jet-induced medium excitation on the jet energy loss were investigated in our earlier work~\cite{He:2018xjv}, where we found including contributions of recoil partons reduces the jet energy loss, while including ``negative" partons increases the jet energy loss.

\begin{figure}[!tbp]
\vspace{-1.8cm}
\includegraphics[width=7.5cm,bb=15 150 585 687]{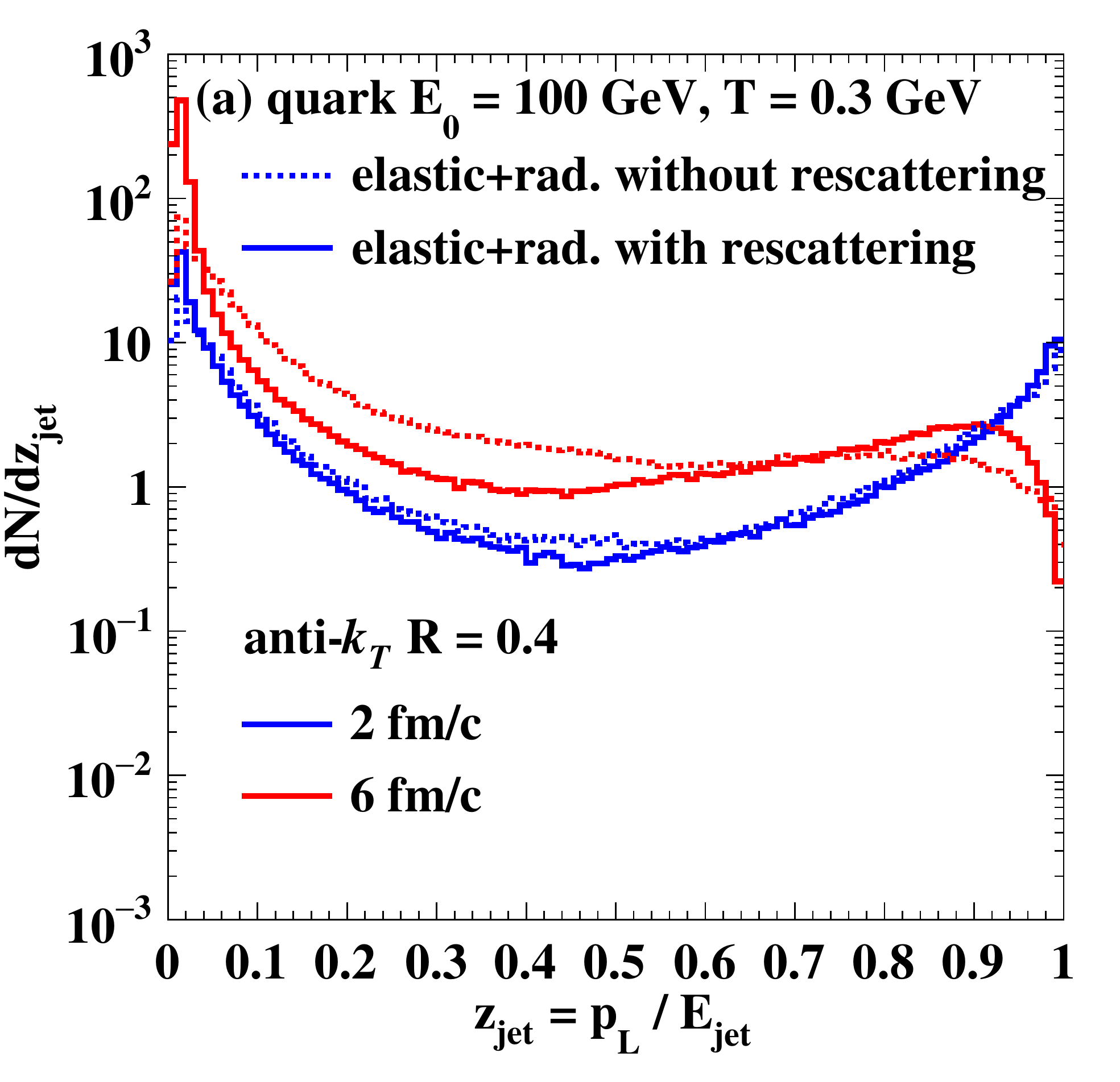}\\
\vspace{-0.82cm}
\includegraphics[width=7.5cm,bb=15 150 585 687]{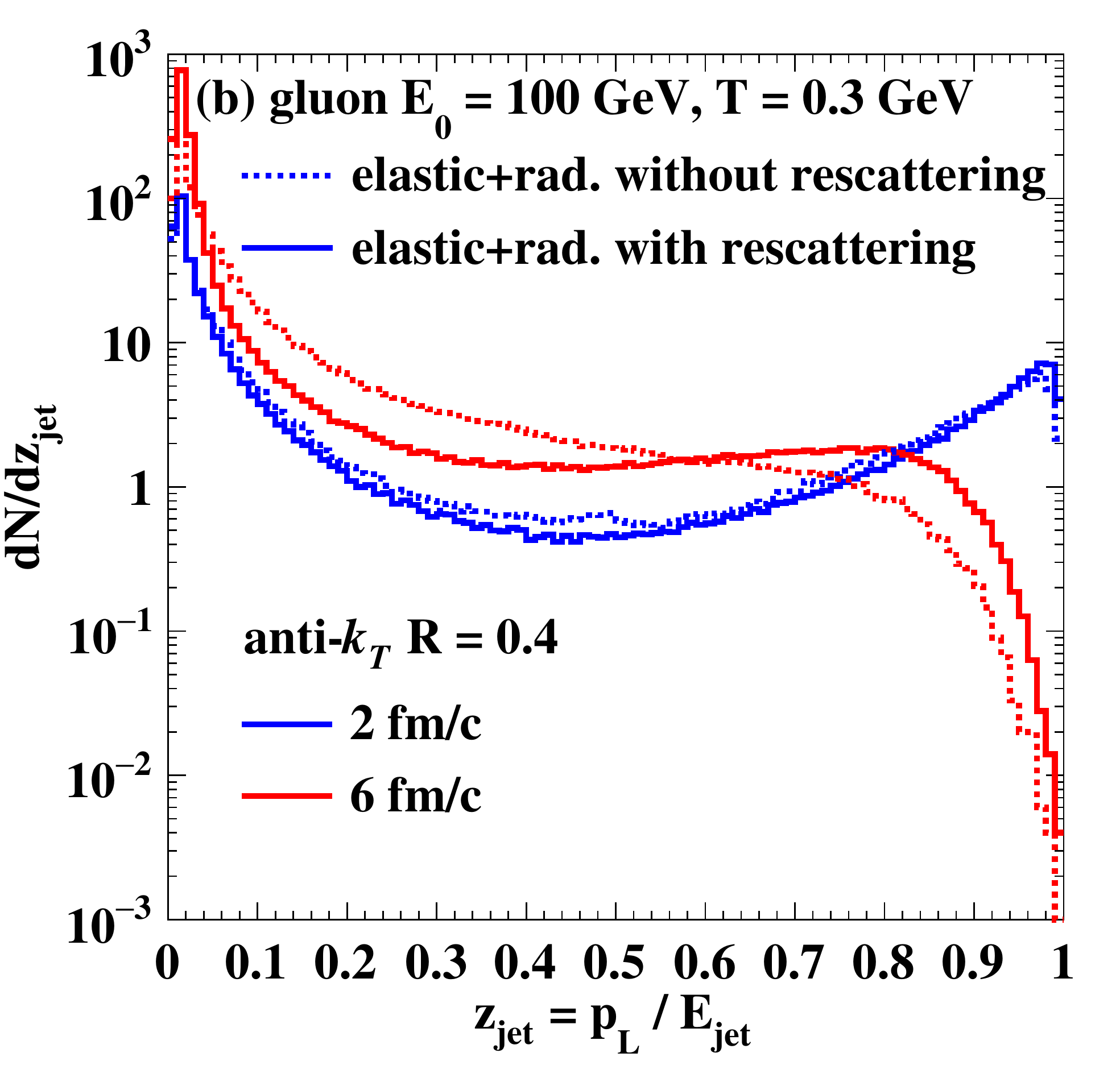}\\
\vspace{1.8cm}
\caption{(Color online) Fractional longitudinal momentum ($z_\mathrm{jet} = p_\mathrm{L}/E_\mathrm{jet}$) distribution of constituent partons within a jet with a cone size $R=0.4$ at different evolution times, with (solid) and without (dashed) rescatterings of radiated gluons and recoil partons. The jet is initiated with a 100~GeV quark (upper panel) or gluon (lower panel), propagating through a static medium at $T = 300$~MeV with both elastic and inelastic scatterings switched on.}
\label{fig:fragFnc}
\end{figure}

\begin{figure}[!tbp]
\vspace{-1.8cm}
\includegraphics[width=7.5cm,bb=15 150 585 687]{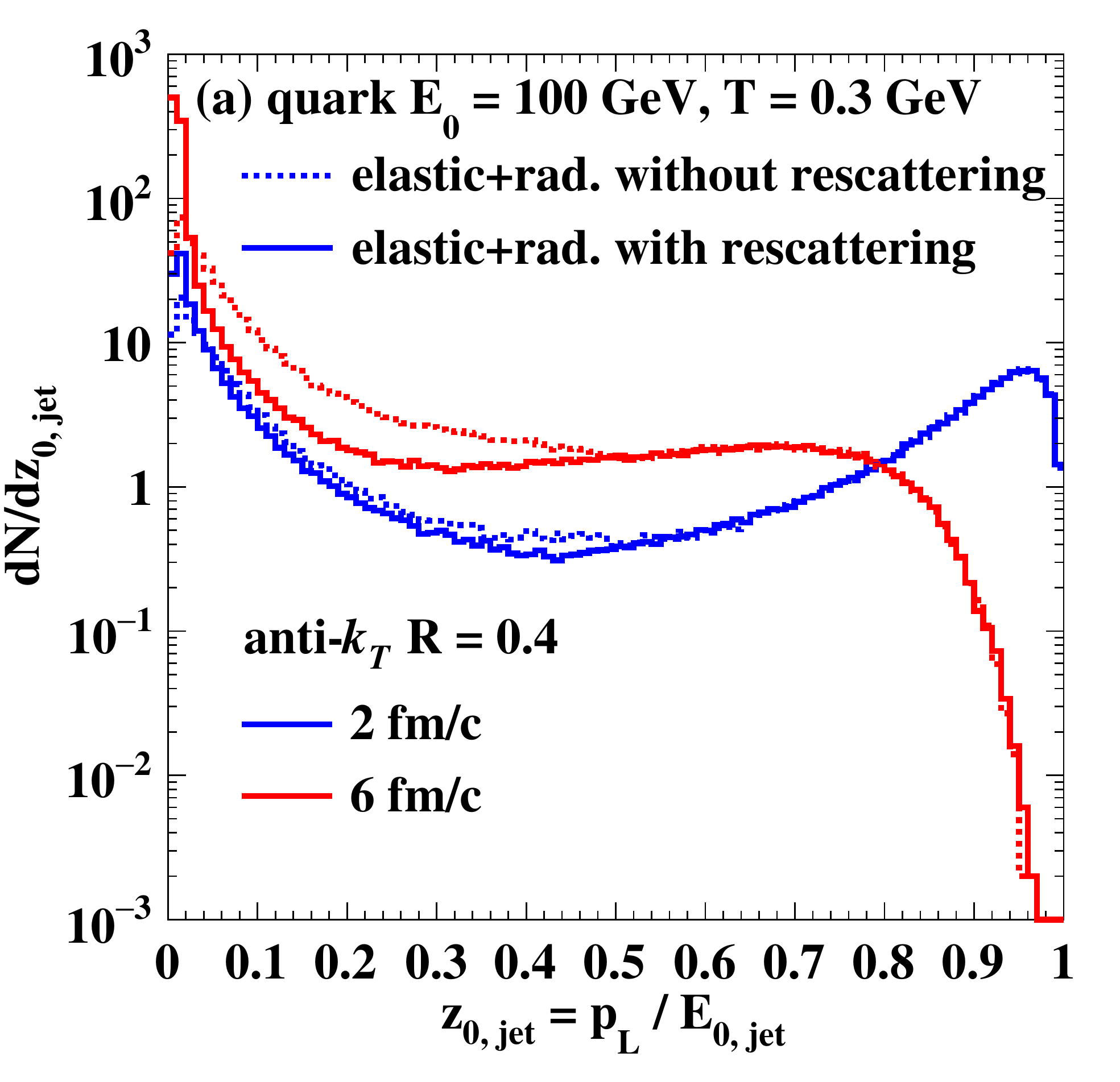}\\
\vspace{-0.82cm}
\includegraphics[width=7.5cm,bb=15 150 585 687]{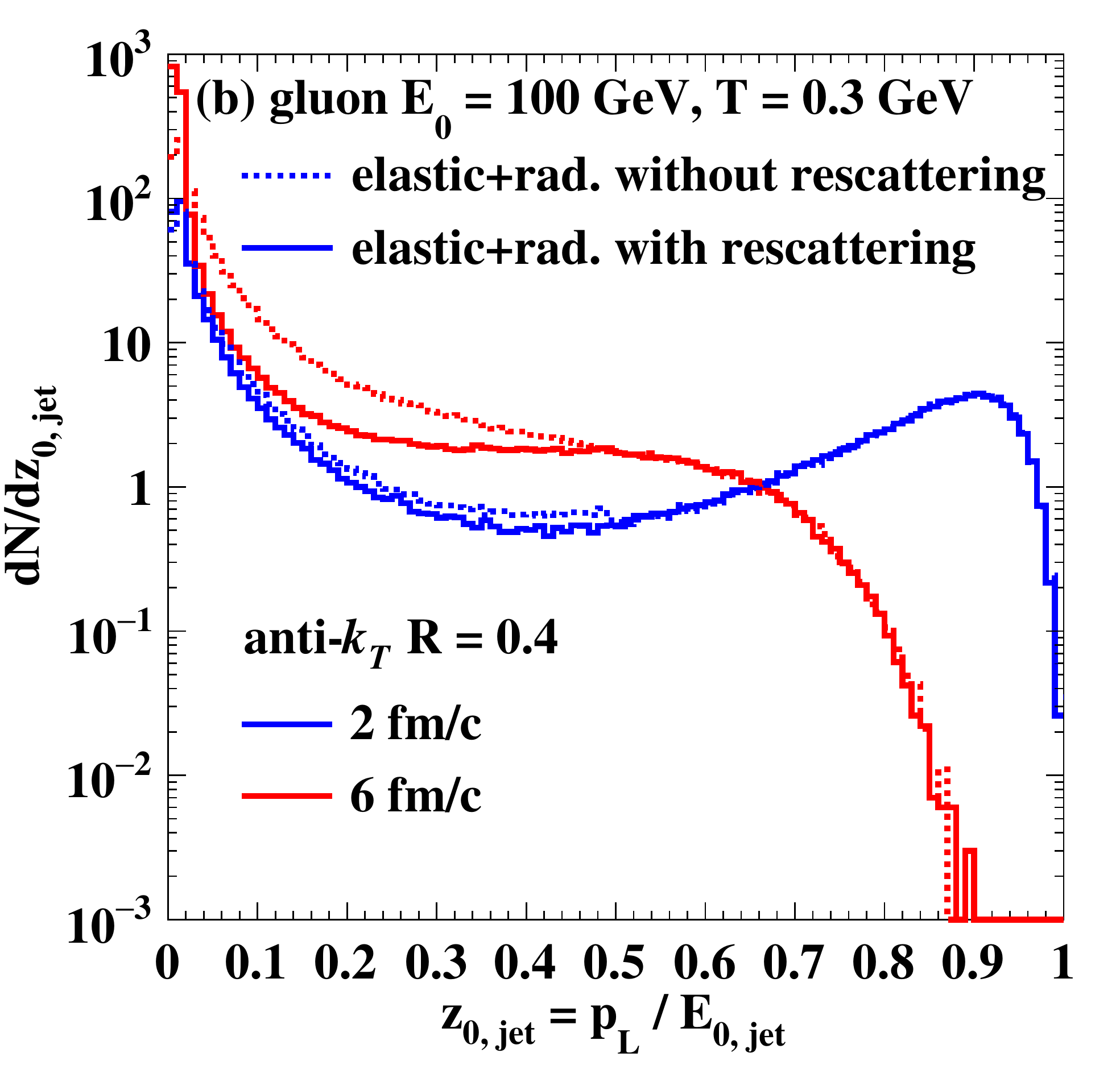}\\
\vspace{1.8cm}
\caption{(Color online) The same as Fig.~\ref{fig:fragFnc} except that the fractional momentum $z_{0,\mathrm{jet}}$ is defined with respect to the initial parton energy.}
\label{fig:FFwithZ0jet}
\end{figure}

\subsection{Jet fragmentation function}

Apart from the energy loss of reconstructed jets, one can also study how the jet energy is redistributed among its constituents -- leading jet parton, radiated gluons, recoil and ``negative" partons -- in jet substructure observables. In Fig.~\ref{fig:fragFnc}, we present the fractional longitudinal momentum ($z_\mathrm{jet}=p_\mathrm{L}/E_{\rm jet}$) distribution within a quark jet (upper panel) and a gluon jet (lower panel). This is also known as the jet fragmentation function, albeit in terms of partons instead of hadrons in experimental measurements. Here, $p_\mathrm{L}$ denotes the longitudinal momentum of a parton inside the jet with respect to the jet direction, and $E_{\rm jet}$ denotes the energy of the reconstructed jet. Since the jet develops from a single parton here, this $z_\mathrm{jet}$ distribution starts with a $\delta$-function at $z_\mathrm{jet}=1$ before any interaction occurs in the medium. As the jet evolves through a medium, a new peak near $z_\mathrm{jet}=0$ appears and continues increasing, due to newly emitted gluons and generated recoil partons at the thermal energy scale. Meanwhile, the peak at $z_\mathrm{jet}=1$ first decreases and then disappears as the leading parton loses its energy. 
A similar pattern was reported in an earlier study on the $\gamma$-triggered jet fragmentation function in realistic heavy-ion collisions~\cite{Wang:2013cia}. The subtraction of the ``negative" partons from both the parton distribution and the reconstructed jet energy reduces the soft parton distribution at small $z_\mathrm{jet}$. We observe that rescattering enhances the fragmentation function at very low $z_{\rm jet}$ due to soft partons from medium response (recoil minus ``negative" partons) as well as at large $z_{\rm jet}$ due to trigger bias. One can also plot the coincident fragmentation function as shown in Fig.~\ref{fig:FFwithZ0jet} in which the momentum fraction $z_{0,{\rm jet}}=p_{\rm L}/E_{0}$ is defined relative to the initial parton energy $E_0$ instead of the final jet energy $E_{\rm jet}$. The enhancement due to rescattering at low $z_{0,{\rm jet}}$ remain approximately the same. However, there is no enhancement at large $z_{0,{\rm jet}}$. Note that the medium modification of the fragmentation function in terms of the momentum fraction relative to the initiating parton energy (or the $\gamma$'s energy in $\gamma$-triggered jets) is more sensitive to jet-medium interactions than that in terms of the momentum fraction relative to the reconstructed jet energy, since in the latter case, the reduced energy of the reconstructed jet will rescale $z_{\rm jet}$ to a larger value and therefore enhances the parton distribution at $z_\mathrm{jet}\approx 1$. Comparing the two panels of Fig.~\ref{fig:fragFnc} or Fig.~\ref{fig:FFwithZ0jet}, we observe that the population of the distribution in the $z_\mathrm{jet}<1$ region is faster for a gluon jet than for a quark jet.  In each panel, we also notice that rescatterings of radiated gluons and recoil partons accelerates the smearing of the distribution function towards the small $z_\mathrm{jet}$ region.

\subsection{Jet shape}

\begin{figure}[!tbp]
\vspace{-1.8cm}
\includegraphics[width=7.5cm,bb=15 150 585 687]{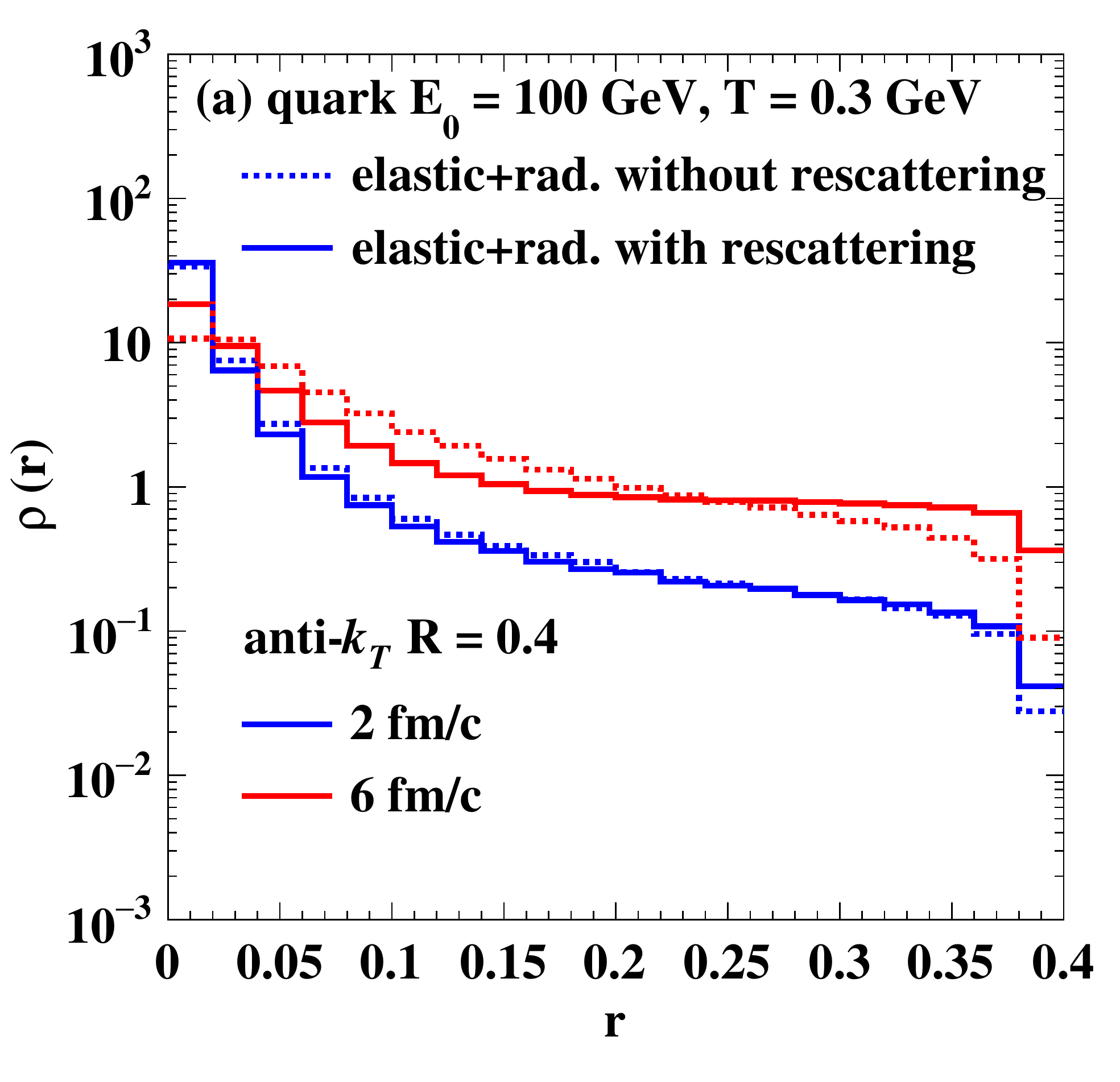}\\
\vspace{-0.84cm}
\includegraphics[width=7.5cm,bb=15 150 585 687]{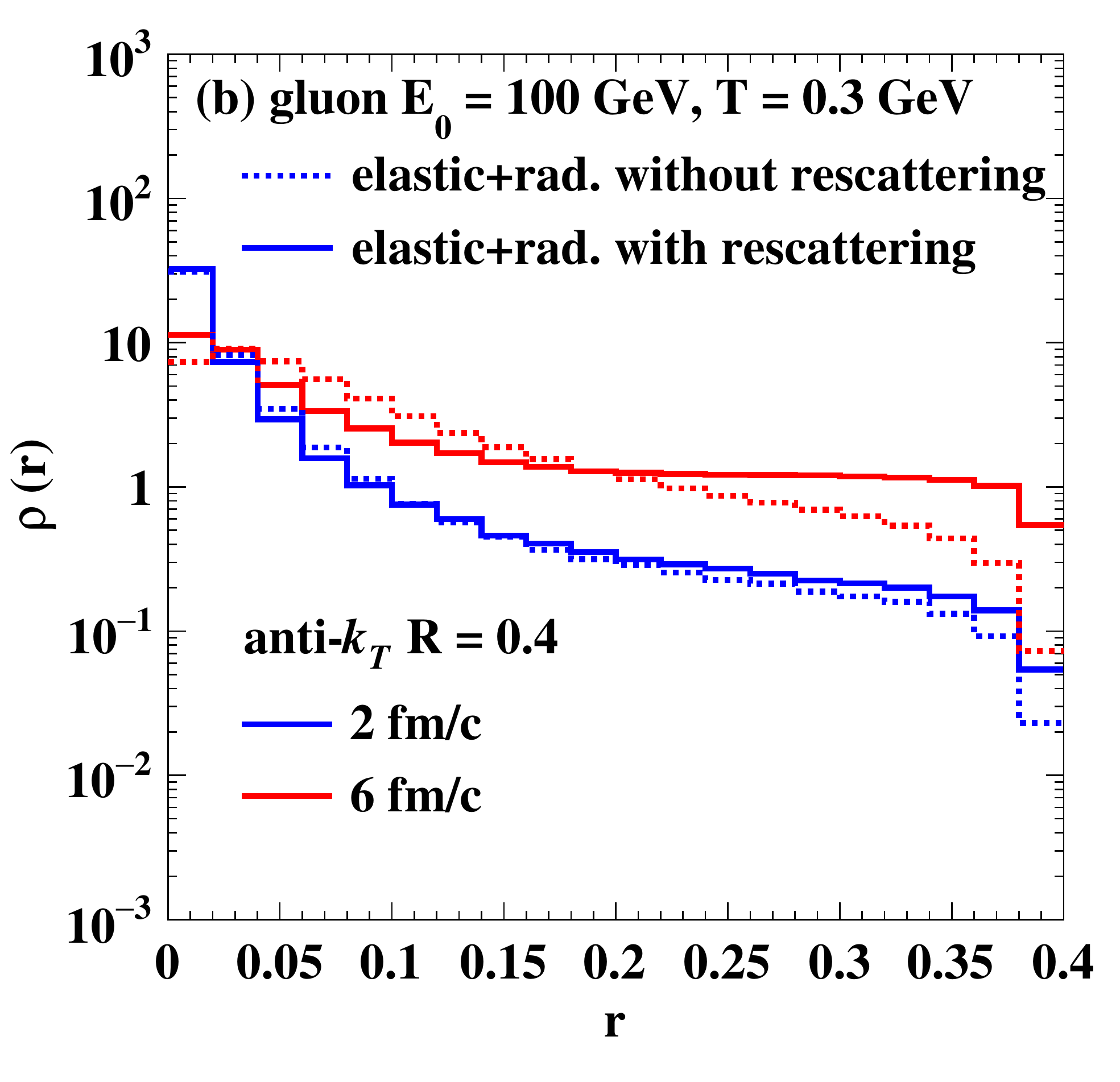}\\
\vspace{2.2cm}
\caption{(Color online) Transverse energy distribution within a jet with a cone size $R=0.4$ at different times during the jet propagation with (solid lines) and without (dashed lines) rescatterings of radiated gluons and recoil partons. The jet is initiated with a 100~GeV quark (upper panel) or gluon (lower panel), propagating through a static medium at $T = 300$~MeV with both elastic and inelastic scatterings switched on.}
\label{fig:shape}
\end{figure}

In addition to the jet fragmentation function, another jet substructure observable is the jet shape. It measures the jet energy distribution transverse to the jet axis, defined as 
\begin{equation}
\rho(r)=\frac{1}{\Delta r}\frac{1}{N^\text{jet}}\sum_\text{jets}\frac{E(r-\Delta r/2,r+\Delta r/2)}{E(0,R)},
\end{equation}
where $r=\sqrt{(\eta-\eta_\mathrm{jet})^2+(\phi-\phi_\mathrm{jet})^2}$ is the radius to the jet axis in the $\eta$-$\phi$ plane, $E(r_1,r_2)$ denotes the energy carried by all partons in the jet shower within the annulus between $r_1$ and $r_2$, $R$ is the jet cone size, and $N^\mathrm{jet}$ is the total number of jets being analyzed. Shown in Fig.~\ref{fig:shape} is the shape functions of a jet at different times during the propagation and evolution in a static medium. The jet originates from a single quark (upper panel) or a gluon (lower panel) with initial energy $E_0=100$~GeV, and is reconstructed with a cone size of $R=0.4$ at a given time. Starting as a $\delta$-function at $r=0$, the jet shape smears towards the large $r$ region as the jet-medium interactions transport the jet energy into large angles. With both elastic and inelastic scatterings switched on in the LBT simulations, we observe that the rescatterings of radiated gluons and recoil partons with the medium accelerate the broadening of the jet. Comparing the upper and lower panel, we see a faster broadening of a gluon jet than a quark jet. The medium modification of the  single inclusive and $\gamma$-triggered jet shapes in realistic heavy-ion collisions were previously explored in Refs.~\cite{Tachibana:2017syd} and \cite{Luo:2018pto}, where the jet-induced medium excitation is shown to significantly enhance the jet shape at large $r$. Considering that the hard core of the jet is less affected by rescattering than the soft partons at larger $r$, and the jet shape is defined as a self-normalized function, introducing rescattering may also enhance the jet shape at very small $r$ when the intermediate $r$ region is depleted.

\section{Nuclear modification of jets in heavy-ion collisions}
\label{sec:resultQGP}

In this section, we use the LBT model to study jet propagation and evolution in an expanding QGP in realistic heavy-ion collisions following the procedure as outlined in Sec.~\ref{sec:LBT} and illustrated by the flow chart in Fig.~\ref{fig:chart}. This is a supplement to our earlier studies using LBT and we compare to experimental data that only become available recently. We will especially focus on the cone-size dependence of the jet modifications.

\subsection{Simulating jet quenching event-by-event}

As the first step in the setup of the LBT model, we use Pythia 8~\cite{Sjostrand:2007gs} to generate the initial jet showers from nucleon-nucleon collisions at the given center-of-mass energy within specified regions of kinematics. To achieve enough statistics, we sometimes generate a given number of events of initial jet production within different bins of kinematics (e.g. transverse momentum exchange in hard scatterings) and then average the final results over these kinematic bins with the corresponding initial jet production cross sections as weights. The initial spatial distribution of these jets in heavy-ion collisions is determined according to the hard scattering locations from the AMPT~\cite{Lin:2004en} model. We also use the AMPT model to consistently generate the initial energy density distributions of the QGP by coarse graining the energy-momentum tensor of produced partons in the initial stage of nuclear collisions~\cite{Pang:2012he}. The subsequent evolution of the QGP is then simulated using the CLVisc hydrodynamic model~\cite{Pang:2012he,Pang:2014ipa,Pang:2018zzo}. With the starting time of the hydrodynamic evolution set as $\tau_0=0.6$~fm/$c$, the specific shear viscosity set as $\eta/s=0.08$ and hadronic freeze-out temperature $T_f=137$ MeV, the hydrodynamic calculation is able to provide a good description of the soft hadron spectra and their anisotropic flow coefficients at both RHIC and LHC. Using the local temperature and flow velocity information given by these hydrodynamic simulations at each time step, we first boost a given hard parton in the jet shower into the local rest frame of the expanding QGP, in which the linear Boltzmann equation [Eq.~(\ref{eq:boltzmann1})] is solved by simulating the parton scattering with its surrounding medium. The final state jet partons, together with its companions (radiated gluons, recoil and ``negative" partons) if scatterings happen during the time interval, are then boosted back to the global frame, where they stream freely to the locations of the next time step. This process is iterated for all hard partons until they travel outside the QGP regime, i.e., to locations where the local temperature drops below the phase transition temperature $T_\mathrm{c}$ (taken as 160~MeV in this work). We neglect the jet-medium interaction in the hadronic phase where the jet transport coefficient becomes negligibly small \cite{Chen:2010te}. To incorporate effects of event-by-event fluctuations, we prepare 200 hydrodynamic events for each collision setup (each centrality for each colliding system), and simulate a large number of events of jet-medium interactions through each hydrodynamic profile until the statistical errors of the extracted observables are sufficiently small. This LBT model has been successfully applied to study the nuclear modification of single inclusive hadrons~\cite{Cao:2017hhk}, jets~\cite{He:2018xjv,He:2022evt} and photon-triggered jets~\cite{Wang:2013cia,Luo:2018pto}. As a complementary study to these earlier work, we present additional results from LBT and compare to new experimental data in this section and discuss how jet-medium interactions affect different jet observables in heavy-ion collisions. 

In some of the published studies with LBT and in the following subsections, we have only considered the first term in Eq.~(\ref{eq:formation-time}) for the formation time of the initial jet shower partons from Pythia simulations. Values of the effective strong coupling constant $\alpha_{\rm s}$ from fitting to experimental data are somewhat small, ranging from 0.15 to 0.2, depending on the colliding energy. If we include the second term in the formation time for the initial jet shower partons, we can still fit the experimental data by readjusting the value of $\alpha_{\rm s}$. Since this will increase the formation time for some of the initial jet shower partons and delay the parton-medium interaction to later times, the effective value of $\alpha_{\rm s}$ will be larger, from $\alpha_{\rm s}=0.2$ to 0.3, in order to describe the data. We should emphasize that this is only an effective and constant coupling constant in LBT model which should be considered as an adjustable parameter. In the most recent extension of the LBT model, the quasi-particle LBT (QLBT) model  \cite{Liu:2021dpm}, one assumes that the QGP consists of thermal quasi-particles with an effective Debye screening mass that depends on a temperature-dependent strong coupling $g(T)$ that also controls the interaction between exchanged gluons and medium partons. The effective strong coupling constant between jet shower partons and the medium through Bayesian analyses of experimental data on heavy flavor mesons ranges from 0.2 to 0.3, depending on the temperature and parton energy. While readjusting $\alpha_\mathrm{s}$ can fit the jet $R_\mathrm{AA}$ data for different applications of the parton formation time, deviations may remain in more detailed substructures of jets that are more sensitive to the evolution of their soft components. Thus, we utilize the full Eq.~(\ref{eq:formation-time}) in our current and future calculations.

\begin{figure}[!tbp]
\vspace{-2.0cm}
\includegraphics[width=7.5cm,bb=15 150 585 687]{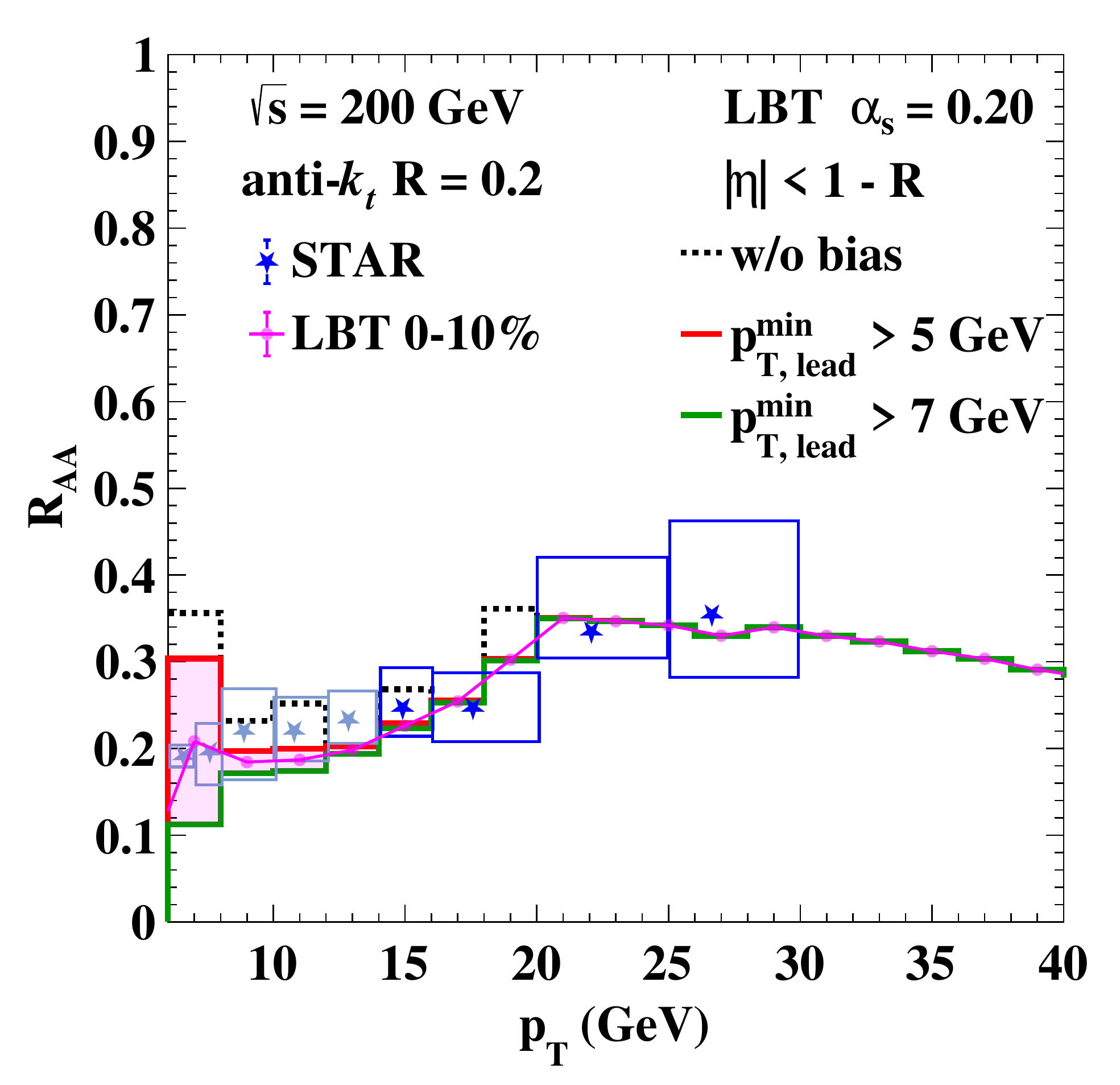}\\
\vspace{-0.84cm}
\includegraphics[width=7.5cm,bb=15 150 585 687]{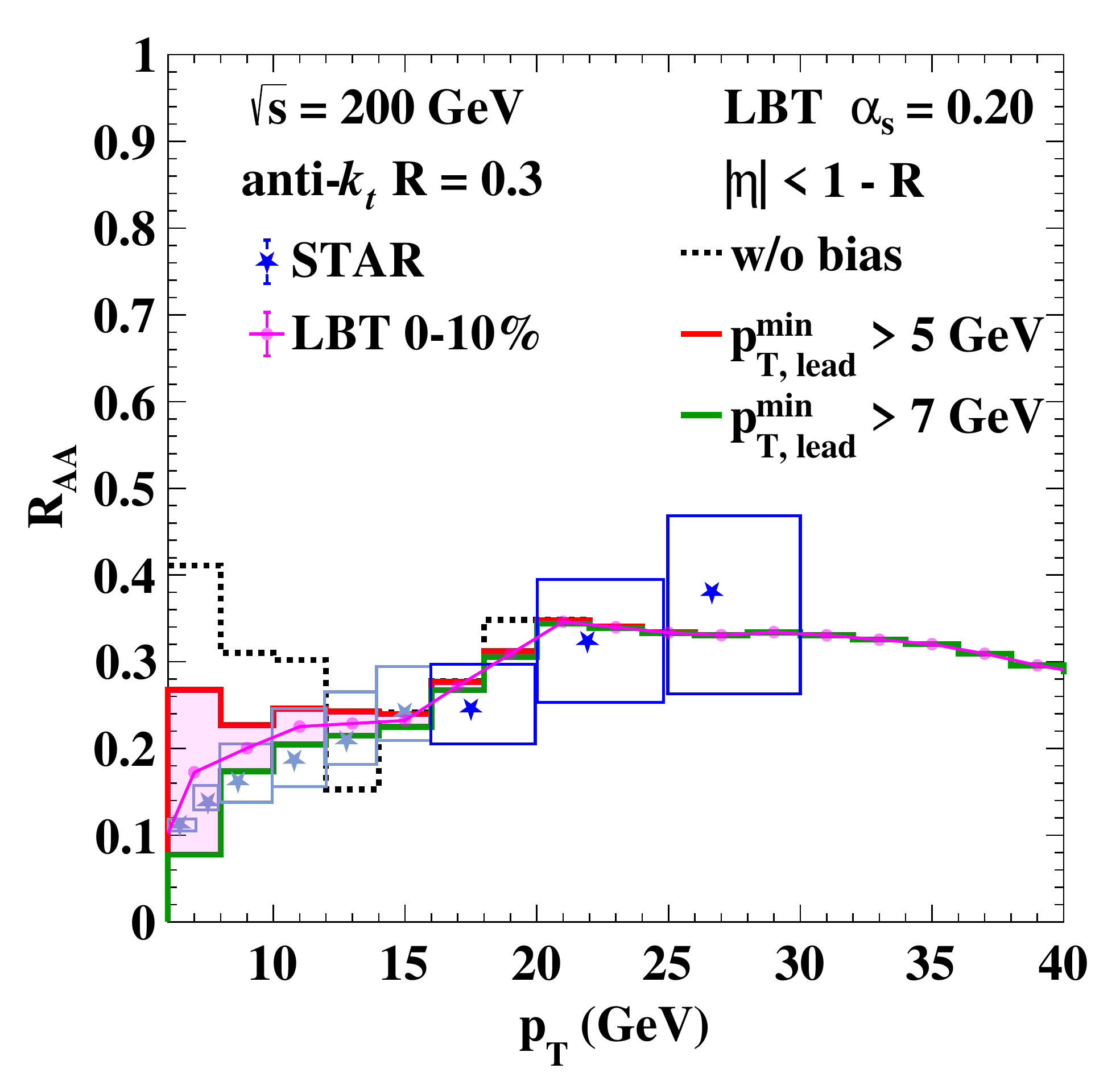}\\
\vspace{-0.84cm}
\includegraphics[width=7.5cm,bb=15 150 585 687]{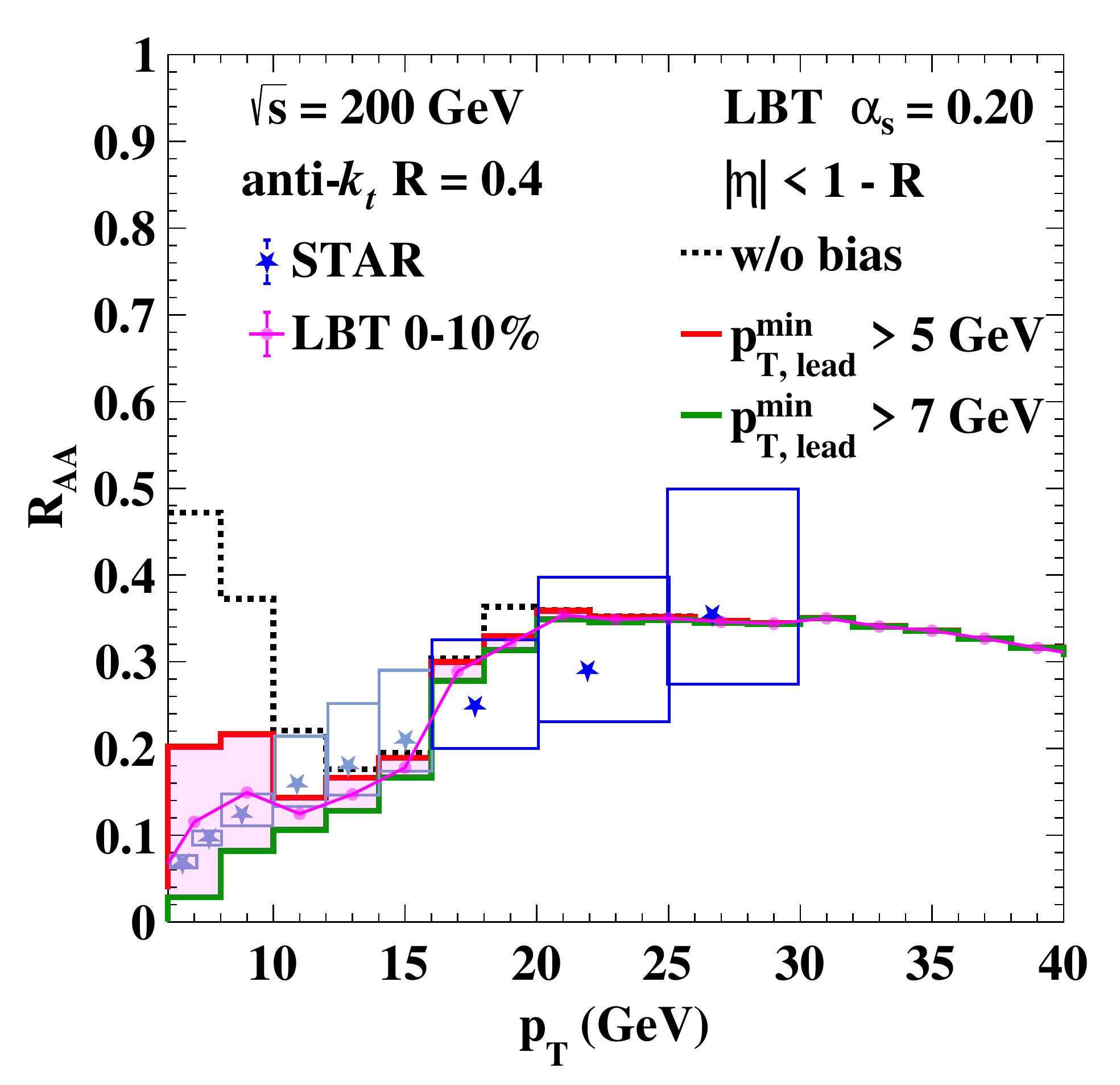}\\
\vspace{1.8cm}
\caption{(Color online) LBT simulations of the nuclear modification factor $R_\mathrm{AA}$ for single inclusive jets in 0-10\% Au+Au collisions at $\sqrt{s_\mathrm{NN}}=200$~GeV with jet cone sizes $R=0.2$ (top panel), 0.3 (middle panel) and 0.4 (bottom panel) as compared to STAR data~\cite{STAR:2020xiv}.}
\label{fig:RHIC-RAA}
\end{figure}

\begin{figure}[!tbp]
\vspace{-2.2cm}
\includegraphics[width=7.5cm,bb=15 150 585 687]{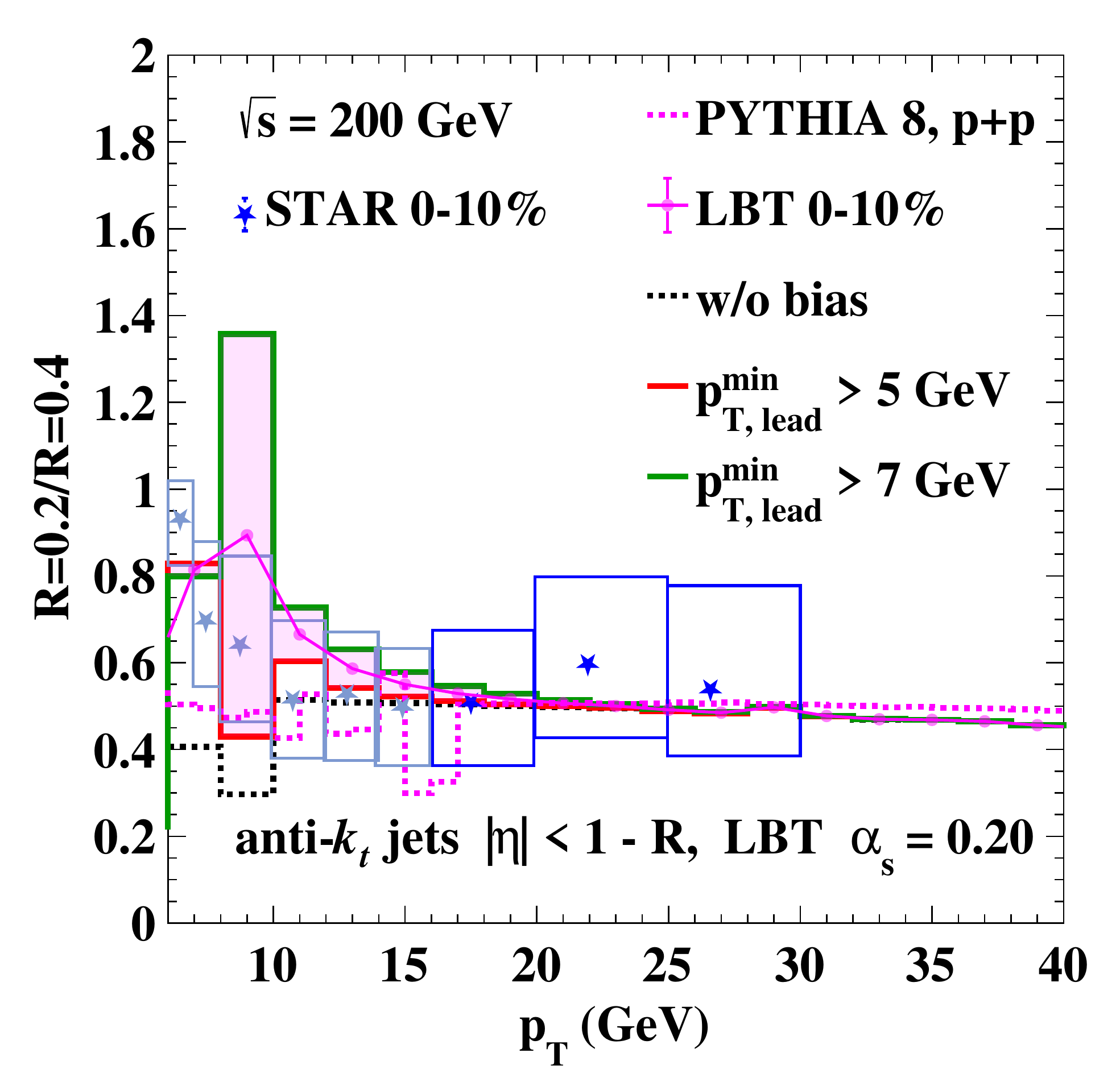}\\
\vspace{1.8cm}
\caption{(Color online) The ratio of the inclusive jet spectra with jet cone size $R=0.2$ and 0.4 in 0-10\% Au+Au collisions at $\sqrt{s_\mathrm{NN}}=200$~GeV from the LBT calculations as compared to the STAR data~\cite{STAR:2020xiv} and the corresponding $p+p$ result from Pythia simulation. The black dashed line, red solid line and green solid line represent LBT results with minimal $p_\mathrm{T}$ trigger bias as 0, 5~GeV/c and 7~GeV/c for the leading parton respectively. The purple solid line represents the average of LBT results between minimal $p_\mathrm{T}$ trigger bias of 5~GeV/c and 7~GeV/c for the leading parton, while the purple dashed line represents the Pythia result for $p+p$ collisions.}
\label{fig:RHIC-RAA_ratio}
\end{figure}

\subsection{Cone-size dependence of single inclusive jet suppression}

The most widely used observable that quantifies the QGP effect on jets is their nuclear modification factor, as first proposed in Ref.~\cite{Wang:1998bha},
\begin{equation}
R_\mathrm{AA}=\frac{d\sigma^{\rm AA}_{\rm jet}}{\langle N_{\rm bin}\rangle d\sigma^{pp}_{\rm jet}},
\end{equation}
where $\langle N_{\rm bin}\rangle$ is the number of binary collisions in the given centrality bin of A+A collisions. The LBT model was shown to describe the centrality and $p_{\rm T}$ dependence of $R_\mathrm{AA}$ at the LHC energies and the effect of jet-induced medium response is very important \cite{He:2018xjv}.  Shown in Fig.~\ref{fig:RHIC-RAA} is the jet $R_\mathrm{AA}$ in 0-10\% central Au+Au collisions at $\sqrt{s_\mathrm{NN}}=200$~GeV as compared to the recent STAR experimental data~\cite{STAR:2020xiv}. From the top to bottom panel, we present results for jets with cone sizes of $R=0.2$, 0.3 and 0.4. In the STAR measurement, a high $p_\mathrm{T}$ hadron (above 5~GeV) inside the jet is required to suppress the background. Since the hadronization process has not been implemented in the LBT calculations presented in this study, we select jet events with energetic partons and vary the $p_\mathrm{T}$ cut to study its effect on the jet $R_\mathrm{AA}$. As shown in the figure, increasing the $p_\mathrm{T}$ cut from 0 (denoted as ``without trigger bias") to 5 and 7~GeV/$c$ excludes certain jet events at low $p_\mathrm{T}$. Since a large number of jet events at low $p_\mathrm{T}$ are generated in A+A collisions after jet-medium interactions, effect of the trigger bias is more significant in A+A than in $p+p$ collisions. This leads to a smaller jet $R_\mathrm{AA}$ at low $p_\mathrm{T}$ when the trigger $p_\mathrm{T}$ cut is increased. However, no visible effect is observed on the jet $R_\mathrm{AA}$ above $p_{\rm T}=20$~GeV/$c$. With a strong coupling constant set as $\alpha_\mathrm{s}=0.2$, the LBT results (shown by the pink band between 5 and 7~GeV/$c$ cuts for the leading hadron $p_\mathrm{T}$) are consistent with the experimental data. Since the leading hadron usually comes from the fragmentation of a higher $p_\mathrm{T}$ parton, a slightly higher $p_\mathrm{T}$ cut for partons is expected  compared to that used for hadrons. 

Shown in Fig.~\ref{fig:RHIC-RAA_ratio} is the ratio between the single inclusive jet spectra  with cone size $R=0.2$ and 0.4 at the RHIC energy. As one uses a larger jet cone, more particles are included in the jet reconstruction and thus the yield of jets increases. Our LBT model result with a lower $p_\mathrm{T}$ cut on the triggered parton between 5 and 7~GeV/$c$ agrees with the STAR data. Comparing $p+p$ and Au+Au collisions, we observe this ratio for the former is slightly higher than the latter at high $p_\mathrm{T}$, while the opposite is seen at low $p_\mathrm{T}$. This cone-size dependence of the jet spectrum in $p+p$ collisions affects the cone-size dependence of the jet $R_\mathrm{AA}$ shown in Fig.~\ref{fig:RHIC-RAA}.

\begin{figure}[!tbp]
\vspace{-2.0cm}
\includegraphics[width=7.5cm,bb=15 150 585 687]{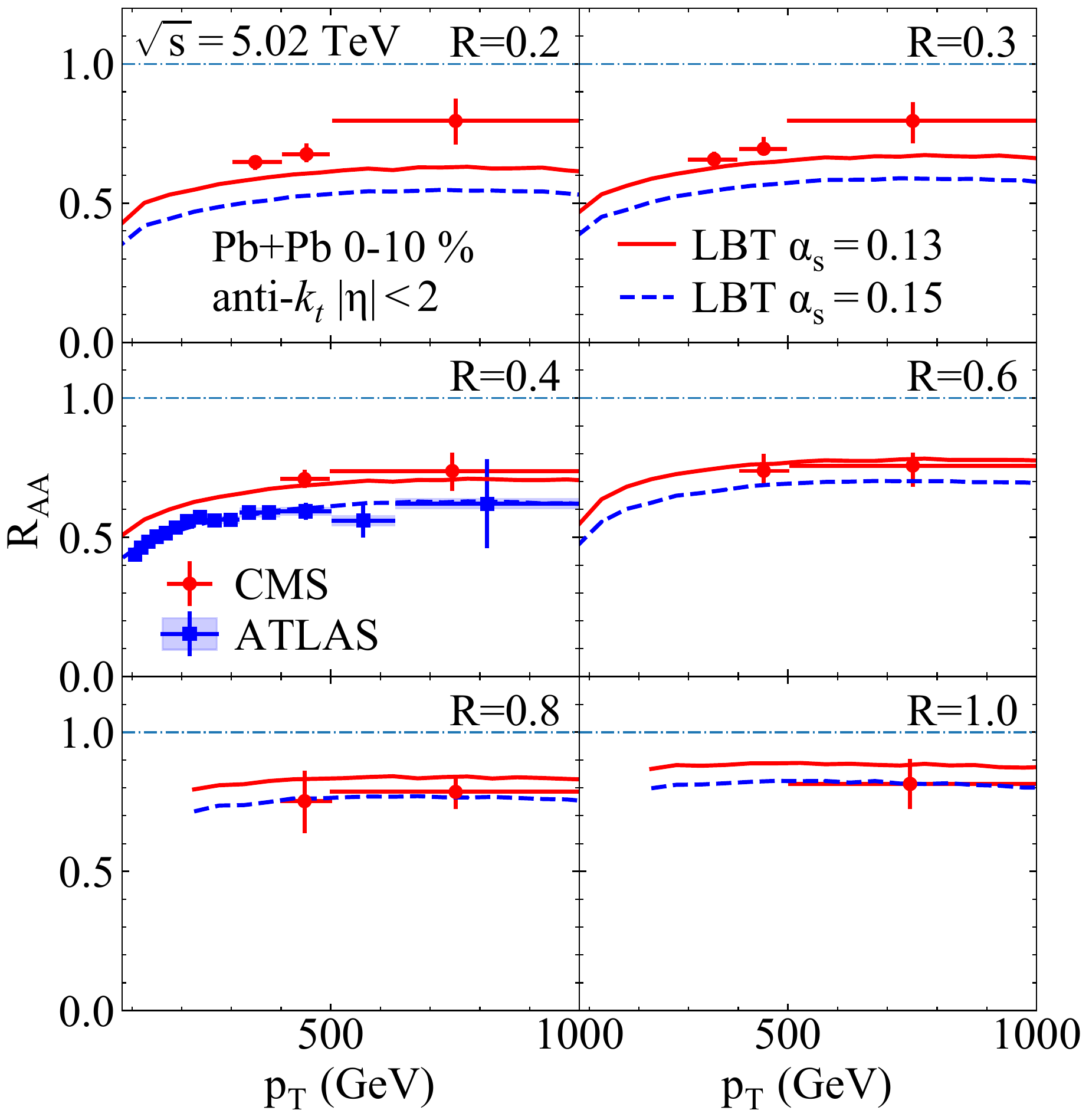}\\
\vspace{1.8cm}
\caption{(Color online) LBT results on the nuclear modification factor $R_\mathrm{AA}$ for single inclusive jets in 0-10\% Pb+Pb collisions at $\sqrt{s_\mathrm{NN}}=5.02$~TeV, with different jet cone sizes and two different values of $\alpha_\mathrm{s}=$ 0.13 (red) and 0.15 (blue) as 
compared to the CMS~\cite{CMS:2021vui} and ATLAS~\cite{ATLAS:2018gwx} data. }
\label{fig:LHC-RAA}
\end{figure}

 CMS experiment ~\cite{CMS:2021vui} has recently measured $R_{\rm AA}$ of large $p_{\rm T}$ single inclusive jets at LHC with very large jet cone size and found a weak jet cone-size dependence of $R_{\rm AA}$ which was not reproduced by many models, including LBT. The jet $R_\mathrm{AA}$ as a function of $p_\mathrm{T}$ from LBT with different jet cone-sizes is shown in Fig.~\ref{fig:LHC-RAA} together with the CMS data for central Pb+Pb collisions at $\sqrt{s_\mathrm{NN}}=5.02$~TeV. The LBT results at LHC with $\alpha_{\rm s}=0.15$ (blue dashed lines) are tuned to reproduce the ATLAS data \cite{ATLAS:2014ipv,ATLAS:2018gwx} on single inclusive jet suppression at both $\sqrt{s_\mathrm{NN}}=2.76$ and 5.02~TeV \cite{He:2018xjv}. This value of $\alpha_\mathrm{s}$ at LHC is smaller than the value we used to fit the STAR data at RHIC. A possible cause of this colliding energy dependence is the lower average temperature of the QGP and thus stronger jet-medium interactions at RHIC than LHC.  With this value of $\alpha_{\rm s}$, LBT however fails to reproduce the CMS data for $R\lesssim0.4$.

In Fig.~\ref{fig:LHC-RAA}, we note that the jet $R_\mathrm{AA}$ from LBT calculations increases as the jet cone becomes larger. This is a general feature due to the jet-induced medium excitation: when more recoil partons scattered into large angles are included in jet reconstruction as one increases the cone size, the jet $R_\mathrm{AA}$ should increase. On the other hand, such cone-size dependence becomes much weaker if one only considers the medium-induced gluon emission process which prefers small angle (shown in Fig.~\ref{fig:dngdtheta}). Similar conclusions have also been drawn by other model calculations, e.g. JEWEL \cite{Zapp:2011ya,Zapp:2012ak,Zapp:2013vla,KunnawalkamElayavalli:2017hxo} and jet-fluid~\cite{Tachibana:2017syd,Chang:2019sae}, that take into account of the effects of medium response induced by jet-medium interaction.

We note that with the jet cone size $R=0.4$, where measurements are available from both CMS and ATLAS experiment, tension exists between CMS and ATLAS data: the jet $R_\mathrm{AA}$ observed by CMS seems larger than ATLAS.  If we adjust the strong coupling constant to $\alpha_\mathrm{s}=0.13$ in LBT to fit CMS data with $R=0.4$ instead, the agreement between LBT and CMS data improves for $R< 0.4$. But LBT results are still about one sigma blow the CMS data for small jet cone size $R=0.2$. It is clear that resolving the tension between ALTAS and CMS data is necessary to draw a more solid conclusion on the cone size dependence of the nuclear modification on jets.

\begin{figure}[!tbp]
\vspace{-0.2cm}
\centering
\includegraphics[width=8.5cm]{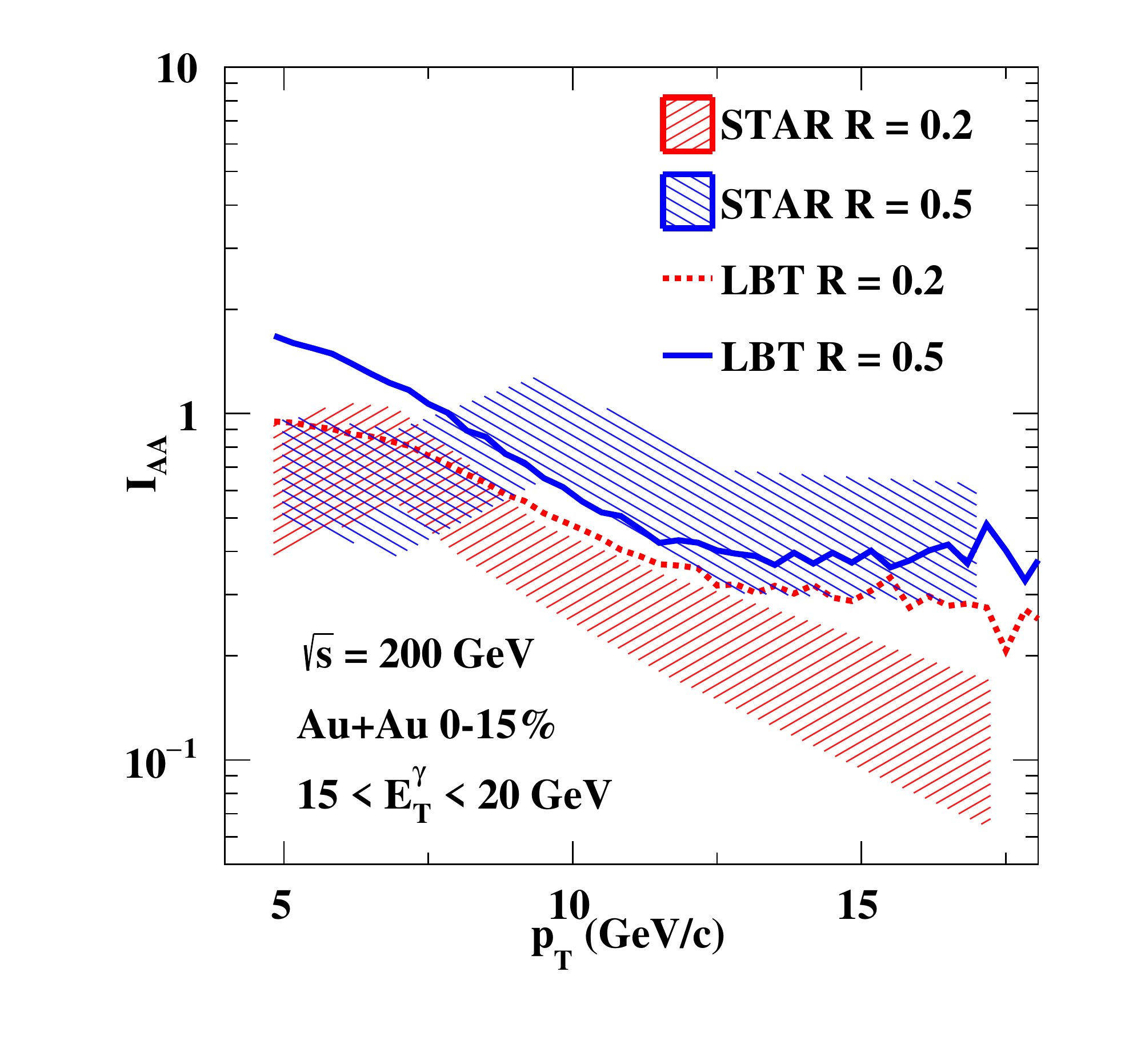} \\
\vspace{-0.4cm}
\caption{(Color online) The nuclear modification factor $I_\mathrm{AA}$ from LBT simulations for $\gamma$-triggered jets in 0-15\% Au+Au collisions at $\sqrt{s_\mathrm{NN}}=200$~GeV with different jet cone sizes as compared to the STAR data~\cite{Anderson:2022nxb}.}
\label{fig:IAA_gammajet_RHIC}
\end{figure}

\subsection{$\gamma$-jets and dijets}

Electro-weak-boson-triggered jets are considered a ``golden channel" to quantify the nuclear modification of jets, since these bosons do not participate in strong interactions with the QGP and thus serve as a reliable reference of the initial energy of its associated jets. Suppression and modification of $\gamma/Z$-triggered jets 
in Pb+Pb collisions at the LHC energies can be described well by the LBT model \cite{Luo:2018pto,Zhang:2018urd}. 
In Fig.~\ref{fig:IAA_gammajet_RHIC}, we study the nuclear modification factor of jets triggered by photons within $15<E_\mathrm{T}^\gamma<20$~GeV ($I_\mathrm{AA}$) in central Au+Au collisions at RHIC. Two cone sizes, $R=0.2$ and 0.5, are used for jet reconstruction. Both the STAR experimental data~\cite{Anderson:2022nxb}. and our LBT calculations indicate that a larger cone size captures more energy in the reconstructed jets, thus results in a higher $I_\mathrm{AA}$. This is consistent with our earlier findings for the suppression factor $R_\mathrm{AA}$ for single inclusive jets in LBT simulations.

\begin{figure}[!tbp]
\vspace{-2.2cm}
\includegraphics[width=7.5cm,bb=15 150 585 687]{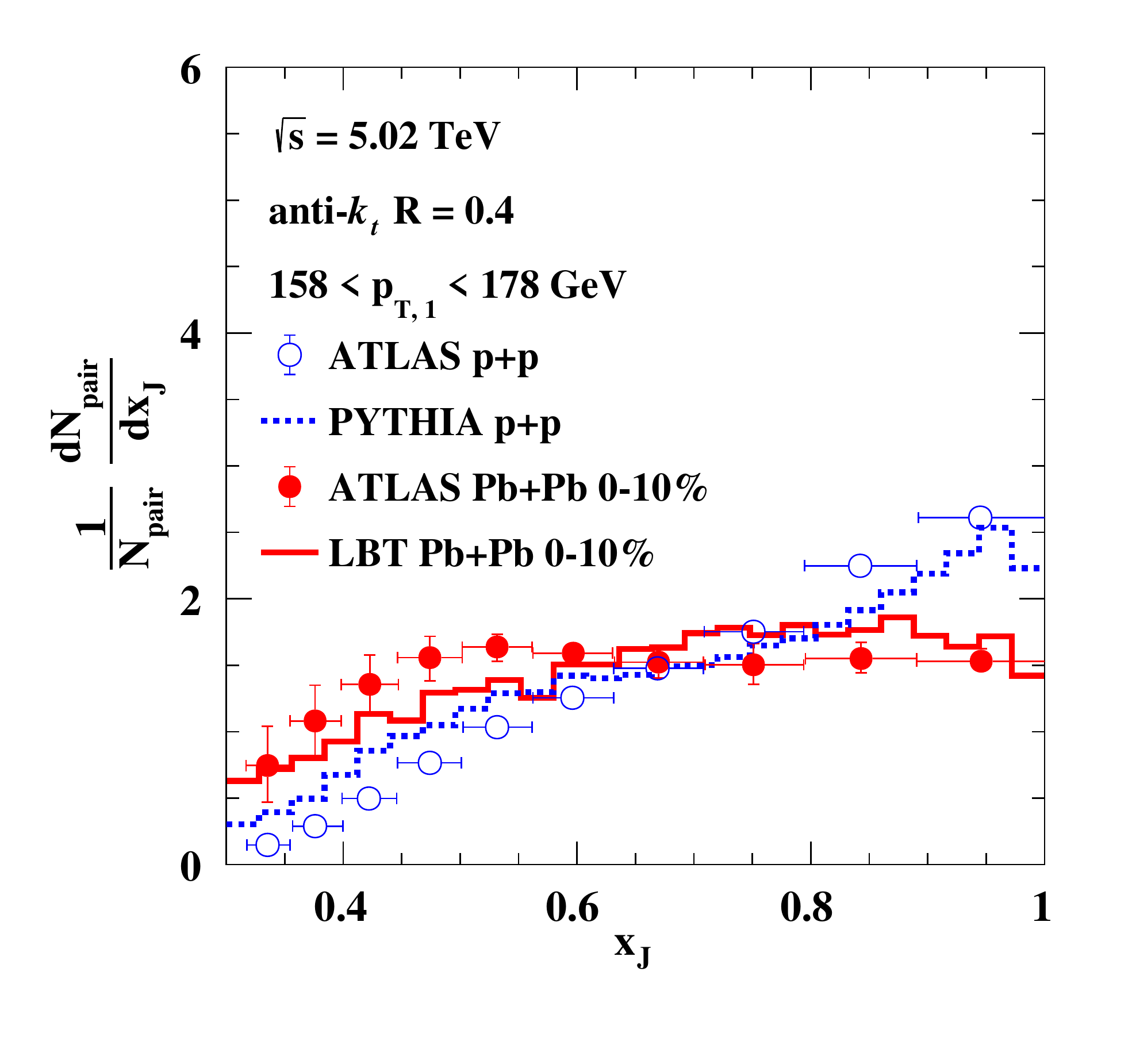}\\
\vspace{-1.3cm}
\includegraphics[width=7.5cm,bb=15 150 585 687]{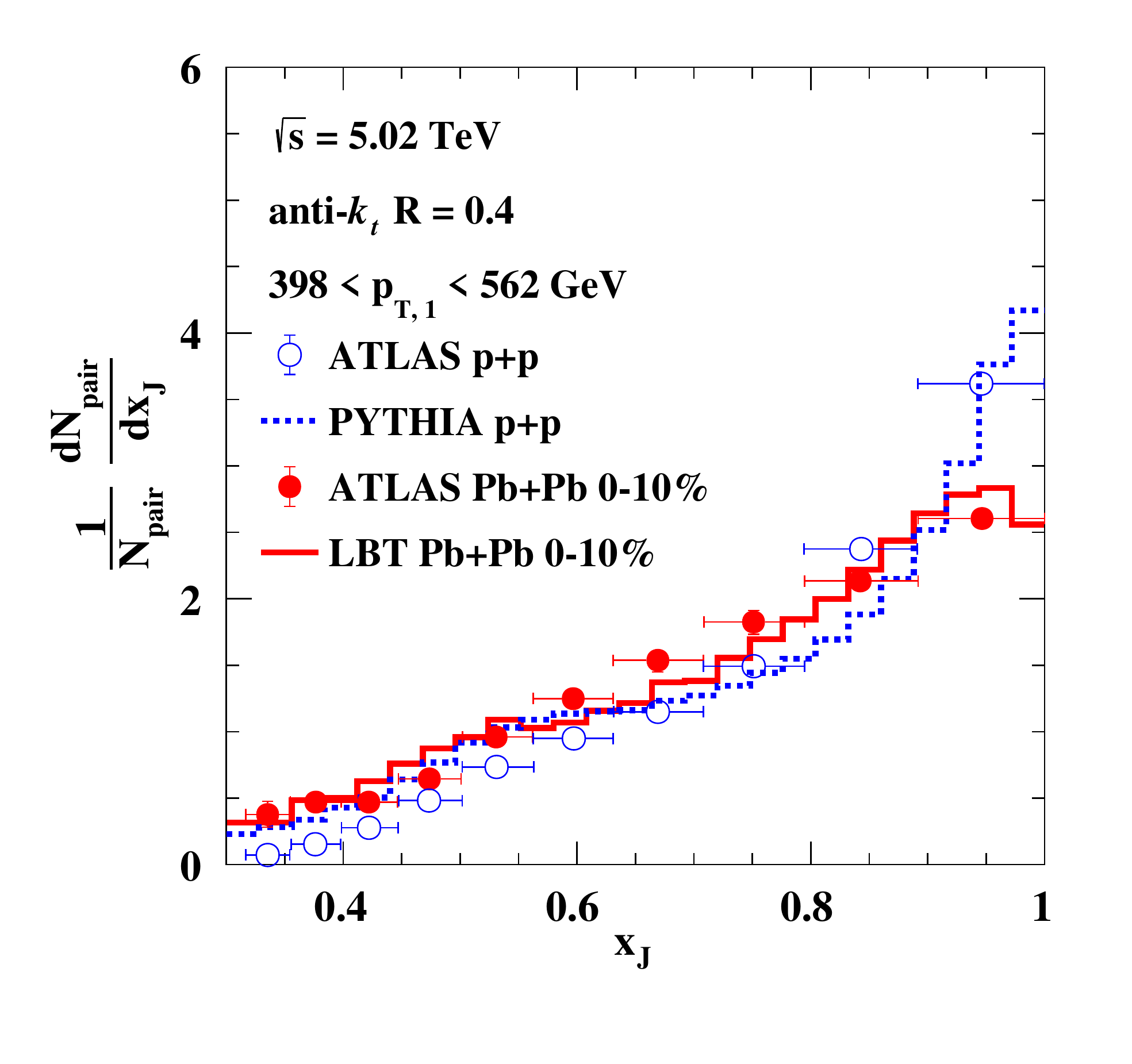}
\vspace{1.6cm}
\caption{(Color online)  Dijet asymmetry ($x_\mathrm{J}$) distribution in $p+p$  (blue) and Pb+Pb collisions (red) at $\sqrt{s_\mathrm{NN}} = 5.02$~TeV from LBT calculations as compared to the ATLAS data~\cite{ATLAS:2022zbu}. The transverse momentum of the leading jet is selected within $158 < p_\mathrm{T, 1} < 178$~GeV/$c$ in the upper panel and $398 < p_\mathrm{T, 1} < 562$~GeV/$c$ in the lower panel.}
\label{fig:atlas_xJ_R4_pT1}
\end{figure}

Similar to photon-triggered jets, one may also quantify the jet energy loss by comparing the transverse momenta of the subleading and leading jet in a dijet event \cite{Qin:2010mn}. In Fig.~\ref{fig:atlas_xJ_R4_pT1}, we present the distribution of subleading-to-leading-jet-$p_\mathrm{T}$ ratio $(x_\mathrm{J}=E_T^{\rm sublead}/E_T^{\rm lead})$ of dijets produced at the LHC energy, compared between $p+p$ and Pb+Pb collisions. One observes that while the $x_\mathrm{J}$ distribution peaks around one in $p+p$ collisions, it shifts to lower $x_\mathrm{J}$ values in Pb+Pb collisions due to the unbalanced energy loss between subleading and leading jets as they travel through the QGP. As one increases the $p_\mathrm{T}$ range of the leading jet (from the upper to the lower panel), dijets appear more symmetric in both $p+p$ and Pb+Pb collisions. Our Pythia simulations for $p+p$ collisions and Pythia+LBT simulation for Pb+Pb collisions agree well with the corresponding data from the ATLAS collaboration.

\section{Code availability}
\label{sec:scode}

The source code of the LBT model can be downloaded from ``https://github.com/lbt-jet",  which includes the main modules, the data tables for the elastic and inelastic scattering rate, a sample data file of the hydrodynamic profile and the modified version of the FASTJET package that can take into account of the ``negative" partons.

The LBT code is also embedded as a jet energy loss module in the JETSCAPE framework, which can be downloaded at ``https://github.com/JETSCAPE". 

\section{Summary}
\label{sec:summary}

We have developed a linear Boltzmann transport (LBT) model for studying jet-medium interaction in relativistic heavy-ion collisions. This article provides a detailed description of the model structure, physics processes, Monte Carlo implementation and numerical validation.

Both elastic and inelastic scatterings between jet partons and the QGP are implemented in LBT. The former is based on the leading-order pQCD scatterings between a jet shower parton and a medium parton at the thermal scale, while the latter is based on the medium-induced gluon emission process within the higher-twist energy loss formalism. With these scattering processes, we track the phase space evolution of not only leading partons and their emitted gluons, but also the thermal partons being scattered out of the QGP background known as recoil partons and the energy-momentum holes left inside the medium (``negative" partons). This guarantees the energy-momentum conservation in each scattering and allows us to study both medium modification of jets and jet-induced medium excitation. All partons from the LBT simulations are fed into the FASTJET package for jet finding in the end for reconstructing partonic jets. The FASTJET routine is modified such that the energy-momentum of ``negative" partons is subtracted from that of regular ones throughout the jet finding procedure. Within this framework, we present a systematical study on the evolution of energy and transverse momentum distributions of single hard partons, jet-induced medium partons, and fully reconstructed jets inside a static and uniform medium at a fixed temperature. 

We show that the accumulated energy loss of a single hard parton increases linearly with time (or path length) in elastic scatterings, but increases quadratically at early times after the gluon emission process is included, as they should according to the formalism implemented in LBT.  As the parton energy decreases along its path, its energy loss rate (per unit length) first increases with time, then saturates and decreases in our simulations, with its inflection point becoming earlier in a hotter medium and for a lower energy parton. These are all the consequences of the energy dependence of the parton energy loss rate. A broadening in the transverse momentum distributions is observed for the hard partons, with the broadening being accelerated after the radiation process is taken into account. By incorporating all partons into the analysis, including leading partons, gluons from medium-induced radiation and jet-induced medium excitation (recoil and ``negative" partons), a Mach-cone-like structure can be clearly observed in the energy density distribution of these partons with an enhanced wave front when rescatterings between recoil partons and radiated gluons with the medium are allowed.
Propagation of radiated gluons and recoil partons and their further interaction with the medium convert the energy deposition from jets into the wave front followed by a diffusion wake in the direction opposite to the jet propagation due to the energy depletion via ``negative" partons. Compared to the pure elastic scattering processes, the Mach-cone-like structure is significantly enhanced when medium-induced gluon emission is introduced. By analyzing the parton spectra within different energy ranges, one can observe the transport of energy from the hard parton at small angle to soft partons at large angle with respect to the initial parton. This energy flow results in the energy loss of fully reconstructed jets, which is found to increase approximately linearly with time. The energy flow also causes the shift of the jet fragmentation function towards the smaller momentum fraction region and the enhancement of the jet shape at large radius.  Rescatterings of recoil partons and radiated gluons with the medium are found essential in the nuclear modification of jets. They significantly broaden the shock wave of the parton energy density distribution, and accelerate the energy transport from hard to soft partons and from small to large angles.
Similar features of nuclear modification are observed for quark jets and gluon jets, except that the modification of the latter is stronger due to its larger color factor in jet-medium interaction. This documentation of the physics implementations in the LBT model, in particular inelastic processes, together with a thorough validation and a detailed study of the contributions from different model components to jet energy loss, momentum broadening and its induced medium excitation, will help to advance our understanding of jet evolution inside a dense QCD medium.

To simulate jet-medium interactions in realistic heavy-ion collisions, the LBT model is further combined with the Pythia 8 generator for the initial jet production in nucleon-nucleon collisions, the AMPT model for the spatial distribution of the hard scattering vertices in nucleus-nucleus collisions, and the CLVisc hydrodynamic model for the space-time evolution profile of the QGP with the initial conditions from the same AMPT model calculations. Complimentary to jet observables presented in our earlier publications, we calculate the jet cone-size dependence of the jet suppression factor in different collision systems at different beam energies, the $\gamma$-jet suppression factor and dijet asymmetry within different kinematic ranges in the present study, and compare to more recent experimental data at RHIC and LHC.

\begin{acknowledgments}
 We thank Wei Chen, Zhong Yang and Weiyao Ke for helpful discussions. This work is supported in part by National Natural Science Foundation of China (NSFC) under Grant Nos.~11935007, 11221504, 11861131009, 11890714, 12075098, 12175122, 2021-867 and 12147134, by EU ERDF and H2020 grant 82409, ERC grant ERC-2018-ADG-835105, Spanish AEI grant PID2020-119632GB-I00 and CEX2020-001035-M, Xunta de Galicia Research Center accreditation 2019-2022, by Guangdong Major Project of Basic and Applied Basic Research No.~2020B0301030008, by Guangdong Basic and Applied Basic Research Foundation No.~2021A1515110817, by Science and Technology Program of Guangzhou No. 2019050001, 
 by the Director, Office of Energy Research, Office of High Energy and Nuclear Physics, Division of Nuclear Physics, of the U.S. Department of Energy under Contract No. DE-AC02- 05CH11231, by the US National Science Foundation under Grant No. OAC-2004571 within the X-SCAPE Collaboration. Computations in this study are performed at the NSC3/CCNU and the National Energy Research Scientific Computing Center (NERSC), a U.S. Department of Energy Office of Science User Facility located at Lawrence Berkeley National Laboratory and operated under Contract No.~DE-AC02-05CH11231.
\end{acknowledgments}

\appendix
\section{Energy-momentum conservation of the $2\rightarrow2+n$ process}
\label{sec:appendix}

In this Appendix, we explain the procedure to guarantee energy-momentum conservation for a general $2\rightarrow 2+n$ process in LBT.

\begin{figure}[!tbh]
\begin{center}
\includegraphics[width=0.45\textwidth]{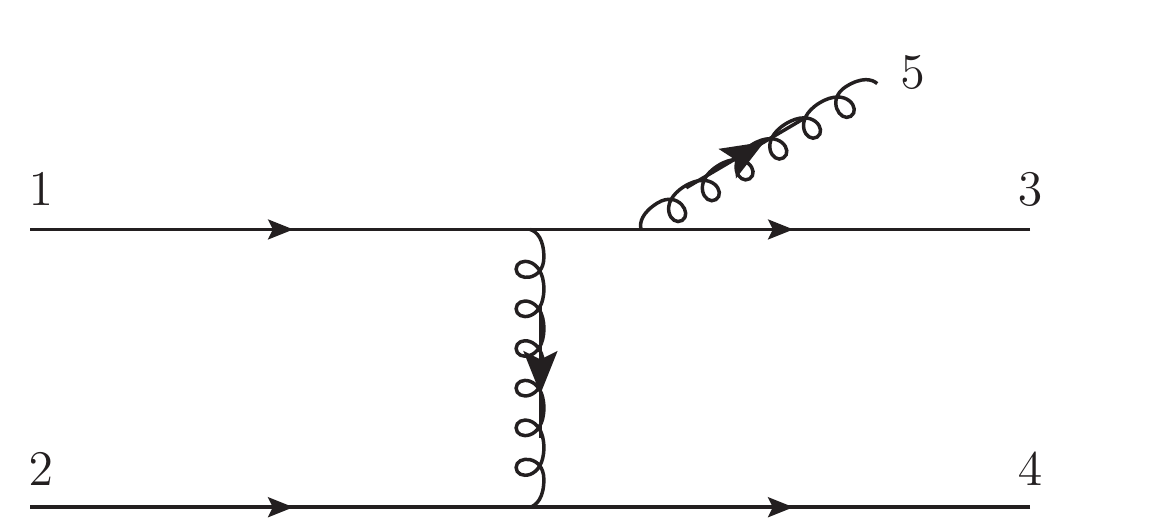}
\end{center}
\caption{Illustration of a $2\rightarrow 3$ process.}
\label{fig:23conservation}
\end{figure}

For a $2\rightarrow 3$ process, as illustrated in Fig.~\ref{fig:23conservation}, we first sample a $2\rightarrow 2$ process according to Eq.~(\ref{eq:rate2}) which by itself respects the energy-momentum conservation. Then we sample a medium-induced gluon ``5" from parton ``3" in Fig.~\ref{fig:23conservation} based on Eq.(\ref{eq:gluondistribution}). To maintain the energy-momentum conservation of the entire $2\rightarrow 3$ process, we choose to adjust the energy-momentum of partons ``3" and ``4". 

For convenience, we assume parton ``1" travels along the $z$-direction with four-momentum $(E_1, p_{1x}=0, p_{1y}=0, p_{1z})$, and fix the momentum of parton ``2" from the $2\rightarrow 2$ process as $(E_2, p_{2x}, p_{2y}, p_{2z})$, and the momentum of ``5" from the medium-induced gluon spectrum as $(k_0, k_x, k_y, k_z)$. The momentum of the exchanged gluon between ``1" and ``2", $(q_0, q_x, q_y, q_z)$, can also be calculated from the $2\rightarrow 2$ process. However, since the direction of this exchanged gluon can be correlated with that of the emitted gluon, we only keep the magnitude $q_\perp=\sqrt{q_x^2+q_y^2}$ and re-sample its direction in the transverse plane with respect to parton ``1".  Components $q_0$ and $q_z$ will be re-evaluated to recover the energy-momentum conservation of this $2\rightarrow 3$ process. With this setup, the momenta of ``3" and ``4" are then
\begin{align}
p_3=\,&(E_1-q_0-k_0,\,p_{1x}-q_x-k_x, \,p_{1y}-q_y-k_y, \nonumber\\
&p_{1z}-q_z-k_z),\\
p_4=\,&(E_2+q_0, \,p_{2x}+q_x, \,p_{2y}+q_y, \,p_{2z}+q_z),
\end{align}
respectively. The two on-shell conditions for ``3" and ``4",
\begin{align}
\label{eq:onshell1}
&p_3^3=m^2,\\
\label{eq:onshell2}
&p_4^2=0,
\end{align}
will be solved to determine $q_0$ and $q_z$. We require that the exchanged gluon is space-like $(q^2<0)$ for the solutions of $q_0$ and $q_z$. If there is no solution for Eqs.~(\ref{eq:onshell1}) and~(\ref{eq:onshell2}), or their solutions do not satisfy the space-like condition, we re-sample the direction of $q_\perp$ or the four-momentum of the emitted gluon ``5". This process is iterated until an appropriate solution is found or the number of trials exceeds a given large value. For the latter case, we consider the gluon emission process is not kinematically allowed by the associated $2\rightarrow 2$ process and reject its formation. Note that there could be two solutions from Eqs.~(\ref{eq:onshell1}) and~(\ref{eq:onshell2}) which both satisfy the space-like condition of $q$. In this situation, we currently select the solution with a smaller $q_0$, assuming that the energy exchange between ``1" and ``2" is small when they exchange momentum. One may verify that the same procedure here is also applicable for the scenario where the momentum exchange ($q$) occurs between the emitted gluon ``5" and the thermal parton ``2".

\begin{figure}[!tbh]
\begin{center}
\includegraphics[width=0.45\textwidth]{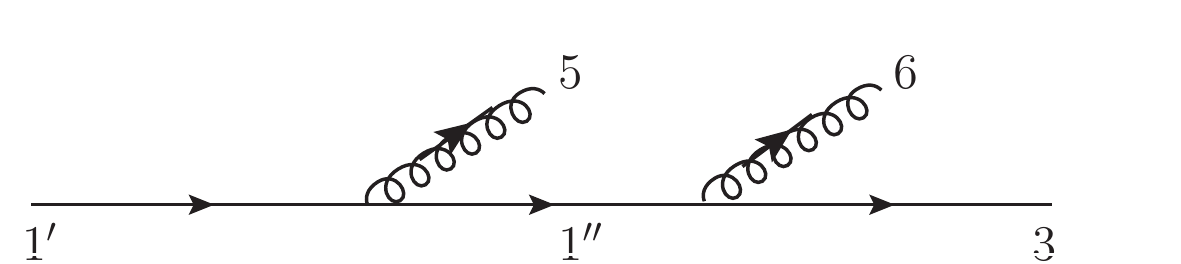}
\end{center}
\caption{Illustration of multiple gluon emissions.}
\label{fig:13conservation}
\end{figure}

For a multiple gluon emission process, we first determine a $2\rightarrow 3$ process as described above and fix the energy-momentum of the recoil parton ``4" in Fig.~\ref{fig:23conservation}. As illustrated by Fig.~\ref{fig:13conservation}, if one more gluon ``6" is emitted, we need to adjust the energy-momentum of the previous gluon ``5" and the final state jet parton ``3" to respect the energy-momentum conservation of this $1\rightarrow 3 $ splitting process. This is sufficient to guarantee the energy-momentum conservation of the entire $2\rightarrow 4$ process. 

We denote the energy-momentum of ``$1'$" (state of the jet parton before emitting the previous gluon ``5") as ($E_0,\vec{p}_{0\perp}=0,p_{0z}$). This is determined from the $2\rightarrow 3$ process as described earlier and rotated into the coordinate where ``$1'$" travels along the $z$-direction for convenience. The energy-momentum of the new gluon ``6" is sampled using Eq.~(\ref{eq:gluondistribution}) from ``3" (before adjusting its momentum here) as $(k_{6e},\vec{k}_{6\perp},k_{6z})$. The energy-momentum of the previous gluon is represented by $(k_{5e},\vec{k}_{5\perp},k_{5z})$. Although it is already given by the $2\rightarrow 3$ process, we choose to only keep the value of $k_{5e}$ but re-calculate the $k_{5z}$ component. The transverse component is then re-evaluated as $k_{5\perp}=\sqrt{k_{5e}^2-k_{5z}^2}$ given $k_{5z}$ and its direction is sampled randomly with respect to $\vec{k}_{6\perp}$. 

Within this setup, the energy-momentum conservation of the $1\rightarrow 3$ splitting process is achieved using the on-shell condition for the final state jet parton ``3",
\begin{eqnarray}
\label{eq:onshell3}
p_3&=&(E_0-k_{5e}-k_{6e},\,-\vec{k}_{5\perp}-\vec{k}_{6\perp}, p_{0z}-k_{5z}-k_{6z}), \nonumber \\
p_3^2&=& m^2
\end{eqnarray}
which can be further expanded as 
\begin{align}
\label{eq:onshell3}
&(E_0-k_{5e}-k_{6e})^2-(k_{5\perp}^2+k_{6\perp}^2+2k_{5\perp}k_{6\perp}\cos\theta) \nonumber\\
&-(p_{0z}-k_{5z}-k_{6z})^2=m^2.
\end{align}
Using a randomly sampled $\theta$, we solve the above equation for $k_{5z}$. Two conditions are examined for the solution: (1) $k_{5e}>|k_{5z}|$ and (2) the intermediate jet parton between the two emitted gluons (``$1''$" in Fig.~\ref{fig:13conservation}) is time-like: $p_1''^2>m^2$. If both solutions of Eq.~(\ref{eq:onshell3}) satisfy these conditions, we choose the solution $k_{5z}$ that is closer to its original value obtained from the previous $2\rightarrow 3$ process. If there is no appropriate solution, we iterate the procedure above by re-sampling the angle $\theta$ or the four-momentum of ``6" until the solution is found or the number of trials exceeds a preset large value.

If more gluons are emitted, we repeat the procedure above to sample the four-momentum of the last gluon from Eq.~(\ref{eq:gluondistribution}) and adjust that of the second to last gluon to ensure the energy-momentum conservation of the entire $2\rightarrow 2+n$ process.

\bibliography{SCrefs}

\end{document}